\documentclass{nature}

\usepackage[british]{babel}
\usepackage[utf8]{inputenc}
\usepackage{babelbib}
\usepackage{url}
\usepackage{graphicx}
\usepackage{comment}
\usepackage{subfig}
\usepackage{calc}
\usepackage{floatflt}
\usepackage{amssymb, amsmath, amsthm}
\usepackage{tabularx}
\usepackage{multirow}
\usepackage{array}
\usepackage{xcolor}
\usepackage{bm}
\usepackage{stmaryrd}
\usepackage{stackrel}
\usepackage{algpseudocode}
\usepackage{algorithm}
\usepackage{rotating}
\usepackage{mwe,tikz}\usepackage[percent]{overpic}
\usepackage{float}
\usepackage{listings}
\usepackage[colorlinks=true,allcolors=blue]{hyperref}

\usepackage{numprint}
\usepackage{memhfixc}
\usepackage{makecell}
\usepackage{siunitx}
\usepackage{textcomp}
\usepackage{arydshln}

\usepackage{amsmath}

\usepackage{mathtools}

\title{De-occluding broadband metalens}

\author{Seungwoo Yoon$^{1*}$, Dohyun Kang$^{2*}$, Eunsue Choi$^{1}$, Sohyun Lee$^{3}$, Seoyeon Kim$^{3}$, Minho Choi$^{4,5}$, Hyeonsu Heo$^{2}$, Dong-Ha Shin$^{3}$,  Suha Kwak$^{3}$, Arka Majumdar$^{4,6}$, Junsuk Rho$^{2,7,8,9,10\dagger}$, Seung-Hwan Baek$^{1\dagger}$}

\usepackage{lipsum}

\hypersetup{pdftitle={De-occluding broadband metalens},pdfauthor={Seungwoo Yoon et al.}}
\begin{document}

\maketitle\par
\begingroup
\small
\setlength{\parskip}{4pt}
\noindent This author manuscript was accepted following peer review.
It precedes the publisher's copyediting and typesetting, omits subsequent
improvements and corrections, and is not the Version of Record.
\par
\noindent For the definitive published article, see:
Yoon, S., Kang, D., Choi, E. et al.
\textit{De-occluding broadband metalens}.
\textit{Nature Communications} \textbf{17}, 9808 (2026).
\href{https://doi.org/10.1038/s41467-026-76606-0}{doi:10.1038/s41467-026-76606-0}.
This accepted manuscript is shared under
\href{https://creativecommons.org/licenses/by-nc-nd/4.0/}{CC BY-NC-ND 4.0}.
Supplementary videos are available with the Version of Record.
\par
\endgroup

\definecolor{brightray}{rgb}{0.8,0.8,0.8}
\definecolor{Gray}{rgb}{0.5,0.5,0.5}
\definecolor{darkblue}{rgb}{0,0,0.7}
\definecolor{orange}{rgb}{1,.5,0} %
\definecolor{red}{rgb}{1,0,0} %
\definecolor{blue}{rgb}{0,0,1} %
\definecolor{darkgreen}{rgb}{0,0.7,0} %
\definecolor{darkred}{rgb}{0.7,0,0} %

\newcommand{\heading}[1]{\noindent\textbf{#1}}
\newcommand{\de}{\mathrm{d}}

\newcommand{\BEAS}{\begin{eqnarray*}}
\newcommand{\EEAS}{\end{eqnarray*}}
\newcommand{\BEA}{\begin{eqnarray}}
\newcommand{\EEA}{\end{eqnarray}}
\newcommand{\BEQ}{\begin{equation}}
\newcommand{\EEQ}{\end{equation}}
\newcommand{\BIT}{\begin{itemize}}
\newcommand{\EIT}{\end{itemize}}
\newcommand{\BNUM}{\begin{enumerate}}
\newcommand{\ENUM}{\end{enumerate}}

\newcommand{\BA}{\begin{array}}
\newcommand{\EA}{\end{array}}

\newcommand{\eg}{{\it e.g.}}
\newcommand{\ie}{{\it i.e.}}
\newcommand{\etc}{{\it etc.}}

\newcommand{\ones}{\mathbf 1}

\newcommand{\reals}{{\mbox{\bf R}}}
\newcommand{\integers}{{\mbox{\bf Z}}}
\newcommand{\eqbydef}{\mathrel{\stackrel{\Delta}{=}}}
\newcommand{\complex}{{\mbox{\bf C}}}
\newcommand{\symm}{{\mbox{\bf S}}}  %

\newcommand{\Span}{\mbox{\textrm{span}}}
\newcommand{\Range}{\mbox{\textrm{range}}}
\newcommand{\nullspace}{{\mathcal N}}
\newcommand{\range}{{\mathcal R}}
\newcommand{\Nullspace}{\mbox{\textrm{nullspace}}}
\newcommand{\Rank}{\mathop{\bf Rank}}
\newcommand{\Tr}{\mathop{\bf Tr}}
\newcommand{\diag}{\mathop{\bf diag}}
\newcommand{\lambdamax}{{\lambda_{\rm max}}}
\newcommand{\lambdamin}{\lambda_{\rm min}}

\newcommand{\Expect}{\mathop{\bf E{}}}
\newcommand{\Prob}{\mathop{\bf Prob}}
\newcommand{\erf}{\mathop{\bf erf}}

\newcommand{\Co}{{\mathop {\bf Co}}}
\newcommand{\co}{{\mathop {\bf Co}}}
\newcommand{\dist}{\mathop{\bf dist{}}}
\newcommand{\Ltwo}{{\bf L}_2}
\newcommand{\QED}{~~\rule[-1pt]{8pt}{8pt}}\def\qed{\QED}
\newcommand{\approxleq}{\mathrel{\smash{\makebox[0pt][l]{\raisebox{-3.4pt}{\small$\sim$}}}{\raisebox{1.1pt}{$<$}}}}
\newcommand{\epi}{\mathop{\bf epi}}

\newcommand{\vol}{\mathop{\bf vol}}
\newcommand{\Vol}{\mathop{\bf vol}}
\newcommand{\Card}{\mathop{\bf card}}

\newcommand{\dom}{\mathop{\bf dom}}
\newcommand{\aff}{\mathop{\bf aff}}
\newcommand{\cl}{\mathop{\bf cl}}
\newcommand{\Angle}{\mathop{\bf angle}}
\newcommand{\intr}{\mathop{\bf int}}
\newcommand{\relint}{\mathop{\bf rel int}}
\newcommand{\bd}{\mathop{\bf bd}}
\newcommand{\vect}{\mathop{\bf vec}}
\newcommand{\dsp}{\displaystyle}
\newcommand{\foequal}{\simeq}
\newcommand{\VOL}{{\mbox{\bf vol}}}
\newcommand{\xopt}{x^{\rm opt}}

\newcommand{\Xb}{{\mbox{\bf X}}}
\newcommand{\xst}{x^\star}
\newcommand{\varphist}{\varphi^\star}
\newcommand{\lambdast}{\lambda^\star}
\newcommand{\Zst}{Z^\star}
\newcommand{\fstar}{f^\star}
\newcommand{\xstar}{x^\star}
\newcommand{\xc}{x^\star}
\newcommand{\lambdac}{\lambda^\star}
\newcommand{\lambdaopt}{\lambda^{\rm opt}}

\newcommand{\geqK}{\mathrel{\succeq_K}}
\newcommand{\gK}{\mathrel{\succ_K}}
\newcommand{\leqK}{\mathrel{\preceq_K}}
\newcommand{\lK}{\mathrel{\prec_K}}
\newcommand{\geqKst}{\mathrel{\succeq_{K^*}}}
\newcommand{\gKst}{\mathrel{\succ_{K^*}}}
\newcommand{\leqKst}{\mathrel{\preceq_{K^*}}}
\newcommand{\lKst}{\mathrel{\prec_{K^*}}}
\newcommand{\geqL}{\mathrel{\succeq_L}}
\newcommand{\gL}{\mathrel{\succ_L}}
\newcommand{\leqL}{\mathrel{\preceq_L}}
\newcommand{\lL}{\mathrel{\prec_L}}
\newcommand{\geqLst}{\mathrel{\succeq_{L^*}}}
\newcommand{\gLst}{\mathrel{\succ_{L^*}}}
\newcommand{\leqLst}{\mathrel{\preceq_{L^*}}}
\newcommand{\lLst}{\mathrel{\prec_{L^*}}}

\newtheorem{theorem}{Theorem}[section]
\newtheorem{corollary}{Corollary}[theorem]
\newtheorem{lemma}[theorem]{Lemma}
\newtheorem{proposition}[theorem]{Proposition}

\newenvironment{algdesc}%
{\begin{quote}}{\end{quote}}

\def\figbox#1{\framebox[\hsize]{\hfil\parbox{0.9\hsize}{#1}}}

\makeatletter
\long\def\@makecaption#1#2{
   \vskip 9pt
   \begin{small}
   \setbox\@tempboxa\hbox{{\bf #1:} #2}
   \ifdim \wd\@tempboxa > 5.5in
        \begin{center}
        \begin{minipage}[t]{5.5in}
        \addtolength{\baselineskip}{-0.95pt}
        {\bf #1:} #2 \par
        \addtolength{\baselineskip}{0.95pt}
        \end{minipage}
        \end{center}
   \else
    \hbox to\hsize{\hfil\box\@tempboxa\hfil}
   \fi
   \end{small}\par
}
\makeatother

\newcounter{oursection}
\newcommand{\oursection}[1]{
 \addtocounter{oursection}{1}
 \setcounter{equation}{0}
 \clearpage \begin{center} {\Huge\bfseries #1} \end{center}
 {\vspace*{0.15cm} \hrule height.3mm} \bigskip
 \addcontentsline{toc}{section}{#1}
}
\newcommand{\oursectionf}[1]{  %
 \addtocounter{oursection}{1}
 \setcounter{equation}{0}
 \foilhead[-.5cm]{#1 \vspace*{0.8cm} \hrule height.3mm }
 \LogoOn
}
\newcommand{\oursectionfl}[1]{  %
 \addtocounter{oursection}{1}
 \setcounter{equation}{0}
 \foilhead[-1.0cm]{#1}
 \LogoOn
}

\newcommand{\Mat}[1]    {{\ensuremath{\mathbf{\uppercase{#1}}}}} %
\newcommand{\Vect}[1]   {{\ensuremath{\mathbf{\lowercase{#1}}}}} %
\newcommand{\Vari}[1]   {{\ensuremath{\mathbf{\lowercase{#1}}}}} %
\newcommand{\Id}				{\mathbb{I}} %
\newcommand{\Diag}[1] 	{\operatorname{diag}\left({ #1 }\right)} %
\newcommand{\Opt}[1] 	  {{#1}_{\text{opt}}} %
\newcommand{\CC}[1]			{{#1}^{*}} %
\newcommand{\Op}[1]     {\Mat{#1}} %
\newcommand{\mini}[1] {{\mbox{argmin}}_{#1} \: \: } %
\newcommand{\argmin}[1] {\underset{{#1}}{\mathop{\rm argmin}} \: \: } %
\newcommand{\argmax}[1] {\underset{{#1}}{\mathop{\rm argmax}} \: \: } %
\newcommand{\minimize}{\mathop{\rm minimize} \: \:}
\newcommand{\minimizeu}[1]{\underset{{#1}}{\mathop{\rm minimize}} \: }
\newcommand{\grad}      {\nabla}
\newcommand{\kron}{\otimes} %

\newcommand{\gradt}     {\grad_\z}
\newcommand{\gradx}     {\grad_\x}
\newcommand{\step}      {\text{\textbf{step}}}
\newcommand{\prox}[1]   {\mathbf{prox}_{#1}}
\newcommand{\ind}[1]    {\operatorname{ind}_{#1}}
\newcommand{\proj}[1]   {\Pi_{#1}}
\newcommand{\pointmult}{\odot} %
\newcommand{\rr}   {\mathcal{R}}

\newcommand{\Basis}{\Mat{D}}         		%
\newcommand{\Corr}{\Mat{C}}             %
\newcommand{\conv}{\ast} %
\newcommand{\meas}{\Vect{b}}            %
\newcommand{\Img}{I}                    %
\newcommand{\img}{\Vect{i}}             %
\newcommand{\vv}{\Vect{v}}
\newcommand{\p}{\Vect{p}}
\newcommand{\Splitvar}{T}                %
\newcommand{\splitvar}{\Vect{t}}         %
\newcommand{\Splitbasis}{J}                %
\newcommand{\splitbasis}{\Vect{j}}         %
\newcommand{\var}{\Vari{z}}

\newcommand{\FT}[1]			{\mathcal{F}\left( {#1} \right)} %
\newcommand{\IFT}[1]			{\mathcal{F}^{-1}\left( {#1} \right)} %

\newcommand{\func}{f}
\newcommand{\fMat}{\Mat{K}}

\newcommand{\avar}{\Vari{v}}
\newcommand{\aspvar}{\Vari{z}}

\newcommand{\mask}{\Mat{M}}

\newcommand{\Pen}      		{F} %
\newcommand{\cardset}     {\mathcal{C}}
\newcommand{\Dat}      		{G} %
\newcommand{\Reg}      		{\Gamma} %

\newcommand{\Trans}{\mathbf{\uppercase{T}}} %
\newcommand{\Ph}{\mathbf{\uppercase{\Phi}}} %

\newcommand{\Tvec}{\Vect{T}} %
\newcommand{\Bvec}{\Vect{B}} %

\newcommand{\Wt}{\Mat{W}} %

\newcommand{\Perm}{\Mat{P}} %

\newcommand{\DiagFactor}[1]     {\Mat{O}_{ #1 }}  %

\newcommand{\Proj}{\Mat{P}}             %

\newcommand{\Vector}[1]{\mathbf{#1}}
\newcommand{\Matrix}[1]{\mathbf{#1}}
\newcommand{\Tensor}[1]{\boldsymbol{\mathscr{#1}}}
\newcommand{\TensorUF}[2]{\Matrix{#1}_{(#2)}}

\newcommand{\MatrixKP}[1]{\Matrix{#1}_{\otimes}}
\newcommand{\MatrixKPN}[2]{\Matrix{#1}_{\otimes}^{#2}}

\newcommand{\MatrixKRP}[1]{\Matrix{#1}_{\odot}}
\newcommand{\MatrixKRPN}[2]{\Matrix{#1}_{\odot}^{#2}}

\newcommand{\HP}{\circ}
\newcommand{\HD}{\oslash}

\newcommand{\leftDB}{\left[ \! \left[}
\newcommand{\rightDB}{\right] \! \right]}

\newcommand{\transpose}{T}

\newcommand*\sstrut[1]{\vrule width0pt height0pt depth#1\relax}

\newcommand{\inlineeqnum}{\refstepcounter{equation}~~\mbox{(\theequation)}}
\newcommand{\eqname}[1]{\tag*{#1~(\theequation)}\refstepcounter{equation}}

\newcommand{\lambdas}{\boldsymbol{\lambda}}
\newcommand{\alb}{\boldsymbol{\alpha}} 	%
\newcommand{\depth}{\boldsymbol{z}} 	%
\newcommand{\albi}{\alpha} 	%
\newcommand{\depthi}{z} 	%
\newcommand{\ambient}{s}
\newcommand{\jitter}{w}
\newcommand{\z}{\Vect{z}} 							%
\newcommand{\x}{\Vect{x}}             	%
\newcommand{\y}{\Vect{y}}             	%
\newcommand{\Kvar}{\Mat{K}}
\newcommand{\lagrangemult}{\boldsymbol{\nu}}
\newcommand{\scaledlagrange}{\Vect{u}}
\newcommand{\eps}{\epsilon}
\newcommand{\vp}{\Vect{v}}

\begin{affiliations}
 \item Department of Computer Science and Engineering, Pohang University of Science and Technology (POSTECH), Pohang 37673, Republic of Korea
 \item Department of Mechanical Engineering, Pohang University of Science and Technology (POSTECH), Pohang 37673, Republic of Korea
 \item Graduate School of Artificial Intelligence, Pohang University of Science and Technology (POSTECH), Pohang 37673, Republic of Korea
 \item Department of Electrical and Computer Engineering, University of Washington, Seattle, 98195, WA, USA
 \item Department of Electrical Engineering, Ulsan National Institute of Science and Technology, Ulsan 44919, Republic of Korea.
 \item Department of Physics, University of Washington, Seattle, 98195, WA, USA
 \item Department of Chemical Engineering, Pohang University of Science and Technology (POSTECH), Pohang 37673, Republic of Korea
  \item Department of Electrical Engineering, Pohang University of Science and Technology (POSTECH), Pohang 37673, Republic of Korea
 \item POSCO-POSTECH-RIST Convergence Research Center for Flat Optics and Metaphotonics, Pohang 37673, Republic of Korea
 \item National Institute of Nanomaterials Technology (NINT), Pohang 37673, Republic of Korea
 \item [$*$] Equal contribution
 \item [$\dagger$] Corresponding author. E-mail: shwbaek@postech.ac.kr, jsrho@postech.ac.kr
\end{affiliations}
\clearpage

\section*{Abstract}
Obstructions such as raindrops, fences, or dust degrade images, especially when cleaning is infeasible. Conventional approaches use bulky compound-lens arrays or computational inpainting, compromising compactness or fidelity. Metalenses promise compact imaging, but achieving both broadband and obstruction-free imaging remains challenging, since a metalens cannot simultaneously focus distant scenes and defocus nearby occlusions across a broadband spectrum. We introduce {a de-occluding broadband metalens} that suppresses obstructions while enabling broadband imaging with a learned phase profile. Our approach divides the spectrum of each color channel into pass and stop bands with multi-band spectral filtering; the metalens focuses light from far objects through pass bands, then light from nearby obstructions falls in the stop bands and is rejected. A neural network further enhances imaging quality. Compared with a conventional hyperbolic metalens under obstructions, our system improves PSNR by 32.29\% and raises performance on object detection and two semantic segmentation benchmarks by 13.54\%, 48.45\% and 20.35\%, respectively. This promises robust obstruction-free sensing and vision for compact systems, including mobile robots, drones, and endoscopes.

\section*{Introduction}
Obstructions such as fences, leaves, dirt, or raindrops frequently occlude distant scenes in everyday imaging\cite{you2013adherent, brophy2023review, uricar2021let, liu2008image, nathan2022through}. These occlusions degrade the imaging quality and reliability of downstream vision applications, especially in autonomous or compact systems where manual removal of obstacles is infeasible--for instance, cameras on drones\cite{nathan2022through, chang2024uav}, mobile robots\cite{uricar2021let, uvrivcavr2019soilingnet}, or endoscopes\cite{nabeel2022effective, wang2018variational}. Obstruction-free imaging aims to recover distant scenes covered by near-depth obstacles.
Existing computational approaches tend to hallucinate the occluded content rather than capture the true scene\cite{hirohashi2020removal, zhang2021automatic, gupta2021robust, hao2019learning}.
Optical solutions using multiple compound lenses, or synthetic apertures, can physically remove near-depth obstructions by depth-of-field control\cite{levoy2004synthetic, joshi2007synthetic, pei2016all, pei2019occluded}.
However, their bulky and costly designs hinder integration into compact imaging systems (Supplementary Note~3).

Flat diffractive optics--particularly metalenses composed of subwavelength meta-atoms--offer a promising alternative to bulky compound lenses, as ultra-thin and lightweight imaging elements\cite{chen2022meta, liu2023underwater, froch2023real, liu2024stereo}. Yet, achieving both broadband imaging and obstruction-free imaging with a single metalens remains fundamentally constrained by depth–wavelength symmetry; the point spread function (PSF) change induced by a wavelength shift is similar to that of an appropriate depth shift of a point light source\cite{colburn2019optical, froch2025beating, tan20213d, xu2025polarization, bayati2022inverse, baek2021single}. 
This depth-wavelength symmetry prevents simultaneous broadband imaging and defocusing for near-depth obstructions. 
While hybrid diffractive-compound refractive design\cite{shi2022seeing, richards2023hybrid} or dispersion engineering\cite{chen2018broadband, shrestha2018broadband, chen2019broadband, hu2023asymptotic, chang2024achromatic, hou2025high} mitigate this phenomenon, bulky refractive elements negate the advantage of compactness, and dispersion control is typically restricted to micrometer-scale apertures or extremely low numerical apertures, falling short of practical imaging needs (Supplementary Note~2).

This work introduces {a de-occluding broadband metalens} for simultaneous {obstruction-free and broadband imaging} (Fig.~\ref{fig:figure1}).
We first derive an analytical model describing the depth–wavelength symmetry in diffractive lenses.
Based on this understanding, we split the spectrum of each RGB channel into pass and stop bands using a multi-band spectral filter. This spectral splitting extends the metalens design space in a depth-wavelength decoupled manner. We then computationally learn a metalens design to leverage these pass bands to focus distant scenes, while blurring near-depth obstructions via stop bands. 

Our {de-occluding broadband metalens} achieves broadband obstruction-free imaging, yielding up to 23.90\% PSNR improvement on unobstructed scenes compared to conventional hyperbolic metalenses, and 32.29\% or 11.45\% PSNR improvement on obstructed scenes compared to hyperbolic or broadband metalenses without a de-occluding approach (broadband baseline), respectively. This capability also enhances downstream performance under obstruction, improving object detection by +13.54\%/+14.12\% in mAP and semantic segmentation by +48.45\,\%/+23.67\,\% in IoU, and +20.35\%/+21.00\% in mIoU, compared to the two baseline metalenses, underscoring the efficacy of our approach. We believe that {the de-occluding broadband metalens} promises robust obstruction-free imaging and perception in compact platforms such as mobile robots, drones, and endoscopes.

\section*{Results}
\subsection{Depth–wavelength symmetry.}

Let $P_{\lambda, z}$ denote the PSF of a diffractive lens for a point light source located at depth $z$ and wavelength of $\lambda$. The {depth–wavelength symmetry} refers to the observation\cite{colburn2018metasurface, froch2025beating, tan20213d, xu2025polarization, bayati2022inverse, baek2021single} that the PSF remains invariant under a coupled shift in wavelength $\Delta\lambda$ and depth $\Delta z$:
\begin{equation}
    P_{\lambda+\Delta\lambda, z} \approx P_{\lambda, z+\Delta z}.
    \label{eq:PSF_depth_wavelength_similarity}
\end{equation}

Here we present the first analytical model of depth–wavelength symmetry that explicitly relates the wavelength shift $\Delta\lambda$ to the corresponding depth shift $\Delta z$.
This model provides the theoretical foundation for the design of our {de-occluding broadband metalens}.
We begin by considering the ideal phase profile that focuses a plane wave to a focal length $f$ for all wavelengths $\lambda$. In practice, this condition is satisfied only at a design wavelength $\lambda_{\mathrm d}$. For wavelengths $\lambda > \lambda_{\mathrm d}$, the resulting mismatched phase can be interpreted, under the paraxial approximation, as the conjugate phase of the spherical phase of a point source located at the corresponding depth $z$. Consequently, if the incident light of wavelength $\lambda$ originates from this depth $z$ rather than infinity, its spherical wavefront counterbalances the chromatic phase mismatch, thereby yielding the sharp focus at $f$, as illustrated in Fig.~\ref{fig:figure2}a. This relationship leads to our depth–wavelength symmetry model,
\begin{equation}
    z = \frac{\lambda_{\mathrm d} f}{\lambda - \lambda_{\mathrm d}}.
    \label{eq:depth_wavelength_relationship}
\end{equation}

The complete derivation is described in Supplementary Note~1. Fig.~\ref{fig:figure2}b shows the focal intensity distribution of the PSFs produced by a hyperbolic metalens designed at $\lambda_{\mathrm{d}} = 450$\,nm, with focal length of 4\,mm and aperture diameter of 2.516\,mm. Using the depth–wavelength symmetry model in Eq.~\eqref{eq:depth_wavelength_relationship}, we overlay the predicted depth–wavelength correspondence, which shows strong agreement between the analytical prediction and the numerically observed focal behavior. We provide additional validation of the model, along with a detailed discussion on its applicability in Supplementary Notes~1 and 2.

Based on the depth–wavelength symmetry model of Eq.~\eqref{eq:depth_wavelength_relationship}, the relationship between the wavelength shift $\Delta\lambda$ and the depth shift $\Delta z$ in Eq.~\eqref{eq:PSF_depth_wavelength_similarity} can be obtained as,
\begin{equation}
    \Delta\lambda
    = \lambda_{\mathrm d} f
    \left(
        \frac{1}{z}
        - \frac{1}{z-\Delta z}
    \right).
    \label{eq:delta_lambda_delta_z}
\end{equation}

These expressions imply that if a metalens produces a sharp PSF for a point light source of wavelength $\lambda$ from a far-depth $z_{\text{far}}$, then the PSF of a point light source corresponding to a nearer depth $z_{\text{near}}$ becomes similar to the far-depth PSF at a shifted wavelength $\lambda + \Delta\lambda$, where the shift $\Delta\lambda$ is obtained by Eq.~\eqref{eq:delta_lambda_delta_z} with $\Delta z = z_{\text{near}} - z_{\text{far}}$.

\subsection{Design of de-occluding broadband metalens.}
De-occluding broadband imaging requires a metalens that produces sharp PSFs for distant targets at far depth, $z_{\text{far}}$ (typically $z_{\text{far}}\ge 0.5\,\text{m}$), while simultaneously inducing strong blur for obstructions located at near depth $z_{\text{near}} \simeq 0.045\,\text{m}$. The normally constant value of $z_\text{near}$ models the prevalent scenario where obstructions adhere directly onto the optical window, such as a cover glass. 
Significantly, the depth-wavelength symmetry fundamentally limits this simultaneous far-depth sharp imaging and near-depth defocusing in the visible.
If a metalens forms a sharp far-depth PSF at wavelength $\lambda$, the corresponding near-depth PSF at the shifted wavelength $\lambda + \Delta\lambda$, also becomes sharp where the wavelength shift $\Delta\lambda$ is given by Eq.~\eqref{eq:delta_lambda_delta_z} with $\Delta z = z_{\text{near}} - z_{\text{far}}$.
If this shifted wavelength $\lambda + \Delta\lambda$ remains within the sensor's spectral response range, light from near-depth obstructions is also sharply captured, preventing de-occlusion.

To overcome this limitation, we introduce a split-spectrum design. 
The spectrum of each RGB channel of the sensor $c \in \{\text{R, G, B}\}$ is divided into a high-transmission pass band $\Lambda_{\text{pass}}^c$ and a zero-transmission stop band $\Lambda_{\text{stop}}^c$ using a multi-band spectral filter, as illustrated in Fig.~\ref{fig:figure3}b,c.
We design our metalens to form focused far-depth PSFs for wavelengths in $\Lambda_{\text{pass}}^c$.
At near depth, these focused PSFs shift to wavelengths $\lambda + \Delta\lambda$ that predominantly fall into $\Lambda_{\text{stop}}^c$ and are therefore rejected (Fig.~\ref{fig:figure3}d). This spectral filtering prevents the sharp far-depth focusing behavior from transferring to the near depth, and thus we can explicitly optimize the metalens to yield strongly defocused near-depth responses over $\Lambda_{\text{pass}}^{c}$. 
This split-spectrum mechanism thus opens up the possibility of achieving sharp far-depth PSFs and defocused near-depth PSFs across the visible spectrum, providing a physical pathway toward de-occluding broadband imaging. Detailed validation and the rationale for band selection are provided in Supplementary Note~1.

To find a de-occluding broadband metalens design that enables obstruction-free broadband imaging, we learn the metalens design map $\theta(x,y)$, representing the orientation of each geometric-phase meta-atom at pupil-plane coordinate $(x,y)$ (Fig.~\ref{fig:figure3}a).
To this end, we construct two differentiable components:
(i) a differentiable PSF simulator,
\begin{equation}
    P_z^c = f_{\text{psf}}(\theta, z, c),
\end{equation}
which generates the PSF at depth~$z$ for a sensor color channel~$c$, and
(ii) a differentiable image simulator,
\begin{equation}
    I_{\text{captured}} = f_{\text{img}}\big(\theta, I_{\text{clean}}, I_{\text{obs}}\big),
\end{equation}
which synthesizes the captured RGB image from an unobstructed far-depth target image $I_{\text{clean}}$ and a near-depth obstruction image $I_{\text{obs}}$. We perform importance sampling over wavelengths to reduce the computational burden of the spectral integration, using a distribution proportional to the system spectral sensitivity for each color channel $c$. 
Both $f_{\text{psf}}$ and $f_{\text{img}}$ are differentiable with respect to $\theta$, enabling gradient-based optimization:
\begin{equation}
    \underset{\theta}{\text{minimize}}\;
        \mathcal{L}_{\text{img}}\!\left(I_{\text{captured}}, I_{\text{clean}}\right)
        + \mathcal{L}_{\text{psf}}\!\left( P_{z_{\text{far}}}\right),
    \label{eq:metalens_optimization}
\end{equation}
where the objective function $\mathcal{L}_{\text{img}}$ enforces similarity of $I_\text{captured}$ to the target far-depth image $I_{\text{clean}}$, and $\mathcal{L}_{\text{psf}}$ encourages sharp and high efficiency far-depth PSFs $P_{z_{\text{far}}}$. We solve Eq.~\eqref{eq:metalens_optimization} via a first-order optimizer using $I_\text{clean}$ randomly sampled from a high-resolution image dataset DIV2K\cite{Agustsson_2017_CVPR_Workshops} and randomly generated obstruction images $I_\text{obs}$. Additional implementation details are provided in Supplementary Note~4.

Fig.~\ref{fig:figure3}e shows the training loss curve with the corresponding simulated images, demonstrating the capability of de-occluding broadband imaging. Fig.~\ref{fig:figure3}d further demonstrates this by showing the $x-\lambda$ scans of the far-depth and near-depth PSFs from the learned design, exhibiting sharp far-depth PSFs and blurry near-depth PSFs for pass-band wavelengths and the opposite behavior for stop-band wavelengths. %

\subsection{Characterization of de-occluding broadband metalens.}  
We employ geometric-phase anisotropic meta-atoms whose structural parameters were optimized via rigorous coupled-wave analysis to maximize the average conversion efficiency across the pass bands of the multi-band spectral filter, centered at $457\,$nm, $530\,$nm, and $628\,$nm. Silicon nitride ($\text{SiN}_x$) is chosen as the structural material owing to its sufficiently high refractive index and low extinction coefficient across the visible spectrum (Supplementary Note~6). The selected geometry--period $p = 395\,$nm, height $h = 700\,$nm, width $w = 305\,$nm, and length $l = 125\,$nm--yields average conversion efficiencies of $77.2\%$, $73.0\%$, and $54.4\%$ across the respective pass bands (Supplementary Note~7). This high efficiency is uniformly maintained, with low standard deviations of the conversion efficiencies across wavelengths within each pass band ($2.2\%$, $0.2\%$, and $4.9\%$, respectively), confirming their broadband characteristics (Supplementary Table S5)\cite{berry1984quantal}. 
Beyond the achieved optical performance, the selected single-cell PB-phase design employs simple meta-atom structures that facilitate fabrication. Supplementary Note~15 provides a detailed discussion of spectral throughput, polarization throughput, sensitivity, robustness, and potential limitations. Based on these meta-atom characteristics, we design and fabricate three types of metalenses shown in Fig.~\ref{fig:figure4}b:
(1) a de-occluding broadband metalens (Ours),
(2) a broadband metalens without the de-occluding strategy (Broadband), and
(3) a hyperbolic-phase metalens designed for $532\,$nm (Hyperbolic). All lenses have the same configuration, with a $4\,$mm focal length and a $2.516\,$mm aperture diameter, and were fabricated using high-speed electron-beam lithography (EBL) (Fig.~\ref{fig:figure4}a; see Methods and Supplementary Note~8).

We captured near-depth and far-depth PSFs across wavelengths from 430 to 645\,nm with a 5\,nm step, for the three metalenses. 
The wavelength-dependent intensity profiles of the focal spots, normalized to the total intensity, are shown in Fig.~\ref{fig:figure4}c.
Across the pass bands of the visible spectrum, the hyperbolic metalens suffers from incomplete broadband focusing capability as it is designed for a single wavelength. 
The broadband metalens without the de-occluding strategy improves broadband focusing relative to the hyperbolic baseline but does not achieve simultaneous near-depth blur, thereby failing obstruction-free imaging.
In contrast, the de-occluding broadband metalens maintains sharp focus for far depth while blurring near depth response.
The corresponding two-dimensional PSF profiles, normalized to their peak intensities, are shown in Fig.~\ref{fig:figure4}d.
Within the pass bands highlighted by the colored boxes, our de-occluding broadband metalens produces sharp focal spots at the far depth while exhibiting strong defocus at the near depth, confirming the effectiveness of the split-spectrum design for de-occluding broadband imaging. More details of PSF measurement are described in the Supplementary Notes~9 and 12. 

\subsection{De-occluding broadband imaging.}
We evaluate the de-occluding and broadband imaging performance of the three metalenses.
To recover high-quality images from the raw optical measurements, we further employ a reconstruction network\cite{Kim_2024_CVPR} trained individually for each metalens.
The reconstruction network used throughout this work is trained separately from the metalens optimization so that the de-occluding behavior is enforced at the raw sensor-capture level, ensuring that obstruction suppression originates from the optical design rather than being compensated by the computational backend. The rationale for this decoupled strategy is discussed in Supplementary Note~4.

Fig.~\ref{fig:figure5} shows both the raw sensor measurements and the reconstructed images.
The hyperbolic-phase metalens exhibits strong chromatic aberration and visible obstruction artifacts, demonstrating its limitation in achieving broadband and obstruction-free imaging. The broadband metalens without the de-occluding strategy improves broadband performance but fails to suppress near-depth obstructions that inevitably arise from the depth-wavelength symmetry. In contrast, the de-occluding broadband metalens leaves only faint obstruction traces in the raw measurements, allowing the neural network to reconstruct underlying structures reliably rather than hallucinating them. 
Quantitatively, the broadband metalens without a de-occluding attains a PSNR of 18.79\,dB, corresponding to an 18.7\,\% improvement over the conventional hyperbolic design under obstruction. The de-occluding broadband metalens achieves a PSNR of 20.94\,dB, equivalent to a 32.29\,\% increase, confirming the advantage of our spectrally separated design. See Supplementary Notes~5, 10, 11, 14, 16, and 17, as well as Supplementary Videos~1--4 for details on the neural network, metric measurements, additional experimental results, dynamic near-depth obstruction videos, and comparisons with computational inpainting baselines.

\subsection{Downstream vision tasks.}
We further evaluate downstream vision tasks on images captured with each fabricated metalens to assess the practical utility of de-occluding broadband imaging.
Off-the-shelf vision models are employed without fine-tuning on our metalens-captured data to demonstrate the universality of our imaging approach. 
We benchmark three representative scenarios: (1) object detection\cite{khanam2024yolov11} on drone-captured aerial imagery using VisDrone\cite{zhu2021detection}, (2) semantic segmentation for medical endoscopy\cite{zhao2026m2snet} on Kvasir-SEG\cite{jha2020kvasir}, and (3) semantic segmentation for autonomous driving\cite{wang2023internimage} on Cityscapes\cite{Cordts2016Cityscapes}. For controlled quantitative evaluation with annotations, these benchmark tasks were performed by optically re-capturing printed validation images from the corresponding datasets through the fabricated metalenses, while complementary real-scene downstream-vision demonstrations are provided in Supplementary Note~17 as well as Supplementary videos~4 and 5. 
As shown in Fig.~\ref{fig:figure6}, our de-occluding broadband metalens consistently achieves the best performance across all tasks: it attains an mAP of 0.1704 on VisDrone (vs. 0.0350/0.0292 for hyperbolic/broadband), an IoU of 0.8317 on Kvasir-SEG (vs. 0.3472/0.5950), and an mIoU of 0.6701 on Cityscapes (vs. 0.4666/0.4601) under task-specific obstruction scenarios. These results establish our de-occluding broadband metalens as a practical imaging solution for embedded and unmanned platforms such as drones, mobile robots, and endoscopic probes, where near-depth obstructions are prevalent and physical cleaning is often infeasible. We refer to Supplementary Note~11 for further details on the vision tasks.

\section*{Discussion}
This work presents a single-shot obstruction-free imaging method using a de-occluding broadband metalens. We derive an analytical model for the depth-wavelength relationship of dispersive metalenses and harness it for broadband obstruction-free imaging through split-spectrum filtering of focused near-depth lights. We learn the metalens design via a differentiable framework based on the extended design space provided by the spectrum splitting, achieving PSNR of 20.94\,dB under obstructed conditions. This corresponds to a 32.29\,\% or 11.45\,\% improvement compared to hyperbolic or broadband metalens without a de-occluding strategy, respectively. Importantly, this obstruction-suppression capability does not come at the cost of the imaging performance; our system maintains high-fidelity imaging even for unobstructed scenes, achieving a PSNR of 23.41\,dB, which corresponds to a 23.90\,\% improvement over the conventional hyperbolic metalens (refer to Supplementary Note~10, 13, and 14). 
Furthermore, our de-occluding imaging directly translates to robust computer vision performance; our method achieves absolute gains of +13.54\,\%/+14.12\,\%\,mAP, 48.45\,\%/23.67\,\%\,IoU, and 20.35\,\%/21.00\,\%\,mIoU compared to the two baseline metalenses.

The present implementation targets a finite near-depth range centered around ($z_\text{near}\simeq0.045\,\text{m}$). This operating depth range can be extended by using a narrower pass-band design, with trade-offs including reduced spectral transmission, as analyzed in Supplementary Note~1. The current metalens was optimized using a normal-incidence PSF model and has a finite field of view, and spatially varying aberrations are mitigated by positional encoding in the reconstruction network. Further details on the field of view and the reconstruction network are discussed in the Methods section, Supplementary Notes~5 and 9.

Several avenues for future work remain.
First, our analytical model for the depth-wavelength symmetry and split-spectrum strategy is also applicable to other diffractive optics applications, potentially providing a mathematical foundation to improve performance in color holography\cite{park202536, meng2025ultranarrow}, depth sensing\cite{shen2023monocular, shen2025extended}, and hyperspectral imaging\cite{jeon2019compact, baek2021single}. 
 Second, while the current implementation employs a commercial multi-band filter, jointly optimizing the metalens and spectral filter—or directly incorporating spectral filtering into meta-atom transmittance—could expand the design space, improve light efficiency, and enable monolithic implementations\cite{lin2025resonant}.
Third, while our current geometric-phase design prioritizes robust large-area manufacturing using simple meta-atom structures\cite{kim2023scalable, so2023multicolor}, it requires additional polarization optics. However, our split-spectrum principle fundamentally relies on depth–wavelength symmetry rather than the geometric-phase mechanism itself; future implementations utilizing propagation-phase metalenses could potentially eliminate these optics and further reduce the overall optical-stack thickness (Supplementary Note~1).
Finally, explicitly incorporating fabrication constraints into the inverse design of complex free-form geometries represents a promising pathway toward maximizing both optical performance and scalability.
We anticipate that our obstruction-free broadband imaging is an important step towards robust sensing and vision, not only for unmanned systems like mobile robots, drones, and endoscopes, where manual obstruction removal is challenging, but also for applications with structure-induced occlusions, such as under-display cameras and shielded cameras.

\begin{methods}

\subsection{Fabrication specification.}  
A 700-nm-thick silicon nitride ($\text{SiN}_x$) film was deposited on a glass substrate, pre-cleaned with deionized (DI) water, acetone, and isopropyl alcohol (IPA), using plasma-enhanced chemical vapor deposition (PECVD, BMR Technology HiDep-SC). The designed nanopattern was subsequently exposed onto a positive photoresist (ZEP 520A) using a high-speed electron beam lithography system (ELS-BODEN, ELIONIX) operated at an acceleration voltage of 50 kV and a beam current of 10 nA. The exposed nanopattern was developed in ZEP developer (ZED-N50) at 0°C for 1 minute. A 60-nm-thick chromium (Cr) layer was then deposited via electron beam evaporation (KVT, KVE-ENS4004), serving as an etching mask. The pattern was transferred into $\text{SiN}_x$ layer by dry etching (DRM85DD, TEL) following a lift-off process. Finally, the remaining Cr mask was removed using a commercial Cr etchant (CR-7).

\subsection{Baseline metalens design.}
The hyperbolic metalens is designed at a wavelength of 532\,nm. The Broadband baseline is trained using the same optimization pipeline as the de-occluding metalens, but without the near-depth obstruction model and without the multi-band bandpass filter. Therefore, it targets sharp far-depth focusing across the broadband visible spectrum. Further details on the metalens training pipeline are described in Supplementary Note~4.

\subsection{Imaging setup.}
For controlled quantitative evaluation, imaging experiments and vision benchmarks in the main paper were performed using a benchtop setup enabling repeatable lens exchange, ground-truth acquisition, and fixed obstruction geometry. Representative compact-prototype results are provided as an integration-feasibility demonstration in Supplementary Notes~14 and 17 as well as Supplementary videos~4 and 5. Details on the benchtop imaging setup and compact camera system specifications—including optical-stack dimensions, estimated housing volume, and power consumption—are provided in Supplementary Notes~9. 

\subsection{Image reconstruction specifications.}
We adopt LocalNet\cite{Kim_2024_CVPR} as the image reconstruction backend. High-resolution printed targets containing rich texture and color variations from the DIV2K dataset\cite{Agustsson_2017_CVPR_Workshops} were used to train the neural network. Obstructed scenes were captured with the metalenses, while reference ground-truth scenes (without obstructions) were acquired with a compound lens. A focal length of $f=8$\,mm (twice that of the metalenses) was used for the ground truth captures to achieve better object space resolution\cite{engelberg2022generalized}. The compound-lens captured ground-truth images serve as supervision to correct residual aberrations and remove remaining obstruction artifacts. We further incorporate positional encoding schemes\cite{vaswani2017attention, mildenhall2021nerf} into the neural network\cite{Kim_2024_CVPR} to address spatially varying aberrations. Specifically, we augment the network input with a fixed 19-channel radial positional encoding vector, which is defined as 
\[
\mathbf{e}_{\mathrm{pos}}(x,y)
=
\left[
\begin{gathered}
\sin(\kappa_0\pi r),\,\ldots,\,\sin(\kappa_7\pi r),\\[4pt]
\cos(\kappa_0\pi r),\,\ldots,\,\cos(\kappa_7\pi r),\\[4pt]
\dfrac{y}{\sqrt{2}},\,
\dfrac{x}{\sqrt{2}},\,
\dfrac{r}{\sqrt{2}}
\end{gathered}
\right]^{\mathsf T}
\in\mathbb{R}^{19}.
\]
where $r=r(x,y)$, $\kappa_i=2^i$, and $i=0,\ldots,7$.

The neural networks take 946 MiB of VRAM and 41.41 milliseconds (24.15 frames per second) for the inference of $\text{512}\times\text{512}$ resolution image on NVIDIA RTX 6000 Ada. The neural network can be further optimized and accelerated for edge deployment using model compression techniques, such as pruning and quantization, or by leveraging hardware accelerators, such as FPGAs and ASICs. Further details on the neural network training scheme are provided in Supplementary Note~5.

\subsection{Data Availability}
The data supporting the findings of this study are publicly available in a Google Drive folder accessible through the project GitHub repository at https://github.com/puostyoon/de-occluding-broadband-metalens-public. The third-party datasets used in this study, including DIV2K, VisDrone, Kvasir-SEG and Cityscapes, are available from their original providers as cited in the manuscript and are subject to their respective terms of use.

\subsection{Code Availability}
The custom code used for training, validation and inference of the metalens and image-reconstruction network is publicly available at https://github.com/puostyoon/de-occluding-broadband-metalens-public. The version used in this study (v1.0.0) has been archived in Zenodo at https://doi.org/10.5281/zenodo.20823451\cite{seungwooyoon_2026_20823451}.

\end{methods}

\bibliographystyle{naturemag}
\bibliography{references}

@InProceedings{Agustsson_2017_CVPR_Workshops,
	author = {Agustsson, Eirikur and Timofte, Radu},
	title = {NTIRE 2017 Challenge on Single Image Super-Resolution: Dataset and Study},
	booktitle = {The IEEE Conference on Computer Vision and Pattern Recognition (CVPR) Workshops},
	month = {July},
	year = {2017}
}

@InProceedings{Kim_2024_CVPR,
    author    = {Kim, Woohyeok and Kim, Geonu and Lee, Junyong and Lee, Seungyong and Baek, Seung-Hwan and Cho, Sunghyun},
    title     = {ParamISP: Learned Forward and Inverse ISPs using Camera Parameters},
    booktitle = {Proceedings of the IEEE/CVF Conference on Computer Vision and Pattern Recognition (CVPR)},
    month     = {June},
    year      = {2024},
    pages     = {26067-26076}
}

@article{loshchilov2017decoupled,
  title={Decoupled weight decay regularization},
  author={Loshchilov, Ilya and Hutter, Frank},
  journal={arXiv preprint arXiv:1711.05101},
  year={2017}
}

@article{loshchilov2016sgdr,
  title={Sgdr: Stochastic gradient descent with warm restarts},
  author={Loshchilov, Ilya and Hutter, Frank},
  journal={arXiv preprint arXiv:1608.03983},
  year={2016}
}

@inproceedings{zhang2018unreasonable,
  title={The unreasonable effectiveness of deep features as a perceptual metric},
  author={Zhang, Richard and Isola, Phillip and Efros, Alexei A and Shechtman, Eli and Wang, Oliver},
  booktitle={Proceedings of the IEEE conference on computer vision and pattern recognition},
  pages={586--595},
  year={2018}
}

@article{tseng2021neural,
  title={Neural nano-optics for high-quality thin lens imaging},
  author={Tseng, Ethan and Colburn, Shane and Whitehead, James and Huang, Luocheng and Baek, Seung-Hwan and Majumdar, Arka and Heide, Felix},
  journal={Nature communications},
  volume={12},
  number={1},
  pages={6493},
  year={2021},
  publisher={Nature Publishing Group UK London}
}

@article{matsushima2009band,
  title={Band-limited angular spectrum method for numerical simulation of free-space propagation in far and near fields},
  author={Matsushima, Kyoji and Shimobaba, Tomoyoshi},
  journal={Optics express},
  volume={17},
  number={22},
  pages={19662--19673},
  year={2009},
  publisher={Optica Publishing Group}
}

@inproceedings{liu2020learning,
  title={Learning to see through obstructions},
  author={Liu, Yu-Lun and Lai, Wei-Sheng and Yang, Ming-Hsuan and Chuang, Yung-Yu and Huang, Jia-Bin},
  booktitle={Proceedings of the IEEE/CVF Conference on Computer Vision and Pattern Recognition},
  pages={14215--14224},
  year={2020}
}

@inproceedings{chugunov2024neural,
  title={Neural spline fields for burst image fusion and layer separation},
  author={Chugunov, Ilya and Shustin, David and Yan, Ruyu and Lei, Chenyang and Heide, Felix},
  booktitle={Proceedings of the IEEE/CVF Conference on Computer Vision and Pattern Recognition},
  pages={25763--25773},
  year={2024}
}

@article{zhang2021automatic,
  title={Automatic object removal with obstructed fa{\c{c}}ades completion using semantic segmentation and generative adversarial inpainting},
  author={Zhang, Jiaxin and Fukuda, Tomohiro and Yabuki, Nobuyoshi},
  journal={IEEE Access},
  volume={9},
  pages={117486--117495},
  year={2021},
  publisher={IEEE}
}

@inproceedings{hirohashi2020removal,
  title={Removal of image obstacles for vehicle-mounted surrounding monitoring cameras by real-time video inpainting},
  author={Hirohashi, Yoshihiro and Narioka, Kenichi and Suganuma, Masanori and Liu, Xing and Tamatsu, Yukimasa and Okatani, Takayuki},
  booktitle={Proceedings of the IEEE/CVF Conference on Computer Vision and Pattern Recognition Workshops},
  pages={214--215},
  year={2020}
}

@article{gupta2021robust,
  title={A robust and efficient image de-fencing approach using conditional generative adversarial networks},
  author={Gupta, Divyanshu and Jain, Shorya and Tripathi, Utkarsh and Chattopadhyay, Pratik and Wang, Lipo},
  journal={Signal, Image and Video Processing},
  volume={15},
  number={2},
  pages={297--305},
  year={2021},
  publisher={Springer}
}

@inproceedings{hao2019learning,
  title={Learning from synthetic photorealistic raindrop for single image raindrop removal},
  author={Hao, Zhixiang and You, Shaodi and Li, Yu and Li, Kunming and Lu, Feng},
  booktitle={Proceedings of the IEEE/CVF International Conference on Computer Vision Workshops},
  pages={0--0},
  year={2019}
}

@article{yu2023inpaint,
  title={Inpaint anything: Segment anything meets image inpainting},
  author={Yu, Tao and Feng, Runseng and Feng, Ruoyu and Liu, Jinming and Jin, Xin and Zeng, Wenjun and Chen, Zhibo},
  journal={arXiv preprint arXiv:2304.06790},
  year={2023}
}

@misc{openai2026systemcard,
  title     = {GPT-5.5 System Card},
  author    = {{OpenAI}},
  year      = {2026},
  howpublished = {\url{https://deploymentsafety.openai.com/gpt-5-5}},
  note      = {Accessed: May 2026}
}

@article{pei2019occluded,
  title={Occluded-object 3D reconstruction using camera array synthetic aperture imaging},
  author={Pei, Zhao and Li, Yawen and Ma, Miao and Li, Jun and Leng, Chengcai and Zhang, Xiaoqiang and Zhang, Yanning},
  journal={Sensors},
  volume={19},
  number={3},
  pages={607},
  year={2019},
  publisher={MDPI}
}

@inproceedings{joshi2007synthetic,
  title={Synthetic aperture tracking: Tracking through occlusions},
  author={Joshi, Neel and Avidan, Shai and Matusik, Wojciech and Kriegman, David J},
  booktitle={2007 IEEE 11th International Conference on Computer Vision},
  pages={1--8},
  year={2007},
  organization={IEEE}
}

@article{levoy2004synthetic,
  title={Synthetic aperture confocal imaging},
  author={Levoy, Marc and Chen, Billy and Vaish, Vaibhav and Horowitz, Mark and McDowall, Ian and Bolas, Mark},
  journal={ACM Transactions on Graphics (ToG)},
  volume={23},
  number={3},
  pages={825--834},
  year={2004},
  publisher={ACM New York, NY, USA}
}

@article{pei2016all,
  title={All-in-focus synthetic aperture imaging using image matting},
  author={Pei, Zhao and Chen, Xida and Yang, Yee-Hong},
  journal={IEEE Transactions on Circuits and Systems for Video Technology},
  volume={28},
  number={2},
  pages={288--301},
  year={2016},
  publisher={IEEE}
}

@article{xue2015computational,
  title={A computational approach for obstruction-free photography},
  author={Xue, Tianfan and Rubinstein, Michael and Liu, Ce and Freeman, William T},
  journal={ACM Transactions on Graphics (TOG)},
  volume={34},
  number={4},
  pages={1--11},
  year={2015},
  publisher={ACM New York, NY, USA}
}

@article{kim2023scalable,
  title={Scalable manufacturing of high-index atomic layer--polymer hybrid metasurfaces for metaphotonics in the visible},
  author={Kim, Joohoon and Seong, Junhwa and Kim, Wonjoong and Lee, Gun-Yeal and Kim, Seokwoo and Kim, Hongyoon and Moon, Seong-Won and Oh, Dong Kyo and Yang, Younghwan and Park, Jeonghoon and others},
  journal={Nature Materials},
  volume={22},
  number={4},
  pages={474--481},
  year={2023},
  publisher={Nature Publishing Group UK London}
}

@inproceedings{jonna2017stereo,
  title={Stereo image de-fencing using smartphones},
  author={Jonna, Sankaraganesh and Satapathy, Sukla and Sahay, Rajiv R},
  booktitle={2017 IEEE international conference on acoustics, speech and signal processing (ICASSP)},
  pages={1792--1796},
  year={2017},
  organization={IEEE}
}

@inproceedings{zhu2023occlusion,
  title={Occlusion-free scene recovery via neural radiance fields},
  author={Zhu, Chengxuan and Wan, Renjie and Tang, Yunkai and Shi, Boxin},
  booktitle={Proceedings of the IEEE/CVF Conference on Computer Vision and Pattern Recognition},
  pages={20722--20731},
  year={2023}
}

@inproceedings{tsogkas2023efficient,
  title={Efficient Flow-Guided Multi-frame De-fencing},
  author={Tsogkas, Stavros and Zhang, Fengjia and Jepson, Allan and Levinshtein, Alex},
  booktitle={Proceedings of the IEEE/CVF Winter Conference on Applications of Computer Vision},
  pages={1838--1847},
  year={2023}
}

@article{jonna2024deep,
  title={Deep Generative Adversarial Network for Occlusion Removal from a Single Image},
  author={Jonna, Sankaraganesh and Medhi, Moushumi and Sahay, Rajiv Ranjan},
  journal={arXiv preprint arXiv:2409.13242},
  year={2024}
}

@article{colburn2019optical,
  title={Optical frontend for a convolutional neural network},
  author={Colburn, Shane and Chu, Yi and Shilzerman, Eli and Majumdar, Arka},
  journal={Applied optics},
  volume={58},
  number={12},
  pages={3179--3186},
  year={2019},
  publisher={Optica Publishing Group}
}

@article{so2023multicolor,
  title={Multicolor and 3D holography generated by inverse-designed single-cell metasurfaces},
  author={So, Sunae and Kim, Joohoon and Badloe, Trevon and Lee, Chihun and Yang, Younghwan and Kang, Hyunjung and Rho, Junsuk},
  journal={Advanced Materials},
  volume={35},
  number={17},
  pages={2208520},
  year={2023},
  publisher={Wiley Online Library}
}

@inproceedings{baek2021single,
  title={Single-shot hyperspectral-depth imaging with learned diffractive optics},
  author={Baek, Seung-Hwan and Ikoma, Hayato and Jeon, Daniel S and Li, Yuqi and Heidrich, Wolfgang and Wetzstein, Gordon and Kim, Min H},
  booktitle={Proceedings of the IEEE/CVF International Conference on Computer Vision},
  pages={2651--2660},
  year={2021}
}

@article{shi2022seeing,
  title={Seeing through obstructions with diffractive cloaking},
  author={Shi, Zheng and Bahat, Yuval and Baek, Seung-Hwan and Fu, Qiang and Amata, Hadi and Li, Xiao and Chakravarthula, Praneeth and Heidrich, Wolfgang and Heide, Felix},
  journal={ACM Transactions on Graphics (TOG)},
  volume={41},
  number={4},
  pages={1--15},
  year={2022},
  publisher={ACM New York, NY, USA}
}

@article{colburn2018metasurface,
  title={Metasurface optics for full-color computational imaging},
  author={Colburn, Shane and Zhan, Alan and Majumdar, Arka},
  journal={Science advances},
  volume={4},
  number={2},
  pages={eaar2114},
  year={2018},
  publisher={American Association for the Advancement of Science}
}

@article{huang2020design,
  title={Design and analysis of extended depth of focus metalenses for achromatic computational imaging},
  author={Huang, Luocheng and Whitehead, James and Colburn, Shane and Majumdar, Arka},
  journal={Photonics Research},
  volume={8},
  number={10},
  pages={1613--1623},
  year={2020},
  publisher={Chinese Laser Press and Optical Society of America}
}

@article{tan20213d,
  title={3D imaging using extreme dispersion in optical metasurfaces},
  author={Tan, Shiyu and Yang, Frank and Boominathan, Vivek and Veeraraghavan, Ashok and Naik, Gururaj V},
  journal={Acs Photonics},
  volume={8},
  number={5},
  pages={1421--1429},
  year={2021},
  publisher={ACS Publications}
}

@article{bayati2022inverse,
  title={Inverse designed extended depth of focus meta-optics for broadband imaging in the visible},
  author={Bayati, Elyas and Pestourie, Rapha{\"e}l and Colburn, Shane and Lin, Zin and Johnson, Steven G and Majumdar, Arka},
  journal={Nanophotonics},
  volume={11},
  number={11},
  pages={2531--2540},
  year={2022},
  publisher={De Gruyter}
}

@inproceedings{xu2025polarization,
  title={A polarization-independent and broadband achromatic gallium nitride metalens based on extended depth of focus},
  author={Xu, Hong and Yi, Jue-Min and Huang, Zengli and Wang, Miao and Xu, Ke and Wang, Chinhua},
  booktitle={Fourth International Computational Imaging Conference (CITA 2024)},
  volume={13542},
  pages={1029--1037},
  year={2025},
  organization={SPIE}
}

@article{froch2025beating,
  title={Beating spectral bandwidth limits for large aperture broadband nano-optics},
  author={Fr{\"o}ch, Johannes E and Chakravarthula, Praneeth and Sun, Jipeng and Tseng, Ethan and Colburn, Shane and Zhan, Alan and Miller, Forrest and Wirth-Singh, Anna and Tanguy, Quentin AA and Han, Zheyi and others},
  journal={Nature communications},
  volume={16},
  number={1},
  pages={3025},
  year={2025},
  publisher={Nature Publishing Group UK London}
}

@article{chen2018broadband,
  title={A broadband achromatic metalens for focusing and imaging in the visible},
  author={Chen, Wei Ting and Zhu, Alexander Y and Sanjeev, Vyshakh and Khorasaninejad, Mohammadreza and Shi, Zhujun and Lee, Eric and Capasso, Federico},
  journal={Nature nanotechnology},
  volume={13},
  number={3},
  pages={220--226},
  year={2018},
  publisher={Nature Publishing Group UK London}
}

@article{shrestha2018broadband,
  title={Broadband achromatic dielectric metalenses},
  author={Shrestha, Sajan and Overvig, Adam C and Lu, Ming and Stein, Aaron and Yu, Nanfang},
  journal={Light: Science \& Applications},
  volume={7},
  number={1},
  pages={85},
  year={2018},
  publisher={Nature Publishing Group UK London}
}

@article{chen2019broadband,
  title={A broadband achromatic polarization-insensitive metalens consisting of anisotropic nanostructures},
  author={Chen, Wei Ting and Zhu, Alexander Y and Sisler, Jared and Bharwani, Zameer and Capasso, Federico},
  journal={Nature communications},
  volume={10},
  number={1},
  pages={355},
  year={2019},
  publisher={Nature Publishing Group UK London}
}

@article{hu2023asymptotic,
  title={Asymptotic dispersion engineering for ultra-broadband meta-optics},
  author={Hu, Yueqiang and Jiang, Yuting and Zhang, Yi and Yang, Xing and Ou, Xiangnian and Li, Ling and Kong, Xianghong and Liu, Xingsi and Qiu, Cheng-Wei and Duan, Huigao},
  journal={nature communications},
  volume={14},
  number={1},
  pages={6649},
  year={2023},
  publisher={Nature Publishing Group UK London}
}

@article{hou2025high,
  title={High-efficiency broadband achromatic metalens in the visible},
  author={Hou, Liang and Zhou, Hongyuan and Zhang, Dandan and Lu, Ganqing and Zhang, Dejian and Liu, Tingting and Xiao, Shuyuan and Yu, Tianbao},
  journal={Applied Physics Letters},
  volume={126},
  number={10},
  year={2025},
  publisher={AIP Publishing}
}

@article{chang2024achromatic,
  title={Achromatic metalenses for full visible spectrum with extended group delay control via dispersion-matched layers},
  author={Chang, Shengyuan and Zhang, Lidan and Duan, Yao and Rahman, Md Tarek and Islam, Abrar and Ni, Xingjie},
  journal={Nature communications},
  volume={15},
  number={1},
  pages={9627},
  year={2024},
  publisher={Nature Publishing Group UK London}
}

@article{presutti2020focusing,
  title={Focusing on bandwidth: achromatic metalens limits},
  author={Presutti, Federico and Monticone, Francesco},
  journal={Optica},
  volume={7},
  number={6},
  pages={624--631},
  year={2020},
  publisher={Optical Society of America}
}

@article{berry1984quantal,
  title={Quantal phase factors accompanying adiabatic changes},
  author={Berry, Michael Victor},
  journal={Proceedings of the Royal Society of London. A. Mathematical and Physical Sciences},
  volume={392},
  number={1802},
  pages={45--57},
  year={1984},
  publisher={The Royal Society London}
}

@article{richards2023hybrid,
  title={Hybrid achromatic microlenses with high numerical apertures and focusing efficiencies across the visible},
  author={Richards, Corey A and Ocier, Christian R and Xie, Dajie and Gao, Haibo and Robertson, Taylor and Goddard, Lynford L and Christiansen, Rasmus E and Cahill, David G and Braun, Paul V},
  journal={Nature communications},
  volume={14},
  number={1},
  pages={3119},
  year={2023},
  publisher={Nature Publishing Group UK London}
}

@article{vaswani2017attention,
  title={Attention is all you need},
  author={Vaswani, Ashish and Shazeer, Noam and Parmar, Niki and Uszkoreit, Jakob and Jones, Llion and Gomez, Aidan N and Kaiser, {\L}ukasz and Polosukhin, Illia},
  journal={Advances in neural information processing systems},
  volume={30},
  year={2017}
}

@article{mildenhall2021nerf,
  title={Nerf: Representing scenes as neural radiance fields for view synthesis},
  author={Mildenhall, Ben and Srinivasan, Pratul P and Tancik, Matthew and Barron, Jonathan T and Ramamoorthi, Ravi and Ng, Ren},
  journal={Communications of the ACM},
  volume={65},
  number={1},
  pages={99--106},
  year={2021},
  publisher={ACM New York, NY, USA}
}

@article{engelberg2022generalized,
  title={Generalized metric for broadband flat lens performance comparison},
  author={Engelberg, Jacob and Levy, Uriel},
  journal={Nanophotonics},
  volume={11},
  number={16},
  pages={3559--3574},
  year={2022},
  publisher={De Gruyter}
}

@inproceedings{you2013adherent,
  title={Adherent raindrop detection and removal in video},
  author={You, Shaodi and Tan, Robby T and Kawakami, Rei and Ikeuchi, Katsushi},
  booktitle={Proceedings of the IEEE Conference on Computer Vision and Pattern Recognition},
  pages={1035--1042},
  year={2013}
}

@article{brophy2023review,
  title={A review of the impact of rain on camera-based perception in automated driving systems},
  author={Brophy, Tim and Mullins, Darragh and Parsi, Ashkan and Horgan, Jonathan and Ward, Enda and Denny, Patrick and Eising, Ciar{\'a}n and Deegan, Brian and Glavin, Martin and Jones, Edward},
  journal={IEEE Access},
  volume={11},
  pages={67040--67057},
  year={2023},
  publisher={IEEE}
}

@inproceedings{uricar2021let,
  title={Let's get dirty: Gan based data augmentation for camera lens soiling detection in autonomous driving},
  author={Uricar, Michal and Sistu, Ganesh and Rashed, Hazem and Vobecky, Antonin and Kumar, Varun Ravi and Krizek, Pavel and Burger, Fabian and Yogamani, Senthil},
  booktitle={Proceedings of the IEEE/CVF winter conference on applications of computer vision},
  pages={766--775},
  year={2021}
}

@inproceedings{liu2008image,
  title={Image de-fencing.},
  author={Liu, Yanxi and Belkina, Tamara and Hays, James and Lublinerman, Roberto},
  booktitle={CVPR},
  pages={1--8},
  year={2008}
}

@article{nathan2022through,
  title={Through-foliage tracking with airborne optical sectioning},
  author={Nathan, Rakesh John Amala Arokia and Kurmi, Indrajit and Schedl, David C and Bimber, Oliver},
  journal={Journal of Remote Sensing},
  year={2022},
  publisher={AAAS}
}

@inproceedings{chang2024uav,
  title={Uav-rain1k: A benchmark for raindrop removal from uav aerial imagery},
  author={Chang, Wenhui and Chen, Hongming and He, Xin and Chen, Xiang and Shen, Liangduo},
  booktitle={Proceedings of the IEEE/CVF Conference on Computer Vision and Pattern Recognition},
  pages={15--22},
  year={2024}
}

@inproceedings{uvrivcavr2019soilingnet,
  title={Soilingnet: Soiling detection on automotive surround-view cameras},
  author={U{\v{r}}i{\v{c}}{\'a}{\v{r}}, Michal and K{\v{r}}{\'\i}{\v{z}}ek, Pavel and Sistu, Ganesh and Yogamani, Senthil},
  booktitle={2019 IEEE Intelligent Transportation Systems Conference (ITSC)},
  pages={67--72},
  year={2019},
  organization={IEEE}
}

@article{nabeel2022effective,
  title={Effective cleaning of endoscopic lenses to achieve visual clarity for minimally invasive abdominopelvic surgery: a systematic review},
  author={Nabeel, Ahmad and Al-Sabah, Salman K and Ashrafian, Hutan},
  journal={Surgical Endoscopy},
  volume={36},
  number={4},
  pages={2382--2392},
  year={2022},
  publisher={Springer}
}

@article{wang2018variational,
  title={Variational based smoke removal in laparoscopic images},
  author={Wang, Congcong and Alaya Cheikh, Faouzi and Kaaniche, Mounir and Beghdadi, Azeddine and Elle, Ole Jacob},
  journal={Biomedical engineering online},
  volume={17},
  number={1},
  pages={139},
  year={2018},
  publisher={Springer}
}

@article{chen2022meta,
  title={Meta-lens in the sky},
  author={Chen, Mu Ku and Chu, Cheng Hung and Liu, Xiaoyuan and Zhang, Jingcheng and Sun, Linshan and Yao, Jin and Fan, Yubin and Liang, Yao and Yamaguchi, Takeshi and Tanaka, Takuo and others},
  journal={IEEE Access},
  volume={10},
  pages={46552--46557},
  year={2022},
  publisher={IEEE}
}

@article{liu2023underwater,
  title={Underwater binocular meta-lens},
  author={Liu, Xiaoyuan and Chen, Mu Ku and Chu, Cheng Hung and Zhang, Jingcheng and Leng, Borui and Yamaguchi, Takeshi and Tanaka, Takuo and Tsai, Din Ping},
  journal={ACS Photonics},
  volume={10},
  number={7},
  pages={2382--2389},
  year={2023},
  publisher={ACS Publications}
}

@article{froch2023real,
  title={Real time full-color imaging in a Meta-optical fiber endoscope},
  author={Fr{\"o}ch, Johannes E and Huang, Luocheng and Tanguy, Quentin AA and Colburn, Shane and Zhan, Alan and Ravagli, Andrea and Seibel, Eric J and B{\"o}hringer, Karl F and Majumdar, Arka},
  journal={elight},
  volume={3},
  number={1},
  pages={13},
  year={2023},
  publisher={Springer}
}

@article{liu2024stereo,
  title={Stereo vision meta-lens-assisted driving vision},
  author={Liu, Xiaoyuan and Li, Wuyang and Yamaguchi, Takeshi and Geng, Zihan and Tanaka, Takuo and Tsai, Din Ping and Chen, Mu Ku},
  journal={ACS Photonics},
  volume={11},
  number={7},
  pages={2546--2555},
  year={2024},
  publisher={ACS Publications}
}

@article{jeon2019compact,
  title={Compact snapshot hyperspectral imaging with diffracted rotation.},
  author={Jeon, Daniel S and Baek, Seung-Hwan and Yi, Shinyoung and Fu, Qiang and Dun, Xiong and Heidrich, Wolfgang and Kim, Min H},
  journal={ACM Trans. Graph.},
  volume={38},
  number={4},
  pages={117--1},
  year={2019}
}

@article{shen2025extended,
  title={Extended monocular 3D imaging via the fusion of diffraction-and polarization-based depth cues},
  author={Shen, Zicheng and Zhao, Feng and Ni, Yibo and Yang, Yuanmu},
  journal={Optica},
  volume={12},
  number={6},
  pages={872--878},
  year={2025},
  publisher={Optica Publishing Group}
}

@article{shen2023monocular,
  title={Monocular metasurface camera for passive single-shot 4D imaging},
  author={Shen, Zicheng and Zhao, Feng and Jin, Chunqi and Wang, Shuai and Cao, Liangcai and Yang, Yuanmu},
  journal={Nature Communications},
  volume={14},
  number={1},
  pages={1035},
  year={2023},
  publisher={Nature Publishing Group UK London}
}

@article{park202536,
  title={36-Channel Spin and Wavelength Co-Multiplexed Metasurface Holography by Phase-Gradient Inverse Design},
  author={Park, Cherry and Jeon, Youngsun and Rho, Junsuk},
  journal={Advanced Science},
  pages={2504634},
  year={2025},
  publisher={Wiley Online Library}
}

@article{meng2025ultranarrow,
  title={Ultranarrow-linewidth wavelength-vortex metasurface holography},
  author={Meng, Weijia and Fr{\"o}ch, Johannes E and Cheng, Ke and Pi, Dapu and Li, Baoli and Majumdar, Arka and Maier, Stefan A and Ren, Haoran and Gu, Min and Fang, Xinyuan},
  journal={Science Advances},
  volume={11},
  number={12},
  pages={eadt9159},
  year={2025},
  publisher={American Association for the Advancement of Science}
}

@book{goodman2005introduction,
  title={Introduction to Fourier optics},
  author={Goodman, Joseph W},
  year={2005},
  publisher={Roberts and Company publishers}
}

@inproceedings{chen2020adversarial,
  title={Adversarial-learned loss for domain adaptation},
  author={Chen, Minghao and Zhao, Shuai and Liu, Haifeng and Cai, Deng},
  booktitle={Proceedings of the AAAI conference on artificial intelligence},
  volume={34},
  number={04},
  pages={3521--3528},
  year={2020}
}

@inproceedings{chen2018encoder,
  title={Encoder-decoder with atrous separable convolution for semantic image segmentation},
  author={Chen, Liang-Chieh and Zhu, Yukun and Papandreou, George and Schroff, Florian and Adam, Hartwig},
  booktitle={Proceedings of the European conference on computer vision (ECCV)},
  pages={801--818},
  year={2018}
}

@article{perlin1985image,
  title={An image synthesizer},
  author={Perlin, Ken},
  journal={ACM Siggraph Computer Graphics},
  volume={19},
  number={3},
  pages={287--296},
  year={1985},
  publisher={ACM New York, NY, USA}
}

@inproceedings{woo2018cbam,
  title={Cbam: Convolutional block attention module},
  author={Woo, Sanghyun and Park, Jongchan and Lee, Joon-Young and Kweon, In So},
  booktitle={Proceedings of the European conference on computer vision (ECCV)},
  pages={3--19},
  year={2018}
}

@article{fischler1981random,
  title={Random sample consensus: a paradigm for model fitting with applications to image analysis and automated cartography},
  author={Fischler, Martin A and Bolles, Robert C},
  journal={Communications of the ACM},
  volume={24},
  number={6},
  pages={381--395},
  year={1981},
  publisher={ACM New York, NY, USA}
}

@inproceedings{potje2024xfeat,
  title={Xfeat: Accelerated features for lightweight image matching},
  author={Potje, Guilherme and Cadar, Felipe and Araujo, Andr{\'e} and Martins, Renato and Nascimento, Erickson R},
  booktitle={Proceedings of the IEEE/CVF Conference on Computer Vision and Pattern Recognition},
  pages={2682--2691},
  year={2024}
}

@article{zhu2021detection,
  title={Detection and tracking meet drones challenge},
  author={Zhu, Pengfei and Wen, Longyin and Du, Dawei and Bian, Xiao and Fan, Heng and Hu, Qinghua and Ling, Haibin},
  journal={IEEE Transactions on Pattern Analysis and Machine Intelligence},
  volume={44},
  number={11},
  pages={7380--7399},
  year={2021},
  publisher={IEEE}
}

@inproceedings{Cordts2016Cityscapes,
title={The Cityscapes Dataset for Semantic Urban Scene Understanding},
author={Cordts, Marius and Omran, Mohamed and Ramos, Sebastian and Rehfeld, Timo and Enzweiler, Markus and Benenson, Rodrigo and Franke, Uwe and Roth, Stefan and Schiele, Bernt},
booktitle={Proc. of the IEEE Conference on Computer Vision and Pattern Recognition (CVPR)},
year={2016}
}

@inproceedings{jha2020kvasir, title={Kvasir-seg: A segmented polyp dataset}, author={Jha, Debesh and Smedsrud, Pia H and Riegler, Michael A and Halvorsen, P{\aa}l and de Lange, Thomas and Johansen, Dag and Johansen, H{\aa}vard D}, booktitle={International Conference on Multimedia Modeling}, pages={451--462}, year={2020}, organization={Springer}
}

@inproceedings{wang2023internimage,
  title={Internimage: Exploring large-scale vision foundation models with deformable convolutions},
  author={Wang, Wenhai and Dai, Jifeng and Chen, Zhe and Huang, Zhenhang and Li, Zhiqi and Zhu, Xizhou and Hu, Xiaowei and Lu, Tong and Lu, Lewei and Li, Hongsheng and others},
  booktitle={Proceedings of the IEEE/CVF conference on computer vision and pattern recognition},
  pages={14408--14419},
  year={2023}
}

@article{zhao2026m2snet,
  title={M2SNet: Multi-scale in multi-scale subtraction network for medical image segmentation},
  author={Zhao, Xiaoqi and Jia, Hongpeng and Pang, Youwei and Lv, Long and Tian, Feng and Zhang, Lihe and Sun, Weibing and Lu, Huchuan},
  journal={Machine Intelligence Research},
  pages={1--23},
  year={2026},
  publisher={Springer}
}

@article{khanam2024yolov11,
  title={Yolov11: An overview of the key architectural enhancements},
  author={Khanam, Rahima and Hussain, Muhammad},
  journal={arXiv preprint arXiv:2410.17725},
  year={2024}
}

@article{vrhel1994measurement,
  title={Measurement and analysis of object reflectance spectra},
  author={Vrhel, Michael J and Gershon, Ron and Iwan, Lawrence S},
  journal={Color Research \& Application},
  volume={19},
  number={1},
  pages={4--9},
  year={1994},
  publisher={Wiley Online Library}
}

@article{westland2000colour,
  title={Colour statistics of natural and man-made surfaces},
  author={Westland, Stephen and Shaw, Julian and Owens, Huw},
  journal={Sensor Review},
  volume={20},
  number={1},
  pages={50--55},
  year={2000},
  publisher={MCB UP Ltd}
}

@InProceedings{Jeon_2024_CVPR,
    author    = {Jeon, Yujin and Choi, Eunsue and Kim, Youngchan and Moon, Yunseong and Omer, Khalid and Heide, Felix and Baek, Seung-Hwan},
    title     = {Spectral and Polarization Vision: Spectro-polarimetric Real-world Dataset},
    booktitle = {Proceedings of the IEEE/CVF Conference on Computer Vision and Pattern Recognition (CVPR)},
    month     = {June},
    year      = {2024},
    pages     = {22098-22108}
}

@article{yasuma2010generalized,
  title={Generalized assorted pixel camera: postcapture control of resolution, dynamic range, and spectrum},
  author={Yasuma, Fumihito and Mitsunaga, Tomoo and Iso, Daisuke and Nayar, Shree K},
  journal={IEEE transactions on image processing},
  volume={19},
  number={9},
  pages={2241--2253},
  year={2010},
  publisher={IEEE}
}

@conference{chakrabarti2011statistics,
  title={{Statistics of Real-World Hyperspectral Images}},
  author={Chakrabarti, A. and Zickler, T.},
  booktitle={Proc.~IEEE Conf.~on Computer Vision and Pattern Recognition (CVPR)},
  pages={193--200},
  year={2011}
}

@article{de2025diffraction,
  title={Diffraction and dispersion effects on the wavelength dependence of the focal length of metalenses},
  author={de Wit, JMM and Cussiol, Francesca and Caglayan, Humeyra and G{\'o}mez Rivas, J},
  journal={Applied Optics},
  volume={64},
  number={13},
  pages={3598--3604},
  year={2025},
  publisher={Optica Publishing Group}
}

@inproceedings{ronneberger2015u,
  title={U-net: Convolutional networks for biomedical image segmentation},
  author={Ronneberger, Olaf and Fischer, Philipp and Brox, Thomas},
  booktitle={International Conference on Medical image computing and computer-assisted intervention},
  pages={234--241},
  year={2015},
  organization={Springer}
}

@article{lin2025resonant,
  title={Resonant Meta-Lens in the Visible},
  author={Lin, Rong and Yao, Jin and Wang, Zhihui and Zhou, Junxiao and Tsai, Din Ping},
  journal={Laser \& Photonics Reviews},
  volume={19},
  number={8},
  pages={2401740},
  year={2025},
  publisher={Wiley Online Library}
}

@software{seungwooyoon_2026_20823451,
  author       = {SeungwooYoon},
  title        = {puostyoon/de-occluding-broadband-metalens-public:
                   v1.0.0
                  },
  month        = jun,
  year         = 2026,
  publisher    = {Zenodo},
  version      = {v1.0.0},
  doi          = {10.5281/zenodo.20823451},
  url          = {https://doi.org/10.5281/zenodo.20823451},
}

\begin{addendum}
	\item [Funding]

S.-H.B. acknowledges the Samsung Research Funding \& Incubation Center for Future Technology grant (SRFC-IT1801-52), the National Research Foundation of Korea (NRF) grants funded by the Korea government (MSIT) (RS-2024-00438532, RS-2023-00211658, RS-2025-00518643) and the Institute of Information \& Communications Technology Planning \& Evaluation (IITP) grants funded by the Korea government (MSIT) (IITP-2026-RS-2024-00437866, RS-2024-0045788, and RS-2019-II191906 for the Artificial Intelligence Graduate School Program at POSTECH).
J.R. acknowledges the Samsung Research Funding \& Incubation Center for Future Technology grant (SRFC-IT1901-52) funded by Samsung Electronics, the POSCO-POSTECH-RIST Convergence Research Center program funded by POSCO, and the NRF grants (RS-2026-25507992, RS-2025-02217649, RS-2024-00356928, RS-2024-00462912) funded by the MSIT of the Korean government.
A.M. and M.C. are supported by National Science Foundation (EFRI-BRAID-2223495). M.C. was supported by National Research Foundation of Korea (RS-2026-25470619).

\item [Author Contributions] S.-H.B. and S.Y. conceived the idea. S.Y. designed the depth-wavelength symmetry model. S.Y., S.-H.B., and H.H. verified the depth-wavelength symmetry model. S.Y., D.K., E.C., H.H., and M.C. performed the metalens design. S.Y. and D.S. implemented the metalens simulation. S.Y. and D.K. implemented the experimental prototype. S.L. and Seoyeon Kim performed computer vision tasks. D.K. fabricated the devices. Suha Kwak supervised the computer vision experiments. A.M. advised the metalens design process. J.R. guided the material characterization and device fabrication. All authors participated in discussions and contributed to writing the manuscript. S.-H.B. and J.R. guided all aspects of the work.
\item [Competing Interests] The authors declare no competing interests.
\end{addendum}

\clearpage
\setcounter{figure}{0}
\renewcommand{\figurename}{Fig.}
\renewcommand{\thefigure}{\arabic{figure}}
\begin{figure}[t]
	\centering
		\includegraphics[width=\columnwidth]
        {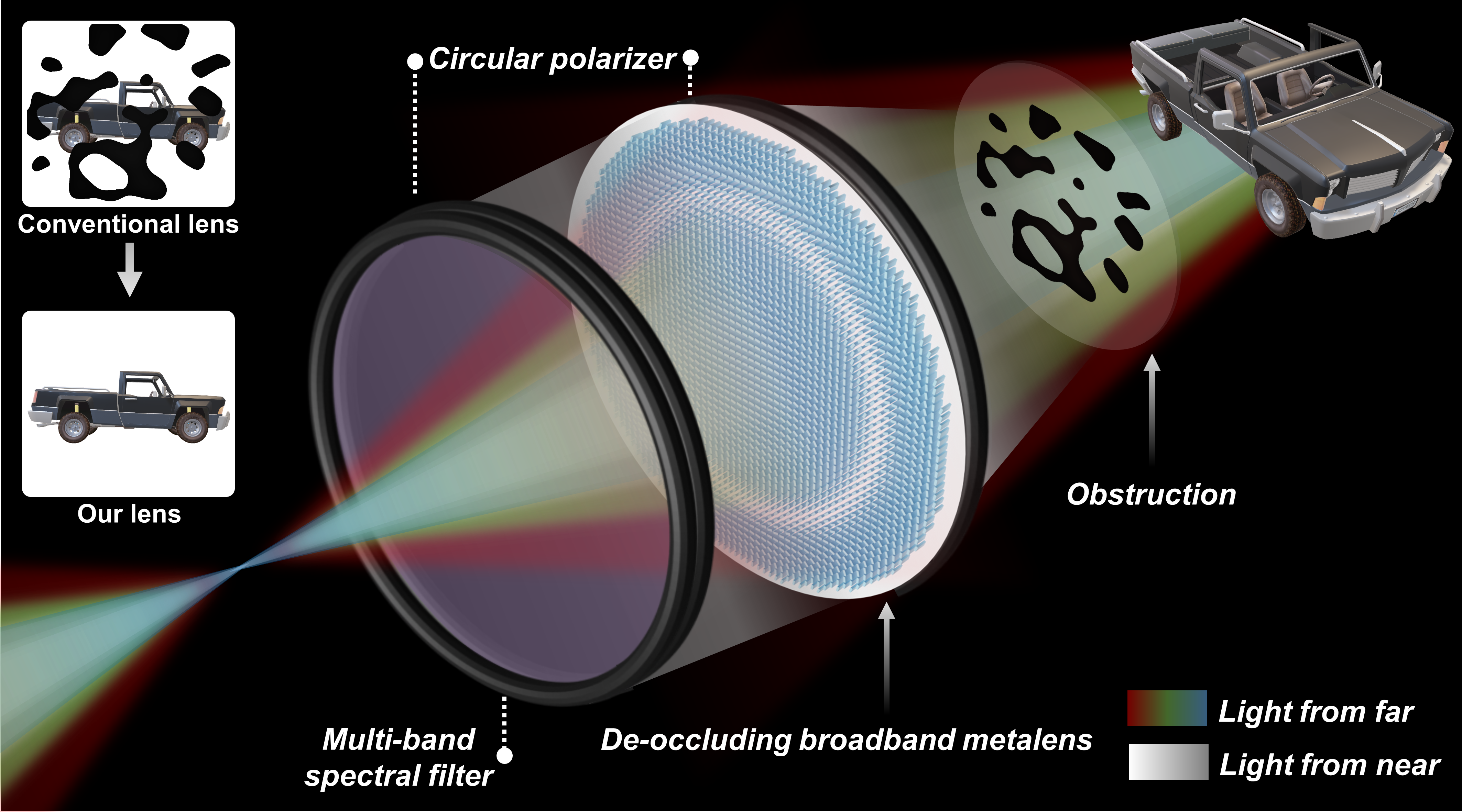}
		\caption{\textbf{De-occluding broadband metalens for obstruction-free broadband imaging.} Our de-occluding broadband metalens enables obstruction-free broadband imaging by filtering out the focused light from near depth with a multi-band spectral filter that selectively transmits far-focused light. An LCP generator and an RCP analyzer are included to select the cross-polarized circular component required for the current geometric-phase implementation.}
		\label{fig:figure1}
\end{figure}

\begin{figure}[t]
	\centering
		\includegraphics[width=\columnwidth]
        {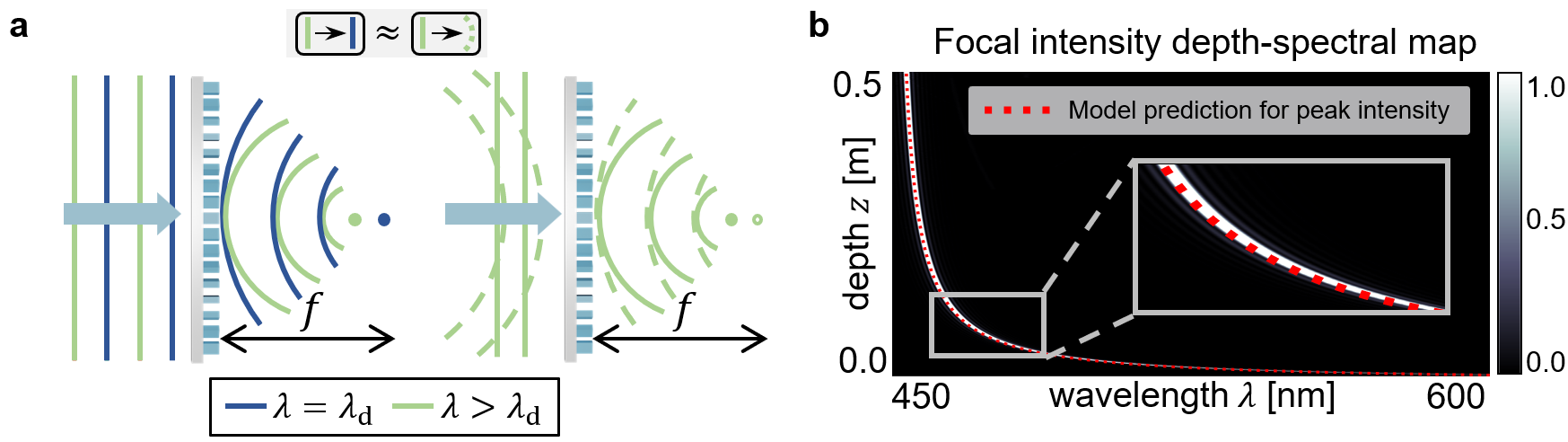}
		\caption{\textbf{Depth-wavelength symmetry.} \textbf{a} Schematic of the depth-wavelength relationship of diffractive lenses. (Left) An incident wavelength $\lambda$ longer than the design wavelength $\lambda_\text{d}$ ($\lambda>\lambda_\text{d}$) of the lens causes a phase mismatch, resulting in a focal front shift. (Right) This focal front shift can be compensated for by the spherical phase of an incident wave originating from a near depth, leading to an opposing focal back shift. Here, $f$ is the focal length. Blue and green denote ($\lambda=\lambda_\text{d}$) and ($\lambda>\lambda_\text{d}$), respectively.
        \textbf{b} Focal point intensity map for point light sources of wavelength, $\lambda$, and depth, $z$, incident on a hyperbolic ($\lambda_\text{d}=450\,$nm) metalens, computed with Rayleigh-Sommerfeld diffraction integral, and overlaid with our depth-wavelength symmetry model.}
		\label{fig:figure2}
\end{figure}

\begin{figure}[t]
	\centering
		\includegraphics[width=0.95\columnwidth]
        {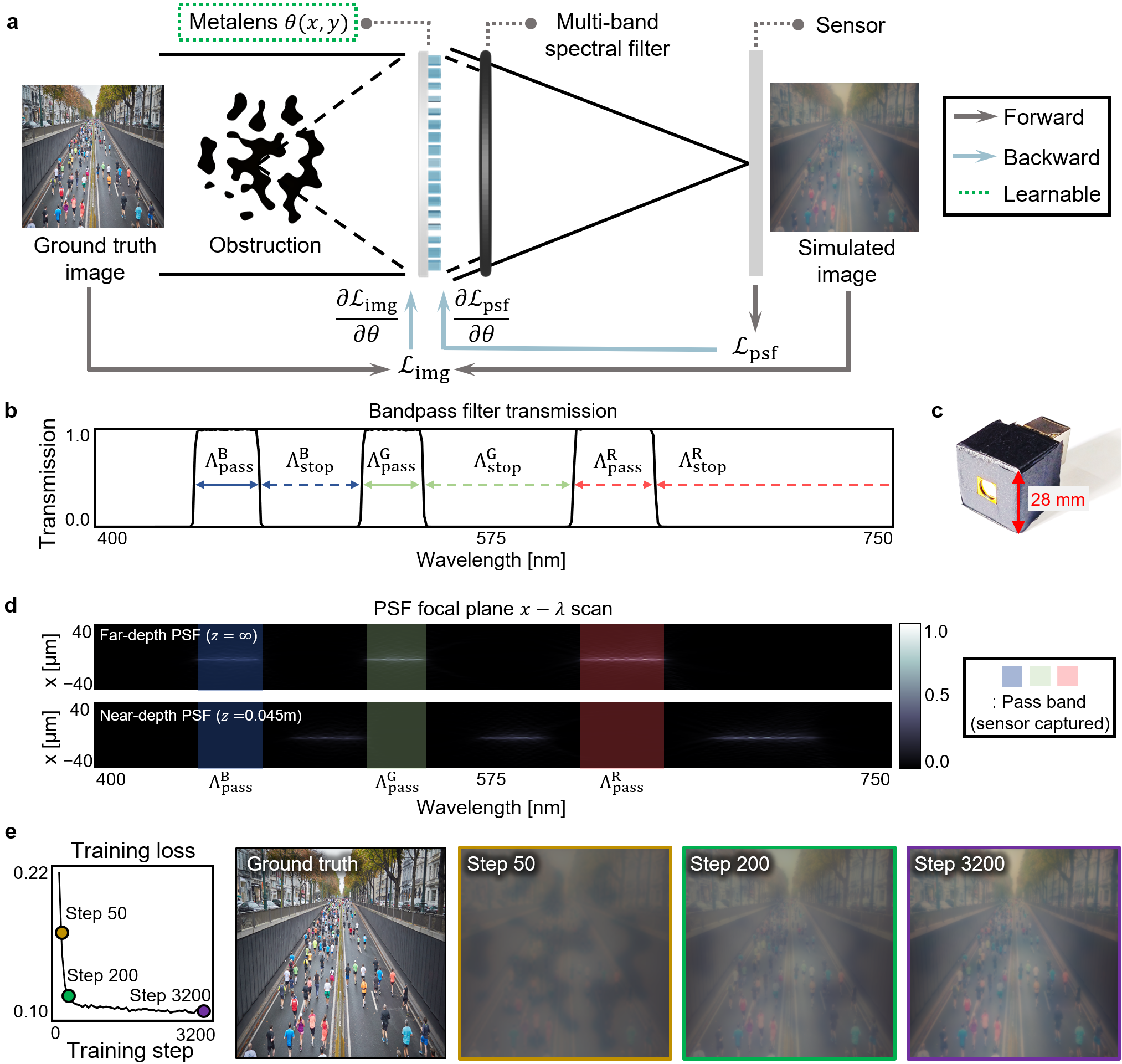}
		\caption{\textbf{Learning de-occluding broadband metalens.} \textbf{a} Overview of the metalens optimization pipeline. The metalens design map $\theta(x,y)$ is learned for the de-occluding broadband imaging. \textbf{b} Transmission of the multi-band spectral filter, separating each color channel into the pass band $\Lambda_\text{pass}^{c}$ and the stop band $\Lambda_\text{stop}^{c}$, providing extended design space for obstruction-free imaging. \textbf{c} Our compact prototype consists of a metalens, a multi-band spectral filter, a color sensor, and polarization optics for the geometric phase modulation. \textbf{d} $x-\lambda$ scan result of the learned metalens for far point spread functions (PSFs) and near PSFs in simulation. \textbf{e} Training loss and visual convergence during learning. The sequence of the simulated sensor images (step 50, 200, 3200) illustrates the learning of the metalens. Initial results exhibit significant blur and interference from obstructions, while the final optimized metalens effectively blurs the near-depth obstructions and maintains a sharp focus on the far-depth scene. The source image shown in a and e is from the DIV2K dataset\cite{Agustsson_2017_CVPR_Workshops}.}
		\label{fig:figure3}
\end{figure}

\begin{figure}[t]
	\centering
		\includegraphics[width=0.95\columnwidth]
        {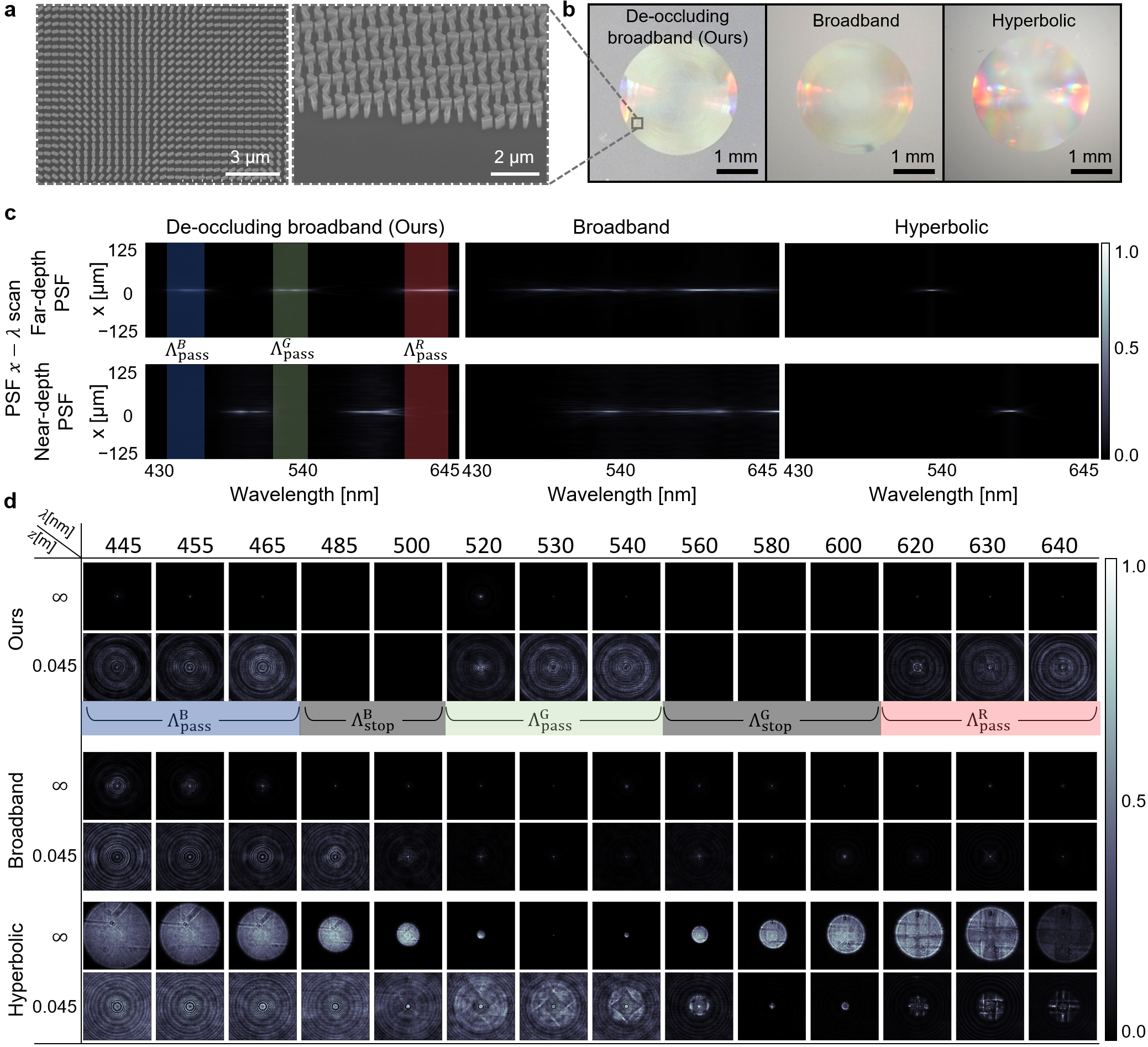}
		\caption{\textbf{Characterization of the de-occluding broadband metalens.} \textbf{a} Scanning electron microscope (SEM) images of the de-occluding broadband metalens with the top-down view (left) and tilted view (right). \textbf{b} Photograph of the three fabricated metalenses. From left to right: the de-occluding broadband metalens (Ours), the broadband metalens without a de-occluding approach (Broadband), and the conventional hyperbolic-phase metalens (Hyperbolic). \textbf{c} Point spread function (PSF) $x-\lambda$ scan results normalized to the total intensity. Pass bands are highlighted. Far-depth focused lights of the obstruction-free metalens (Ours) can be transmitted and captured on the sensor, while near-depth focused lights are filtered out, enabling obstruction-free imaging. \textbf{d} 2D PSFs of metalenses for different depths, normalized to their peak intensities. The stop band PSFs of Ours are marked with black boxes. \textbf{c, d} Our far-depth PSFs remain sharp in the pass bands, enabling clear imaging of the originally occluded scene, while the near-depth PSFs are blurry. The hyperbolic metalens suffers from severe chromatic aberrations. The broadband metalens fails to achieve the near-depth blur required for obstruction-free imaging.} 
		\label{fig:figure4}
\end{figure}

\begin{figure}[t]
	\centering
		\includegraphics[width=0.90\columnwidth]
        {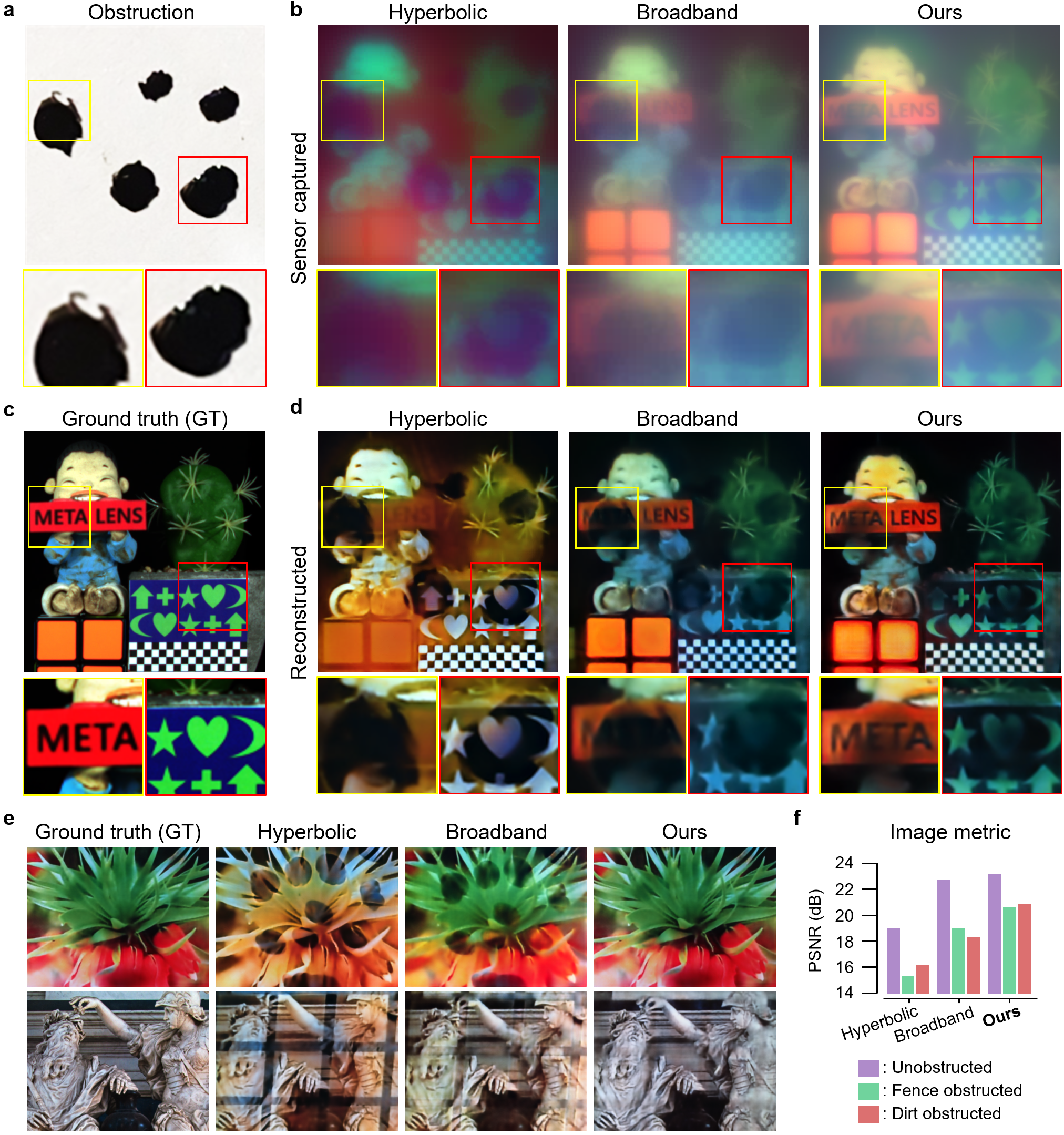}
		\caption{\textbf{Experimental evaluation of de-occluding broadband imaging.} \textbf{a} Near-depth obstruction pattern used for the experiment, placed at $z_{\text{near}}\simeq0.045\,\text{m}$. Objects and 2D images in the scene are placed at $z_{\text{far}}\simeq0.6\,\text{m}$. \textbf{b} Raw sensor captured images under the obstruction for the three metalens designs. Insets provide a magnified view of the regions highlighted by yellow and red boxes. \textbf{c} Ground-truth (GT) unobstructed reference image captured with a compound lens with $f=8\,$mm (two times longer than those of metalenses) for better object plane resolution. \textbf{d} Reconstructed images using the neural network trained for each metalens. \textbf{e} Reconstructed image comparisons of representative printed DIV2K\cite{Agustsson_2017_CVPR_Workshops} images under fence and dirt obstructions. \textbf{f} PSNRs measured on the printed DIV2K images\cite{Agustsson_2017_CVPR_Workshops}. \textbf{b, d, e, f} Our design suppresses obstructions better than other metalens baselines, achieving superior performance across all conditions.}
		\label{fig:figure5}
\end{figure}

\begin{figure}[t]
	\centering
		\includegraphics[width=0.90\columnwidth]
        {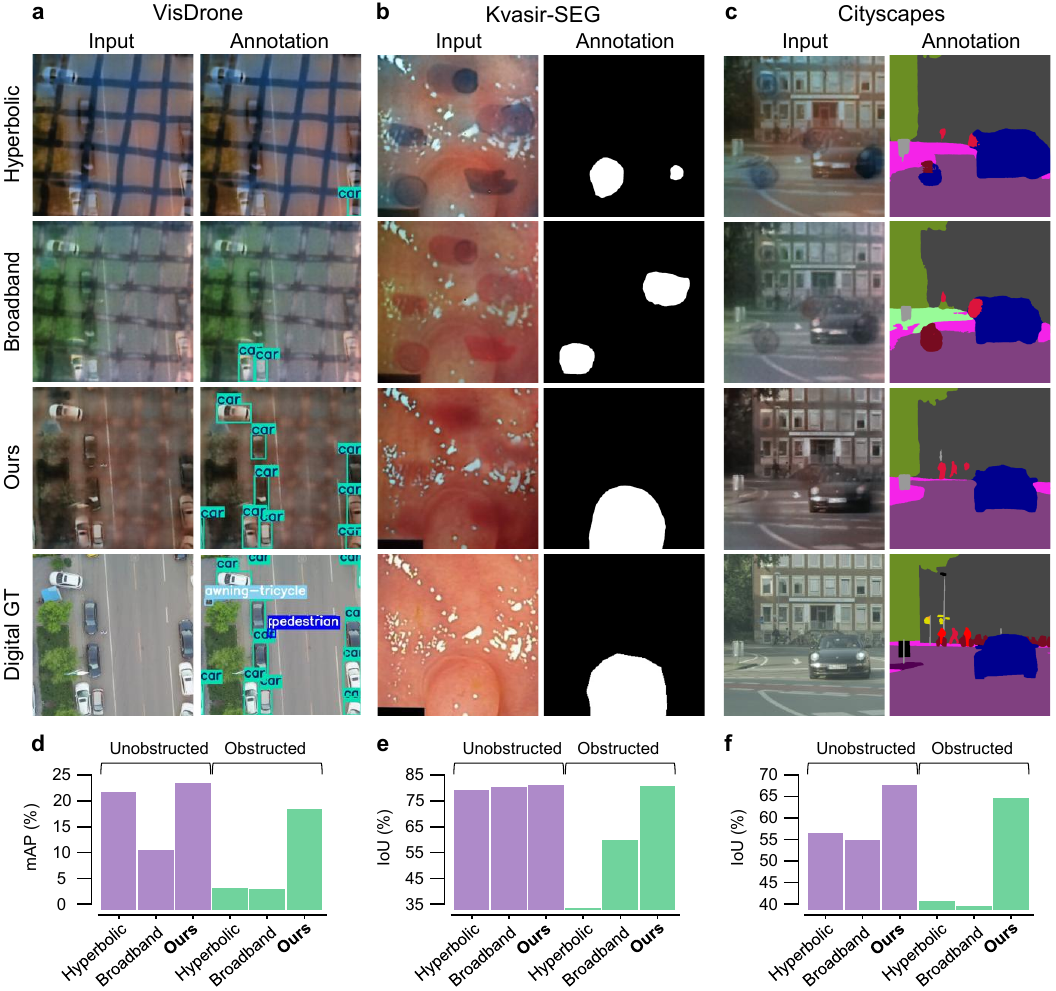}
		\caption{\textbf{Obstruction-free visual perception.} \textbf{a--c} Input images captured with each fabricated metalens design acquired by imaging the printed Digital GT, together with the corresponding original digital images (rows: Hyperbolic, Broadband, Ours, Digital GT) under different obstructions (fence, blood drop, dirt) and the corresponding downstream task outputs (Annotation) predicted by computer vision models on three benchmarks and ground-truth label: \textbf{a} UAV-view object detection\cite{khanam2024yolov11} on VisDrone\cite{zhu2021detection}, \textbf{b} Endoscopy polyp segmentation\cite{zhao2026m2snet} on Kvasir-SEG\cite{jha2020kvasir}, and \textbf{c} Driving scene semantic segmentation\cite{wang2023internimage} on Cityscapes\cite{Cordts2016Cityscapes}. The bottom row (Digital GT) shows the original digital images and their ground-truth labels for reference. \textbf{d--f} Quantitative performance under unobstructed (purple) and obstructed (green) imaging conditions: \textbf{d} detection mean average precision (mAP) on VisDrone, \textbf{e} intersection-over-union (IoU) on Kvasir-SEG, and \textbf{f} mean IoU (mIoU) on Cityscapes over multiple classes. Across all tasks, our de-occluding broadband metalens exhibits minimal performance degradation under obstructed conditions. Additional real-scene downstream-vision demonstrations are provided in Supplementary Note~17, as well as Supplementary Videos~4 and 5.}
		\label{fig:figure6}
\end{figure}

\end{document}


\maketitle\par
{\large\bfseries Supplementary Information\par}
\begingroup
\small
\setlength{\parskip}{4pt}
\noindent This author manuscript was accepted following peer review.
It precedes the publisher's copyediting and typesetting, omits subsequent
improvements and corrections, and is not the Version of Record.
\par
\noindent For the definitive published article, see:
Yoon, S., Kang, D., Choi, E. et al.
\textit{De-occluding broadband metalens}.
\textit{Nature Communications} \textbf{17}, 9808 (2026).
\href{https://doi.org/10.1038/s41467-026-76606-0}{doi:10.1038/s41467-026-76606-0}.
This accepted manuscript is shared under
\href{https://creativecommons.org/licenses/by-nc-nd/4.0/}{CC BY-NC-ND 4.0}.
Supplementary videos are available with the Version of Record.
\par
\endgroup

\begin{affiliations}
 \item Department of Computer Science and Engineering, Pohang University of Science and Technology (POSTECH), Pohang 37673, Republic of Korea
 \item Department of Mechanical Engineering, Pohang University of Science and Technology (POSTECH), Pohang 37673, Republic of Korea
 \item Graduate School of Artificial Intelligence, Pohang University of Science and Technology (POSTECH), Pohang 37673, Republic of Korea
 \item Department of Electrical and Computer Engineering, University of Washington, Seattle, 98195, WA, USA
 \item Department of Electrical Engineering, Ulsan National Institute of Science and Technology, Ulsan 44919, Republic of Korea.
 \item Department of Physics, University of Washington, Seattle, 98195, WA, USA
 \item Department of Chemical Engineering, Pohang University of Science and Technology (POSTECH), Pohang 37673, Republic of Korea
  \item Department of Electrical Engineering, Pohang University of Science and Technology (POSTECH), Pohang 37673, Republic of Korea
 \item POSCO-POSTECH-RIST Convergence Research Center for Flat Optics and Metaphotonics, Pohang 37673, Republic of Korea
 \item National Institute of Nanomaterials Technology (NINT), Pohang 37673, Republic of Korea
  \item [$*$] Equal contribution
 \item [$\dagger$] Corresponding author. E-mail: shwbaek@postech.ac.kr, jsrho@postech.ac.kr
\end{affiliations}

\definecolor{brightray}{rgb}{0.8,0.8,0.8}
\definecolor{Gray}{rgb}{0.5,0.5,0.5}
\definecolor{darkblue}{rgb}{0,0,0.7}
\definecolor{orange}{rgb}{1,.5,0} %
\definecolor{red}{rgb}{1,0,0} %
\definecolor{blue}{rgb}{0,0,1} %
\definecolor{darkgreen}{rgb}{0,0.7,0} %
\definecolor{darkred}{rgb}{0.7,0,0} %

\newcommand{\heading}[1]{\noindent\textbf{#1}}
\newcommand{\de}{\mathrm{d}}

\newcommand{\BEAS}{\begin{eqnarray*}}
\newcommand{\EEAS}{\end{eqnarray*}}
\newcommand{\BEA}{\begin{eqnarray}}
\newcommand{\EEA}{\end{eqnarray}}
\newcommand{\BEQ}{\begin{equation}}
\newcommand{\EEQ}{\end{equation}}
\newcommand{\BIT}{\begin{itemize}}
\newcommand{\EIT}{\end{itemize}}
\newcommand{\BNUM}{\begin{enumerate}}
\newcommand{\ENUM}{\end{enumerate}}

\newcommand{\BA}{\begin{array}}
\newcommand{\EA}{\end{array}}

\newcommand{\eg}{{\it e.g.}}
\newcommand{\ie}{{\it i.e.}}
\newcommand{\etc}{{\it etc.}}

\newcommand{\ones}{\mathbf 1}

\newcommand{\reals}{{\mbox{\bf R}}}
\newcommand{\integers}{{\mbox{\bf Z}}}
\newcommand{\eqbydef}{\mathrel{\stackrel{\Delta}{=}}}
\newcommand{\complex}{{\mbox{\bf C}}}
\newcommand{\symm}{{\mbox{\bf S}}}  %

\newcommand{\Span}{\mbox{\textrm{span}}}
\newcommand{\Range}{\mbox{\textrm{range}}}
\newcommand{\nullspace}{{\mathcal N}}
\newcommand{\range}{{\mathcal R}}
\newcommand{\Nullspace}{\mbox{\textrm{nullspace}}}
\newcommand{\Rank}{\mathop{\bf Rank}}
\newcommand{\Tr}{\mathop{\bf Tr}}
\newcommand{\diag}{\mathop{\bf diag}}
\newcommand{\lambdamax}{{\lambda_{\rm max}}}
\newcommand{\lambdamin}{\lambda_{\rm min}}

\newcommand{\Expect}{\mathop{\bf E{}}}
\newcommand{\Prob}{\mathop{\bf Prob}}
\newcommand{\erf}{\mathop{\bf erf}}

\newcommand{\Co}{{\mathop {\bf Co}}}
\newcommand{\co}{{\mathop {\bf Co}}}
\newcommand{\dist}{\mathop{\bf dist{}}}
\newcommand{\Ltwo}{{\bf L}_2}
\newcommand{\QED}{~~\rule[-1pt]{8pt}{8pt}}\def\qed{\QED}
\newcommand{\approxleq}{\mathrel{\smash{\makebox[0pt][l]{\raisebox{-3.4pt}{\small$\sim$}}}{\raisebox{1.1pt}{$<$}}}}
\newcommand{\epi}{\mathop{\bf epi}}

\newcommand{\vol}{\mathop{\bf vol}}
\newcommand{\Vol}{\mathop{\bf vol}}
\newcommand{\Card}{\mathop{\bf card}}

\newcommand{\dom}{\mathop{\bf dom}}
\newcommand{\aff}{\mathop{\bf aff}}
\newcommand{\cl}{\mathop{\bf cl}}
\newcommand{\Angle}{\mathop{\bf angle}}
\newcommand{\intr}{\mathop{\bf int}}
\newcommand{\relint}{\mathop{\bf rel int}}
\newcommand{\bd}{\mathop{\bf bd}}
\newcommand{\vect}{\mathop{\bf vec}}
\newcommand{\dsp}{\displaystyle}
\newcommand{\foequal}{\simeq}
\newcommand{\VOL}{{\mbox{\bf vol}}}
\newcommand{\xopt}{x^{\rm opt}}

\newcommand{\Xb}{{\mbox{\bf X}}}
\newcommand{\xst}{x^\star}
\newcommand{\varphist}{\varphi^\star}
\newcommand{\lambdast}{\lambda^\star}
\newcommand{\Zst}{Z^\star}
\newcommand{\fstar}{f^\star}
\newcommand{\xstar}{x^\star}
\newcommand{\xc}{x^\star}
\newcommand{\lambdac}{\lambda^\star}
\newcommand{\lambdaopt}{\lambda^{\rm opt}}

\newcommand{\geqK}{\mathrel{\succeq_K}}
\newcommand{\gK}{\mathrel{\succ_K}}
\newcommand{\leqK}{\mathrel{\preceq_K}}
\newcommand{\lK}{\mathrel{\prec_K}}
\newcommand{\geqKst}{\mathrel{\succeq_{K^*}}}
\newcommand{\gKst}{\mathrel{\succ_{K^*}}}
\newcommand{\leqKst}{\mathrel{\preceq_{K^*}}}
\newcommand{\lKst}{\mathrel{\prec_{K^*}}}
\newcommand{\geqL}{\mathrel{\succeq_L}}
\newcommand{\gL}{\mathrel{\succ_L}}
\newcommand{\leqL}{\mathrel{\preceq_L}}
\newcommand{\lL}{\mathrel{\prec_L}}
\newcommand{\geqLst}{\mathrel{\succeq_{L^*}}}
\newcommand{\gLst}{\mathrel{\succ_{L^*}}}
\newcommand{\leqLst}{\mathrel{\preceq_{L^*}}}
\newcommand{\lLst}{\mathrel{\prec_{L^*}}}

\newtheorem{theorem}{Theorem}[section]
\newtheorem{corollary}{Corollary}[theorem]
\newtheorem{lemma}[theorem]{Lemma}
\newtheorem{proposition}[theorem]{Proposition}

\newenvironment{algdesc}%
{\begin{quote}}{\end{quote}}

\def\figbox#1{\framebox[\hsize]{\hfil\parbox{0.9\hsize}{#1}}}

\makeatletter
\long\def\@makecaption#1#2{
   \vskip 9pt
   \begin{small}
   \setbox\@tempboxa\hbox{{\bf #1:} #2}
   \ifdim \wd\@tempboxa > 5.5in
        \begin{center}
        \begin{minipage}[t]{5.5in}
        \addtolength{\baselineskip}{-0.95pt}
        {\bf #1:} #2 \par
        \addtolength{\baselineskip}{0.95pt}
        \end{minipage}
        \end{center}
   \else
    \hbox to\hsize{\hfil\box\@tempboxa\hfil}
   \fi
   \end{small}\par
}
\makeatother

\newcounter{oursection}
\newcommand{\oursection}[1]{
 \addtocounter{oursection}{1}
 \setcounter{equation}{0}
 \clearpage \begin{center} {\Huge\bfseries #1} \end{center}
 {\vspace*{0.15cm} \hrule height.3mm} \bigskip
 \addcontentsline{toc}{section}{#1}
}
\newcommand{\oursectionf}[1]{  %
 \addtocounter{oursection}{1}
 \setcounter{equation}{0}
 \foilhead[-.5cm]{#1 \vspace*{0.8cm} \hrule height.3mm }
 \LogoOn
}
\newcommand{\oursectionfl}[1]{  %
 \addtocounter{oursection}{1}
 \setcounter{equation}{0}
 \foilhead[-1.0cm]{#1}
 \LogoOn
}

\newcommand{\Mat}[1]    {{\ensuremath{\mathbf{\uppercase{#1}}}}} %
\newcommand{\Vect}[1]   {{\ensuremath{\mathbf{\lowercase{#1}}}}} %
\newcommand{\Vari}[1]   {{\ensuremath{\mathbf{\lowercase{#1}}}}} %
\newcommand{\Id}				{\mathbb{I}} %
\newcommand{\Diag}[1] 	{\operatorname{diag}\left({ #1 }\right)} %
\newcommand{\Opt}[1] 	  {{#1}_{\text{opt}}} %
\newcommand{\CC}[1]			{{#1}^{*}} %
\newcommand{\Op}[1]     {\Mat{#1}} %
\newcommand{\mini}[1] {{\mbox{argmin}}_{#1} \: \: } %
\newcommand{\argmin}[1] {\underset{{#1}}{\mathop{\rm argmin}} \: \: } %
\newcommand{\argmax}[1] {\underset{{#1}}{\mathop{\rm argmax}} \: \: } %
\newcommand{\minimize}{\mathop{\rm minimize} \: \:}
\newcommand{\minimizeu}[1]{\underset{{#1}}{\mathop{\rm minimize}} \: }
\newcommand{\grad}      {\nabla}
\newcommand{\kron}{\otimes} %

\newcommand{\gradt}     {\grad_\z}
\newcommand{\gradx}     {\grad_\x}
\newcommand{\step}      {\text{\textbf{step}}}
\newcommand{\prox}[1]   {\mathbf{prox}_{#1}}
\newcommand{\ind}[1]    {\operatorname{ind}_{#1}}
\newcommand{\proj}[1]   {\Pi_{#1}}
\newcommand{\pointmult}{\odot} %
\newcommand{\rr}   {\mathcal{R}}

\newcommand{\Basis}{\Mat{D}}         		%
\newcommand{\Corr}{\Mat{C}}             %
\newcommand{\conv}{\ast} %
\newcommand{\meas}{\Vect{b}}            %
\newcommand{\Img}{I}                    %
\newcommand{\img}{\Vect{i}}             %
\newcommand{\vv}{\Vect{v}}
\newcommand{\p}{\Vect{p}}
\newcommand{\Splitvar}{T}                %
\newcommand{\splitvar}{\Vect{t}}         %
\newcommand{\Splitbasis}{J}                %
\newcommand{\splitbasis}{\Vect{j}}         %
\newcommand{\var}{\Vari{z}}

\newcommand{\FT}[1]			{\mathcal{F}\left( {#1} \right)} %
\newcommand{\IFT}[1]			{\mathcal{F}^{-1}\left( {#1} \right)} %

\newcommand{\func}{f}
\newcommand{\fMat}{\Mat{K}}

\newcommand{\avar}{\Vari{v}}
\newcommand{\aspvar}{\Vari{z}}

\newcommand{\mask}{\Mat{M}}

\newcommand{\Pen}      		{F} %
\newcommand{\cardset}     {\mathcal{C}}
\newcommand{\Dat}      		{G} %
\newcommand{\Reg}      		{\Gamma} %

\newcommand{\Trans}{\mathbf{\uppercase{T}}} %
\newcommand{\Ph}{\mathbf{\uppercase{\Phi}}} %

\newcommand{\Tvec}{\Vect{T}} %
\newcommand{\Bvec}{\Vect{B}} %

\newcommand{\Wt}{\Mat{W}} %

\newcommand{\Perm}{\Mat{P}} %

\newcommand{\DiagFactor}[1]     {\Mat{O}_{ #1 }}  %

\newcommand{\Proj}{\Mat{P}}             %

\newcommand{\Vector}[1]{\mathbf{#1}}
\newcommand{\Matrix}[1]{\mathbf{#1}}
\newcommand{\Tensor}[1]{\boldsymbol{\mathscr{#1}}}
\newcommand{\TensorUF}[2]{\Matrix{#1}_{(#2)}}

\newcommand{\MatrixKP}[1]{\Matrix{#1}_{\otimes}}
\newcommand{\MatrixKPN}[2]{\Matrix{#1}_{\otimes}^{#2}}

\newcommand{\MatrixKRP}[1]{\Matrix{#1}_{\odot}}
\newcommand{\MatrixKRPN}[2]{\Matrix{#1}_{\odot}^{#2}}

\newcommand{\HP}{\circ}
\newcommand{\HD}{\oslash}

\newcommand{\leftDB}{\left[ \! \left[}
\newcommand{\rightDB}{\right] \! \right]}

\newcommand{\transpose}{T}

\newcommand*\sstrut[1]{\vrule width0pt height0pt depth#1\relax}

\newcommand{\inlineeqnum}{\refstepcounter{equation}~~\mbox{(\theequation)}}
\newcommand{\eqname}[1]{\tag*{#1~(\theequation)}\refstepcounter{equation}}

\newcommand{\lambdas}{\boldsymbol{\lambda}}
\newcommand{\alb}{\boldsymbol{\alpha}} 	%
\newcommand{\depth}{\boldsymbol{z}} 	%
\newcommand{\albi}{\alpha} 	%
\newcommand{\depthi}{z} 	%
\newcommand{\ambient}{s}
\newcommand{\jitter}{w}
\newcommand{\z}{\Vect{z}} 							%
\newcommand{\x}{\Vect{x}}             	%
\newcommand{\y}{\Vect{y}}             	%
\newcommand{\Kvar}{\Mat{K}}
\newcommand{\lagrangemult}{\boldsymbol{\nu}}
\newcommand{\scaledlagrange}{\Vect{u}}
\newcommand{\eps}{\epsilon}
\newcommand{\vp}{\Vect{v}}

\newpage
\begin{center}
    {\Large \textbf{Table of contents}}
\end{center}
\vspace{0.5cm} %
\textbf{Supplementary Note 1.} Derivation and validation of the depth-wavelength symmetry model. \\ \\
\textbf{Supplementary Note 2.} Broadband metalens design in the visible. \\ \\
\textbf{Supplementary Note 3.} Comparison with other obstruction-free imaging methods. \\ \\
\textbf{Supplementary Note 4.} Details on image simulation and metalens learning. \\ \\
\textbf{Supplementary Note 5.} Details on image reconstruction neural network. \\ \\
\textbf{Supplementary Note 6.} Measured complex refractive index of the $\text{SiN}_x$ film. \\ \\
\textbf{Supplementary Note 7.} Simulated conversion efficiency under variation of geometric parameters. \\ \\
\textbf{Supplementary Note 8.} Fabrication process of metalenses. \\ \\
\textbf{Supplementary Note 9.} Details on the experimental setup. \\ \\
\textbf{Supplementary Note 10.} Details on the image quality assessment. \\ \\
\textbf{Supplementary Note 11.} Details on the downstream vision tasks. \\ \\
\textbf{Supplementary Note 12.} Metalens point spread functions and optical MTFs. \\ \\
\textbf{Supplementary Note 13.} Slanted edge MTFs. \\ \\
\textbf{Supplementary Note 14.} Additional experimental imaging results. \\ \\
\textbf{Supplementary Note 15.} Robustness to illumination spectrum, polarization, and intensity. \\ \\
\textbf{Supplementary Note 16.} Comparison with computational inpainting baselines. \\ \\
\textbf{Supplementary Note 17.} Additional real-scene downstream vision demonstrations.

\newpage 

\section*{Supplementary Note 1: Derivation and validation of the depth-wavelength symmetry model}
\label{supp:note1}
\subsection{Derivation.}
We start with the ideal broadband focusing phase profile:
\begin{align}
\phi(x,y,\omega) &= -\frac{\omega}{c}(\sqrt{x^2+y^2+f^2}-f), 
\label{eq:brodaband_hyperbolic_phase_profile}
\end{align}
where $x, y$ denote pupil-plane coordinate, $\omega$ is the angular frequency, $c$ is the speed of light, and $f$ is the focal length.

The ideal lens profile is expanded around a certain design angular frequency $\omega_{\text{d}}$ as
\begin{align}
\phi(x,y, \omega) = \phi(x,y,\omega_{\text{d}})+\frac{\partial\phi(x,y,\omega)}{\partial\omega }\Bigr|_{\omega=\omega_{\text{d}}}(\omega-\omega_{\text{d}}),
\label{eq:brodaband_hyperbolic_phase_profile_expanded}
\end{align}
 The residual first-order derivative term $\frac{\partial\phi(x,y,\omega)}{\partial\omega }\Bigr|_{\omega=\omega_{\text{d}}}(\omega-\omega_{\text{d}})$ induces chromatic aberration and our intuition is that the first-order derivative term can be matched with the conjugate phase of the spherical phase of a point source. Specifically, the spherical phase profile of a monochromatic point light source of frequency $\omega_{\text{t}}$ positioned at depth $z$ and normally incident on a metalens is defined by the following:
\begin{align}
\phi_{\text{S}}(x, y, z, \omega_{\text{t}})=\frac{\omega_{\text{t}}}{c}\sqrt{x^2+y^2+z^2}.
\label{eq:point_light_source}
\end{align}

We simplify Equation~\eqref{eq:point_light_source} using the paraxial approximation:
\begin{align}
\phi_{\text{S}}(x, y, z, \omega_{\text{t}})=\frac{\omega_{\text{t}}z}{c}\sqrt{\frac{x^2+y^2}{z^2}+1}\simeq\frac{\omega_{\text{t}}}{c}(z+\frac{x^2+y^2}{2z})=\frac{\omega_{\text{t}}}{c}\frac{x^2+y^2}{2z},
\label{eq:point_light_source_approximated}
\end{align}
where the global constant phase term $\frac{\omega_{\text{t}}}{c}z$ is ignored. 

For the derivation of Equation~\eqref{eq:point_light_source_approximated}, the paraxial approximation used is as follows
\begin{align}
\sqrt{1+b}=1+\frac{b}{2}-\frac{b^2}{8} + \dotsm \simeq 1+\frac{b}{2}.
\label{eq:binomial expansion}
\end{align}

The approximated spherical wave term in Equation~\eqref{eq:point_light_source_approximated} is now equated with the first-order derivative term of the phase profile $\frac{\partial\phi(x,y,\omega)}{\partial\omega }\Bigr|_{\omega=\omega_{\text{d}}}(\omega-\omega_{\text{d}})$ as
\begin{align}
\left.\frac{\partial \phi(x,y,\omega)}{\partial \omega}\right|_{\omega=\omega_{\mathrm{d}}}
(\omega_{\mathrm{t}}-\omega_{\mathrm{d}})
= \frac{\omega_{\mathrm{t}}}{c}
   \frac{x^{2}+y^{2}}{2z},
\label{eq:dispersion_compensation}
\end{align}
where $\frac{\partial\phi(x,y,\omega)}{\partial\omega }\Bigr|_{\omega=\omega_{\text{d}}}=-\frac{1}{c}(\sqrt{x^2+y^2+f^2}-f)$. Solving Equation~\eqref{eq:dispersion_compensation} with the paraxial approximation ($\sqrt{x^2+y^2+f^2} = f\sqrt{\frac{x^2+y^2}{f^2}+1} \simeq \frac{x^2+y^2}{2f}+f$) yields the following relationship:
\begin{align}
z &= \frac{x^{2}+y^{2}}
        {\displaystyle
         \frac{2c}{\omega_{\mathrm{t}}}
         \left.\frac{\partial \phi(x,y,\omega)}{\partial \omega}\right|_{\omega=\omega_{\mathrm{d}}}
         (\omega_{\mathrm{t}}-\omega_{\mathrm{d}})} \notag\\
&= \frac{(x^{2}+y^{2})\lambda_{\mathrm{d}}}
        {2\bigl(\sqrt{x^{2}+y^{2}+f^{2}}-f\bigr)(\lambda_{\text{t}}-\lambda_{\text{d}})}
\simeq \frac{1}{2}\frac{(x^{2}+y^{2})\lambda_{\mathrm{d}}}
        {\left(\dfrac{x^{2}+y^{2}}{2f}\right)(\lambda_{\text{t}}-\lambda_{\text{d}})} = \frac{\lambda_{\mathrm{d}}f}{(\lambda_{\text{t}}-\lambda_{\text{d}})},
\label{eq:depth_wavelength_relationship}
\end{align}
where $\omega = \frac{2\pi c}{\lambda}$. Finally, the relationship between $\Delta\lambda$ and $\Delta z$ of the depth-wavelength symmetry is derived from Equation~\eqref{eq:depth_wavelength_relationship} by defining $\Delta \lambda = \lambda_1-\lambda_2$, and $\Delta z = z_1 -z_2$:
\begin{align}
\Delta \lambda = \lambda_1 - \lambda _2 &= \lambda_\text{d}f\left(\frac{1}{z_\text{1}}-\frac{1}{z_\text{2}}\right)=\lambda_\text{d}f\left(\frac{1}{z_1}-\frac{1}{z_1-\Delta z}\right).
\label{eq:depth_wavelength_symmetry}
\end{align}

\subsection{Validation of the approximations.}
In the derivation process, we have applied paraxial approximation to the spherical wave and hyperbolic lens phase profiles. While using paraxial approximation is a common strategy in many derivation processes in optics, we evaluate the phase and diffraction angle errors to justify the paraxial approximations in our optical configuration ($f= 4$\,mm and aperture size of 2.516\,mm). We compute the diffraction angle map using the phase map gradients\cite{matsushima2009band, goodman2005introduction} as
\begin{equation}
(\theta_x, \theta_y)=\left(\text{arcsin}\left(\frac{\lambda}{2\pi}\cdot \frac{\partial }{\partial x}\phi(x,y)\right), \text{arcsin}\left(\frac{\lambda}{2\pi}\cdot \frac{\partial }{\partial y}\phi(x,y)\right)\right).
\end{equation}

\begin{figure}[h]
	\centering
		\includegraphics[width=0.94\textwidth]{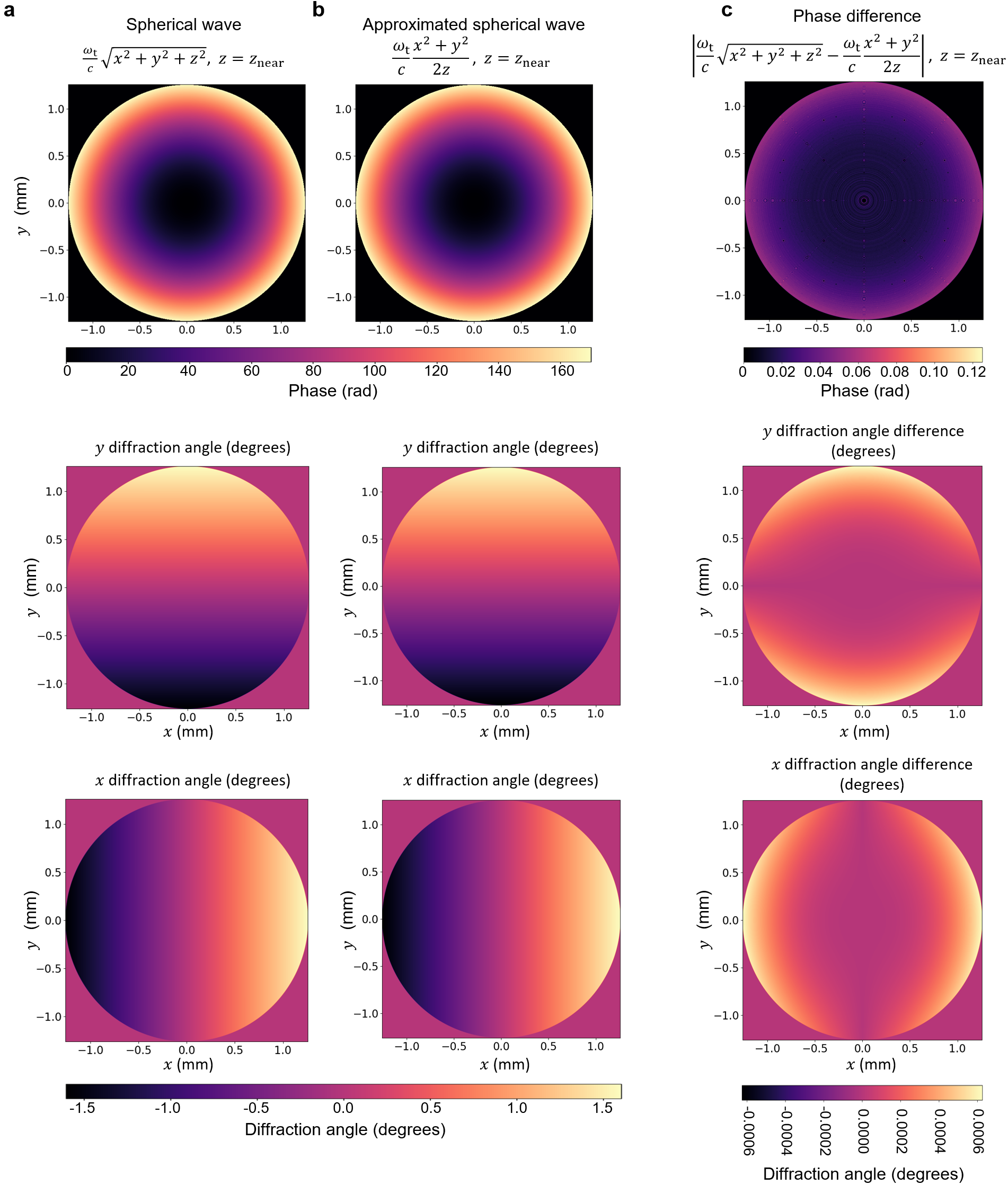}
		\caption{\textbf{Validation of the paraxial approximation for the spherical light source.} First row: Phase maps, second row: $y$-axis diffraction angle maps from the phase maps, third row: $x$-axis diffraction angle maps. \textbf{a} The exact phase map and diffraction angle map of a spherical wave emitted from $z_\text{near}=0.045\,$m, and \textbf{b} its paraxial approximation. \textbf{c} The absolute difference between the exact and the approximated maps.}
		\label{fig:supp_spherical_wave_phase}
\end{figure}

Figure~\ref{fig:supp_spherical_wave_phase}a,b,c visualizes the phase map of the spherical light source $\phi_{\text{S}}(x, y, z, \omega_{\text{t}})$ emitted from a depth of $z=0.045\,\text{m}$, its approximation $\frac{\omega_{\text{t}}}{c}\frac{x^2+y^2}{2z}$, and the difference between the two, along with the diffraction angle map calculated from the phase map and their differences. The maximum value of the diffraction angle map error ratio $\frac{|\text{O}-\text{A}|}{|\text{O}|},$ ($\text{O}$: original, $\text{A}$: approximated) is of 0.03911\,\%. The negligible magnitude of the error validates the use of the approximation for the spherical wavefront. 

\begin{figure}[h]
	\centering
		\includegraphics[width=0.94\textwidth]{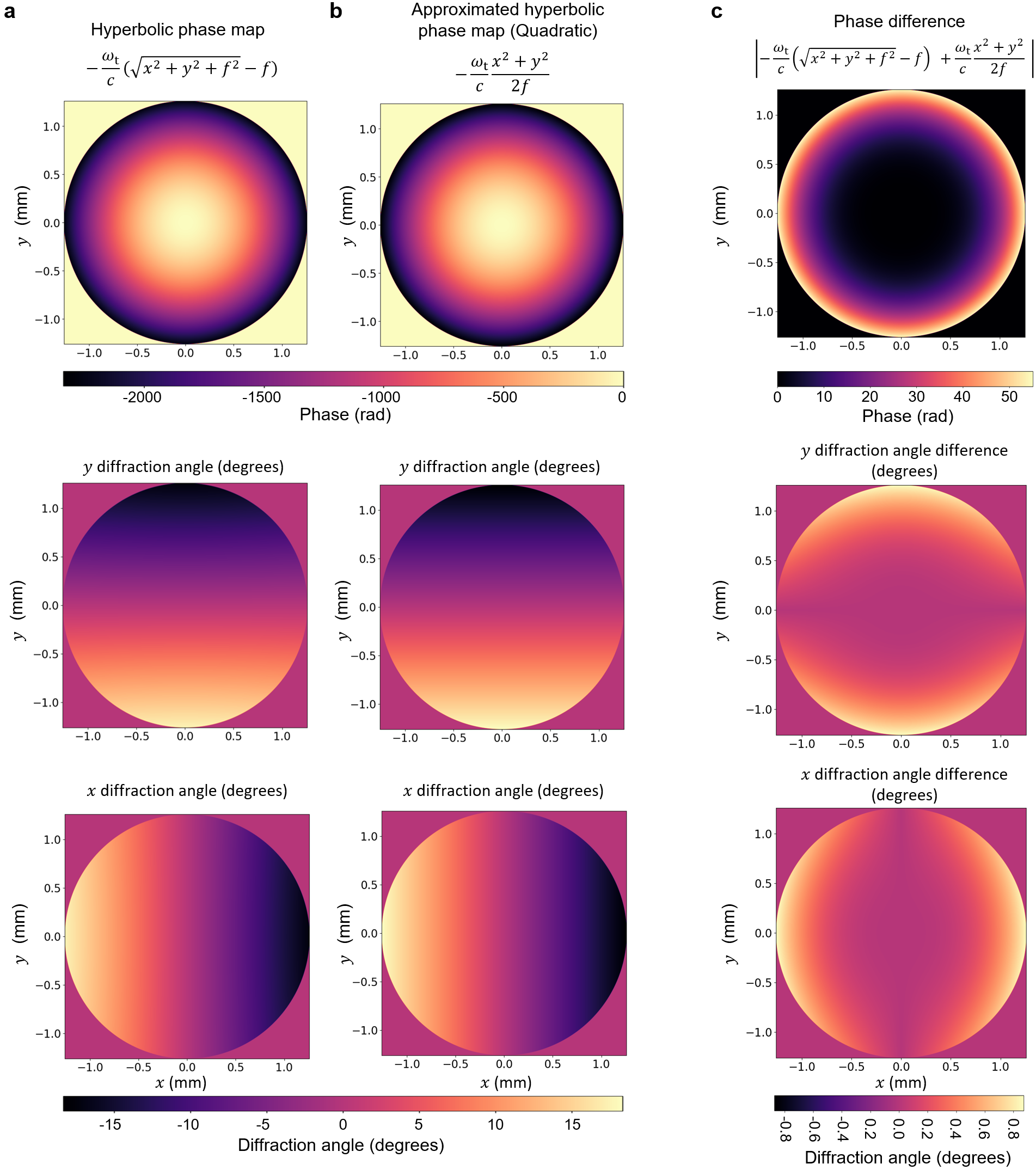}
		\caption{\textbf{Validation of the paraxial approximation for the hyperbolic phase.} First row: Phase maps, second row: $y$-axis diffraction angle maps from the phase maps, third row: $x$-axis diffraction angle maps. \textbf{a} The exact phase map and diffraction angle map of a hyperbolic phase profile designed for a focal length $f=4\,$mm, at $\lambda=532\,$nm and \textbf{b} its paraxial approximation. \textbf{c} The absolute difference between the exact and the approximated maps.}
		\label{fig:supp_hyperbolic_phase}
\end{figure}

\begin{figure}[h]
	\centering
		\includegraphics[width=\textwidth]{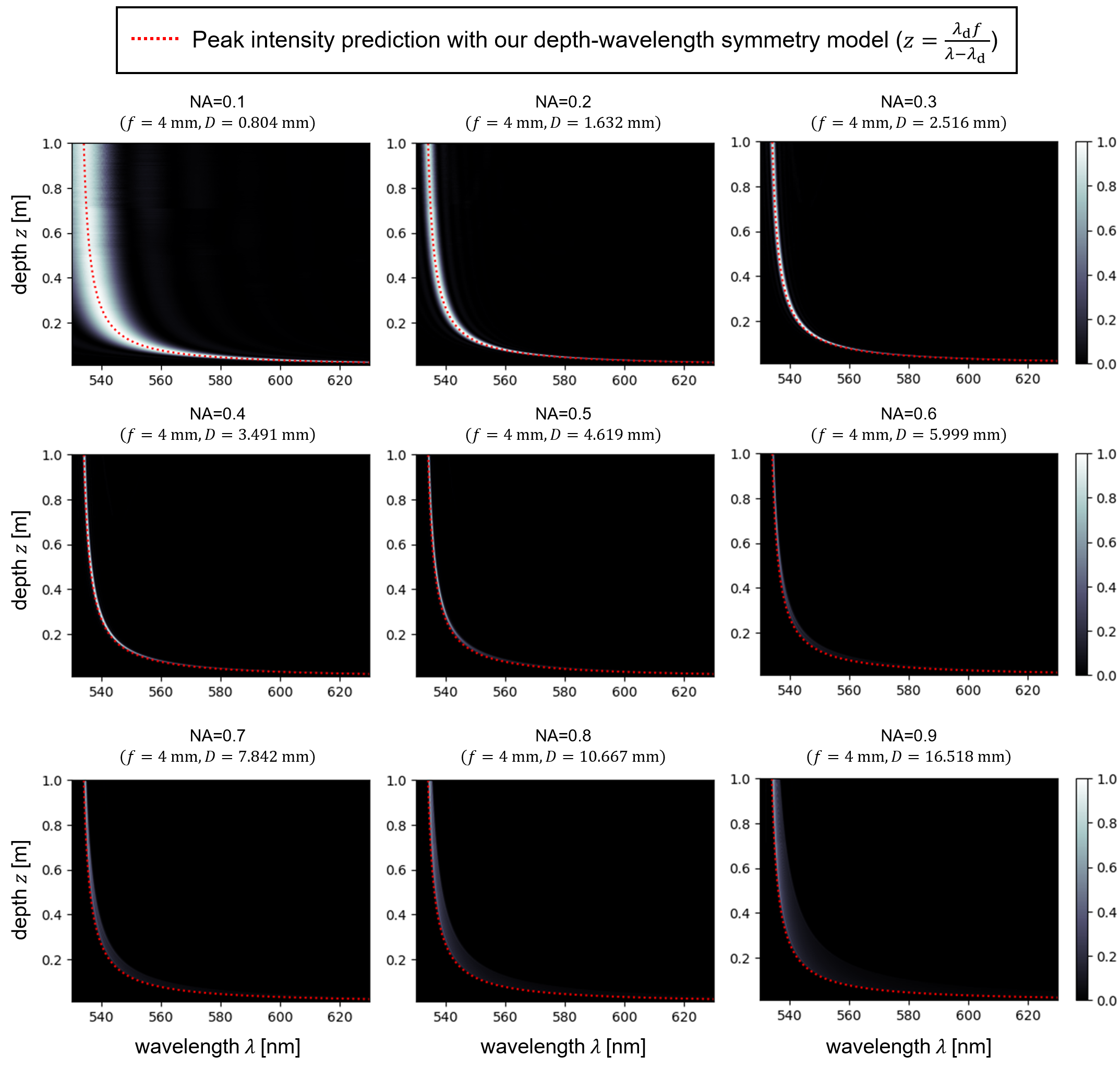}
		\caption{\textbf{Robustness of depth-wavelength symmetry model across various Numerical Apertures (NA).} The focal point intensity depth-spectral maps of hyperbolic phase metalenses are simulated for different NA configurations by varying the lens aperture size while fixing the focal length. The dashed lines indicate the theoretical focal shifts predicted by our depth-wavelength symmetry model. The close agreement between the simulated high-intensity regions and the theoretical curves confirms that the proposed model accurately predicts the focal shift regardless of the approximation errors in the high-NA regime.}
		\label{fig:supp_depth_wavelength_symmetry_per_NAs}
\end{figure}

Next, we examine the approximation of the metalens phase profile. Note that the approximation ($\sqrt{x^2+y^2+f^2} = f\sqrt{\frac{x^2+y^2}{f^2}+1} \simeq \frac{x^2+y^2}{2f}+f$) is approximating the hyperbolic phase map ($-\frac{\omega}{c}(\sqrt{x^2+y^2+f^2}-f)$) to the quadratic phase map ($-\frac{\omega}{c}\frac{x^2+y^2}{2f}$) using the paraxial approximation\cite{goodman2005introduction}. Figure~\ref{fig:supp_hyperbolic_phase}a,b,c compare the exact hyperbolic profile at $\lambda=532\,\text{nm}$ with its quadratic approximation. In this case, the maximum value of the diffraction angle map error ratio increases to 4.997\,\%, indicating the presence of spherical aberration in the quadratic phase map. However, the goal of our model is to predict the focal point shift, rather than the exact PSF structure. Since both the hyperbolic and quadratic phase profiles are defined by the same focal length $f$, the axial positions of their focal points remain almost consistent despite the aberration introduced by the approximation. Consequently, the approximation does not compromise the validity of our depth-wavelength symmetry model. We further validate this claim in Figure~\ref{fig:supp_depth_wavelength_symmetry_per_NAs}, which demonstrates that the predicted focal shifts of hyperbolic phase metalenses align well with numerical simulations across various numerical apertures (NA).

\subsection{Depth-wavelength symmetry model for propagation-phase metalenses.}
While our main implementation uses geometric-phase meta-atoms, propagation-phase (dynamic-phase) meta-atoms are another widely used route to realize metasurfaces. The phase delay of a propagation-phase meta-atom can be written as
\[
\phi(x,y,\omega)=\frac{\omega}{c}\,n_{\mathrm{eff}}(x,y,\omega)\,h \quad (\mathrm{mod}\ 2\pi),
\]
where $n_\text{eff}(x,y,\omega)$ denotes the effective refractive index of the meta-atom at $(x,y)$ and $h$ is the meta-atom height. The hyperbolic phase metalens can be implemented by the propagation phase meta-atoms, and for the design angular frequency $\omega_\text{d}$, the following holds:
\[
\phi(x,y,\omega_\text{d})=\frac{\omega_\text{d}}{c}n_\text{eff}(x,y,\omega_\text{d})h=-\frac{\omega_\text{d}}{c}(\sqrt{x^2+y^2+f^2}-f)\quad (\mathrm{mod}\ 2\pi).
\]
The corresponding first-order derivative (group-delay term) at the design frequency $\omega_\text{d}$ is defined by the following:
\begin{align}
\left.\frac{\partial\phi(x,y,\omega)}{\partial\omega}\right|_{\omega=\omega_\text{d}}
=&\frac{1}{c}n_\text{eff}(x,y,\omega_\text{d})h
+\frac{\omega_\text{d}}{c}\frac{\partial n_\text{eff} (x,y,\omega)}{\partial\omega }\Bigr|_{\omega=\omega_{\text{d}}}h. \notag
\end{align}
The first term corresponds to the $\omega$-dependent optical path length (OPL) contribution to the group delay, whereas the second term captures the additional contribution from the dispersion of $n_{\mathrm{eff}}$.

Similar to the geometric-phase case, standard propagation-phase metalenses without dispersion engineering do not provide the required group-delay profile for broadband achromatic focusing\cite{chang2024achromatic}, leading to chromatic focal shifts. 
Practically, the OPL-related term provides the dominant spatial trend of the group-delay profile, and the $n_{\mathrm{eff}}$-dispersion term acts as a minor correction\cite{de2025diffraction}.
Motivated by this observation, one can match the residual first-order term with the spherical-wave phase contribution of a finite-depth point source to obtain an approximation of the depth-wavelength symmetry mapping in Equation~\eqref{eq:depth_wavelength_relationship}, given a meta-atom library (or a surrogate model) for $n_\text{eff}$ and $h$.

\begin{figure}[h]
	\centering
		\includegraphics[width=\textwidth]{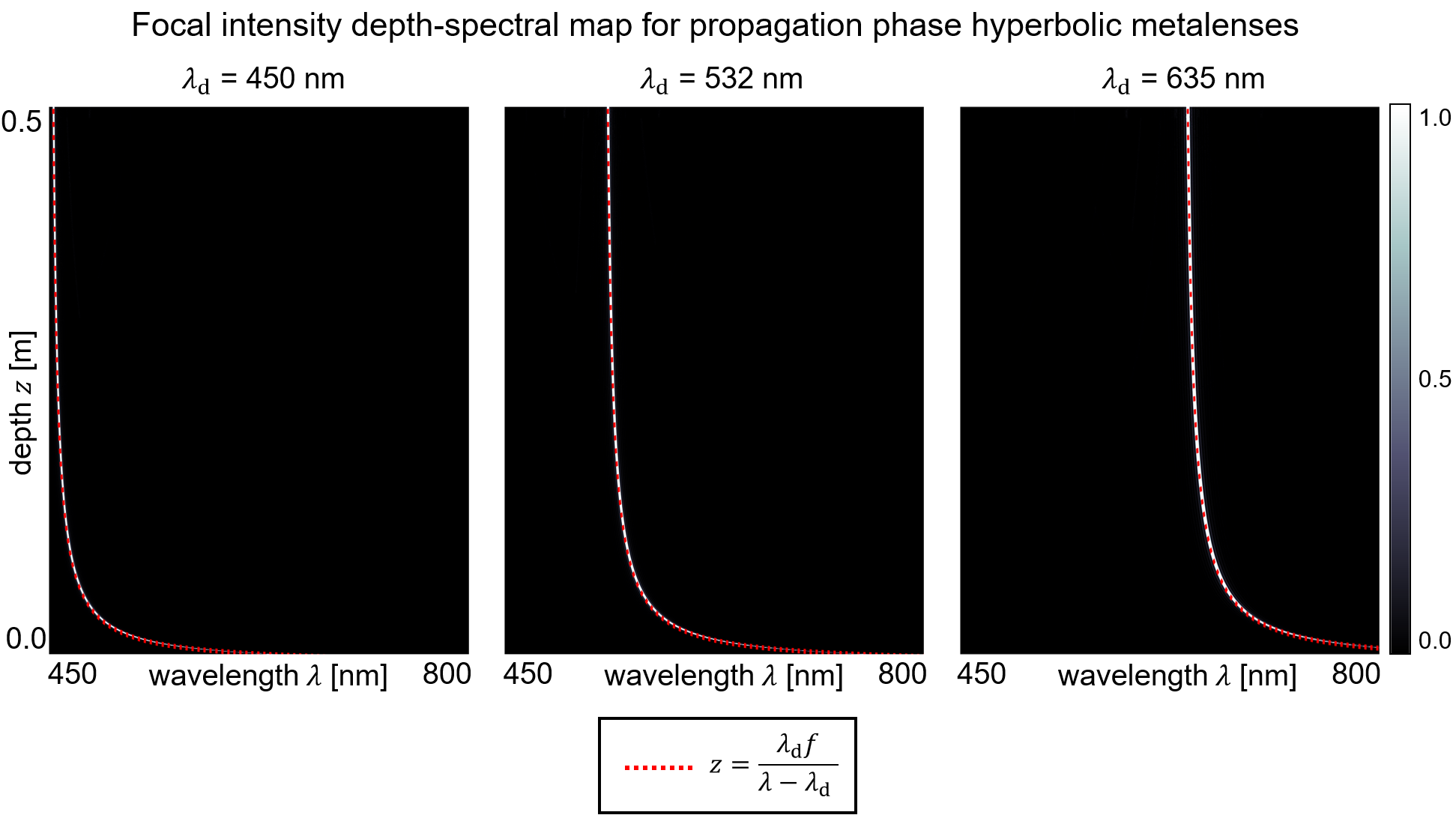}
        \caption{\textbf{Depth-wavelength symmetry in propagation-phase hyperbolic metalenses.}
        Focal intensity depth-spectral maps of propagation-phase hyperbolic metalenses designed at three different design wavelengths, \(\lambda_\text{d}=450\)~nm (left), \(532\)~nm (middle), and \(635\)~nm (right), computed under the same configuration as in the main text (focal length, \(f=4\)~mm and aperture diameter, \(\text{D}=2.516\)~mm). The predicted depth-wavelength symmetry curve (red dotted) from the equation.~\eqref{eq:depth_wavelength_relationship} closely matches the high-intensity regions.}
\label{fig:supp_depth_wavelength_symmetry_propagation_phase}
\end{figure}

We empirically validate this behavior using a publicly available proxy model for propagation-phase meta-atoms\cite{tseng2021neural} in Figure~\ref{fig:supp_depth_wavelength_symmetry_propagation_phase} using the same configuration as in the main text ($\text{D}=2.516\,$mm, $f=4\,$mm), where the model prediction closely agrees with the numerically computed focal shift maps. 
These results provide a useful leading-order description of chromatic focal shifts in propagation-phase metalenses, supporting the applicability of our model beyond geometric-phase implementations, while the accuracy may depend on the strength of $n_{\mathrm{eff}}$ dispersion and the operating bandwidth.

\subsection{Expansion of the depth-wavelength symmetry model to the broadband case.}
The relationship given in Equation~\eqref{eq:depth_wavelength_relationship} implies that a phase map designed to focus a plane light source of $\lambda_{\text{d}}$ also focuses a spherical light source of $\lambda_{\text{t}}$ from depth $z=\frac{\lambda_{\text{d}}f}{\lambda_{\text{t}}-\lambda_{\text{d}}}$ at same focal plane distant at $f$. We extend this principle to a broadband diffractive lens, which is designed to focus plane waves across a continuous spectrum $[\lambda_\text{min}, \lambda_\text{max}]$. This is achieved by spatially multiplexing the phase profiles $\phi(x, y, \omega_\text{d})$ corresponding to $\lambda_\text{d} \in [\lambda_\text{min}, \lambda_\text{max}]$. Consequently, such a lens is expected to focus light from a finite depth $z$ over a shifted broadband spectrum $[\frac{\lambda_\text{min}f}{z}+\lambda_\text{min}, \frac{\lambda_\text{max}f}{z}+\lambda_\text{max}]$ following the depth-wavelength symmetry. To demonstrate this behavior, we define a broadband phase profile by spatially interleaving hyperbolic phase maps:
\begin{equation}
\begin{aligned}
\phi(x, y)_{\text{broadband}} &= \sum_{i=0}^{N-1} \phi_i(x, y) \cdot \mathbb{I}_i(r), \\
\text{where } \phi_i(x, y) &= \frac{2\pi}{\lambda_i} \left( \sqrt{x^2 + y^2 + f^2} - f \right), \\
\lambda_i &= 450\,\text{nm} + i \cdot \Delta \lambda, \quad (N=201, \ \Delta \lambda = 1\,\text{nm}), \\
\mathbb{I}_i(r) &= 
\begin{cases} 
1, & \lfloor r / \Delta r \rfloor \equiv i \pmod{N} \\
0, & \text{otherwise}.
\end{cases}
\end{aligned}
\label{eq:broadband_phase}
\end{equation} 
The phase profiles are configured with a focal length $f = 4\,\text{mm}$ and lens diameter $D = 2.516\,\text{mm}$. The design spectrum $\lambda \in [450\,\text{nm}, 650\,\text{nm}]$ is sampled at $1\,\text{nm}$ intervals, 
resulting in $N=201$ interleaved hyperbolic profiles. Figure~\ref{fig:supp_depth_wavelength_symmetry_broadband_bands}a visualizes the simulated focal point intensity map of this broadband lens. The overlaid analytical curves derived from our depth-wavelength symmetry model accurately predict the high-intensity focal trajectories of the broadband lens. This agreement confirms that our depth-wavelength symmetry model is applicable to diffractive lenses with arbitrary design spectra, effectively predicting the focal shift behavior essential for our split-spectrum strategy and the extended depth-of-field behaviors observed in other studies\cite{colburn2019optical, huang2020design, froch2025beating, tan20213d, xu2025polarization, bayati2022inverse}.
 
The focal intensity maps of the phase profiles in the main text and Figure~\ref{fig:supp_depth_wavelength_symmetry_broadband_bands}a are computed using a circularly symmetric Rayleigh-Sommerfeld integral\cite{froch2025beating}:
\begin{align}
E(r,z) = 2\pi\int_{0}^{R}E(r', 0)\frac{\text{e}^{ikr}}{r}\frac{z}{r}(1+\frac{i}{kr})r'dr',
\end{align}
where $r=\sqrt{(x-x')^2+(y-y')^2+z^2}$.

The core mechanism of our obstruction-free imaging is to ensure that the sharp PSFs, optimized for far-depth targets within the pass bands, are spectrally shifted into the stop bands when the source is located at near depth. This configuration guarantees that the sharp focal components are rejected by the filter; consequently, the remaining optical response for near-depth obstructions within the pass bands becomes dominated by the defocused profiles, effectively rendering the obstructions blurry.

\subsection{Selection of pass bands and stop bands.}
\begin{figure}[h]
	\centering
		\includegraphics[width=0.85\textwidth]{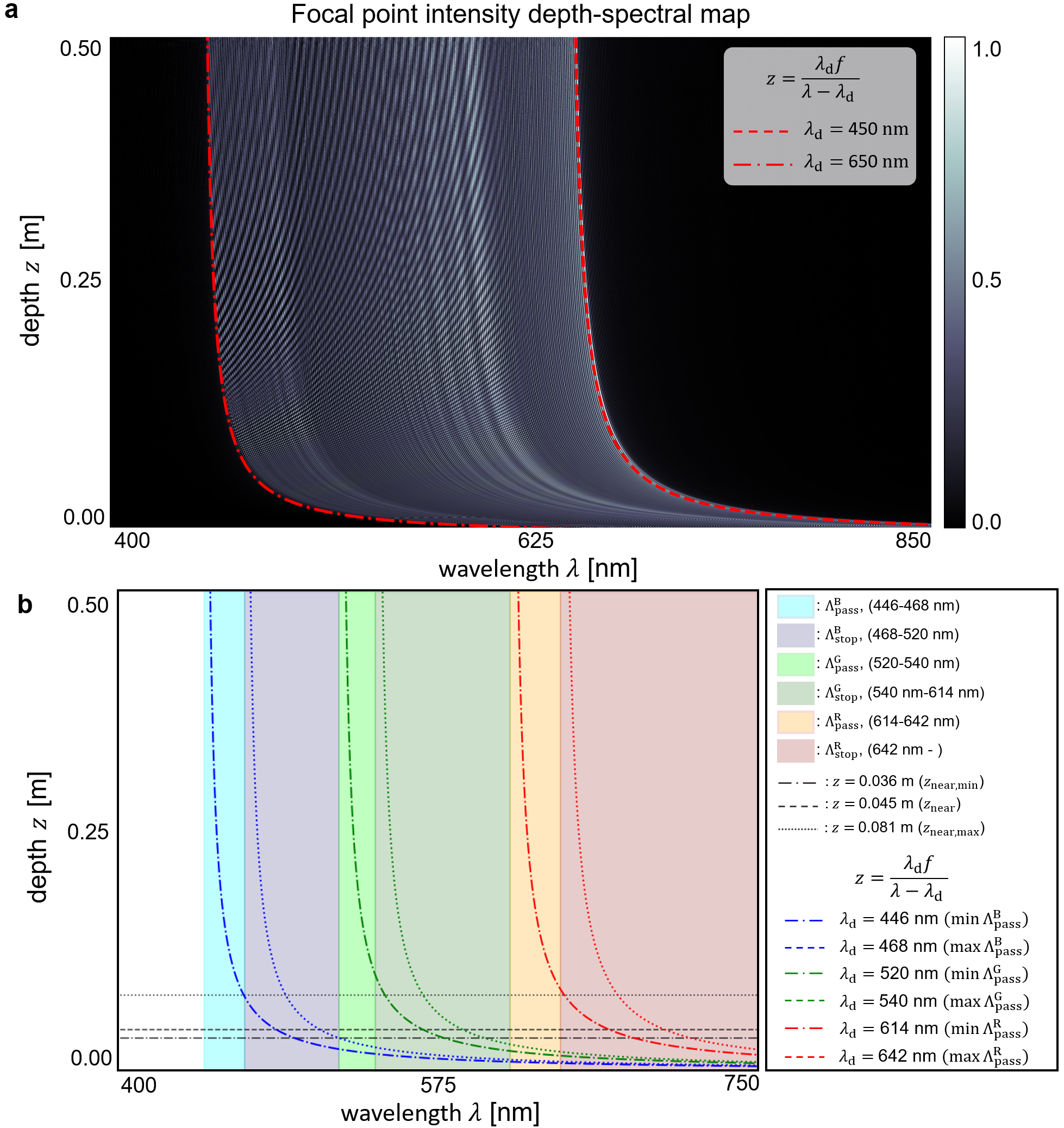}
		\caption{\textbf{Validation of the depth-wavelength symmetry model and band selection for obstruction-free imaging.} \textbf{a} The simulated focal point intensity map of the broadband metalens plotted against wavelength and depth. The overlaid dashed red curves represent the theoretical focal shifts predicted by our depth-wavelength symmetry model for lower and upper spectral limits of the design spectrum, which align with the leftmost and rightmost curves of the simulated high-intensity regions. \textbf{b} The spectral configuration of pass bands and stop bands determined by the multi-band spectral filter. The dashed curves indicate the shifted sharp response wavelengths for the lower and upper limits of the pass bands. The horizontal line at \(z_{\mathrm{near}}=0.045~\mathrm{m}\) denotes the experimental near-depth setting, and the two additional horizontal lines denote the analytic near-depth rejection window, \(z_{\mathrm{near},\min}\simeq0.036~\mathrm{m}\) and \(z_{\mathrm{near},\max}\simeq0.081~\mathrm{m}\). Within this window, the shifted sharp near-depth responses fall into the stop bands or outside the sensor-sensitive/filter-transmitted spectral range.}
		\label{fig:supp_depth_wavelength_symmetry_broadband_bands}
\end{figure}

The mechanism of our obstruction-free imaging is to ensure that the sharp PSFs, designed for far-depth targets within the pass bands, are spectrally shifted into the sensor's stop bands when the source is located at near depth; consequently, these sharp components are rejected, and provide the flexibility to enforce strong blur for near-depth obstructions within the pass bands. We use a commercially available multi-band spectral filter (Edmund Optics, 457, 530 \& 628\,nm, 50\,mm Dia., Tri-Band Filter) whose transmission bandwidths are 22\,nm, 20\,nm, and 28\,nm at center wavelengths 457\,nm (B), 530\,nm (G), and 628\,nm (R). This determines the placement of pass bands, $\Lambda_{\text{pass}}^{c}$ and stop band, $\Lambda_{\text{stop}}^{c}$, which is illustrated in Figure~\ref{fig:supp_depth_wavelength_symmetry_broadband_bands}b. Based on the depth-wavelength symmetry, the condition for obstruction removal requires that the focused near-depth PSFs are spectrally shifted to fall within the stop bands for every color channel. This can be formalized using Equation~\eqref{eq:depth_wavelength_relationship} as a set of inequalities:
\begin{equation}
\begin{aligned}
    f\cdot \frac{\min{\Lambda_\text{pass}^\text{b}}}{z_\text{near}} + \min{\Lambda_\text{pass}^\text{b}} > \min{\Lambda_\text{stop}^\text{b}},  \\
    f\cdot \frac{\max{\Lambda_\text{pass}^\text{b}}}{z_\text{near}} + \max{\Lambda_\text{pass}^\text{b}} < \min{\Lambda_\text{pass}^\text{g}},  \\
    f\cdot \frac{\min{\Lambda_\text{pass}^\text{g}}}{z_\text{near}} + \min{\Lambda_\text{pass}^\text{g}} > \min{\Lambda_\text{stop}^\text{g}}, \\
    f\cdot \frac{\max{\Lambda_\text{pass}^\text{g}}}{z_\text{near}} + \max{\Lambda_\text{pass}^\text{g}} < \min{\Lambda_\text{pass}^\text{r}}, \\
    f\cdot \frac{\min{\Lambda_\text{pass}^\text{r}}}{z_\text{near}} + \min{\Lambda_\text{pass}^\text{r}} > \min{\Lambda_\text{stop}^\text{r}}.
\end{aligned}
\label{eq:bands_conditions}
\end{equation}
As depicted in Figure~\ref{fig:supp_depth_wavelength_symmetry_broadband_bands}b, our chosen configuration with ($z_\text{near}=0.045\,\text{m}$, $f=4\,\text{mm}$) satisfies these conditions. The dashed horizontal line at $z=0.045\,$m intersects the spectral curves in regions that align with the gray stop bands, or outside the sensor sensitivity. This ensures that the sharp focal components corresponding to the far-depth focus are spectrally shifted into the stop bands and rejected, thereby granting us the design freedom to explicitly optimize the remaining pass-band response for strong defocus at near depth.

Note that since Equation~\eqref{eq:depth_wavelength_relationship} contains an asymptote, fixing $z_{\text{near}}$ does not uniquely determine $z_{\text{far}}$; we set the threshold of the $z_{\text{far}}$ as $0.5\,$m, and depth value $z$ greater than $0.5\,$m will satisfy the relationship, and the working range for sharp imaging is not limited. The configuration $z_\text{near}$ and $f$ can be flexibly determined using Equation~\eqref{eq:depth_wavelength_relationship}. Custom-designing the multi-band spectral filter would offer additional degrees of freedom to select the placement and bandwidth of each band. Specifically, we can split the spectra of each color channel into more than two bands and interleave the pass bands. 

\subsection{Finite near-depth rejection window and bandwidth-dependent trade-off.}
Although the experimental characterization was performed at \(z_{\mathrm{near}}=0.045~\mathrm{m}\), the image-formation model used during metalens learning sampled the near-depth obstruction from \(z_{\mathrm{near}}\in[0.04,0.05]~\mathrm{m}\). Therefore, the demonstrated system could operate over a finite near-depth range centered around the target obstruction depth.

This finite operating range can be estimated analytically from the depth-wavelength symmetry model. By solving Equation~\eqref{eq:bands_conditions}, we obtain
\[
z_{\mathrm{near},\min}\simeq0.036~\mathrm{m},
\qquad
z_{\mathrm{near},\max}\simeq0.081~\mathrm{m}.
\]
The experimental setting \(z_{\mathrm{near}}=0.045~\mathrm{m}\) and the training range \(z_{\mathrm{near}}\in[0.04,0.05]~\mathrm{m}\) lie within this analytic rejection window, as indicated in Figure~\ref{fig:supp_depth_wavelength_symmetry_broadband_bands}b.

Table~\ref{tab:supp_bandwidth_depth_range} further shows how this analytic near-depth rejection window changes when the pass-band bandwidth is varied while keeping the band centers fixed at 457, 530, and 628 nm. Narrower pass bands increase the depth interval over which the shifted sharp near-depth responses remain rejected by the stop bands. However, this extension comes at the cost of reduced photon throughput and reduced spectral information, which can degrade signal-to-noise ratio and color fidelity. Conversely, wider pass bands increase photon throughput but narrow the valid rejection window, and sufficiently wide pass bands can eliminate a continuous rejection window because the shifted sharp-response wavelengths overlap with neighboring pass bands. 

Therefore, range-dependent near-depth occlusions can be addressed by co-designing the metalens and spectral filter for a target near-depth distribution, for example, by narrowing or interleaving pass bands and by sampling \(z_{\mathrm{near}}\) from the desired depth distribution during optimization. A tunable or application-specific spectral filter could also shift or reconfigure the rejection window for different expected occlusion-depth ranges, provided that the filter response remains co-designed with the metalens pass-band optimization. However, very broad volumetric occlusions, such as dense airborne rain distributed over a large depth range, remain outside the intended operating regime of the present single-shot single-view implementation and would likely require additional cues such as temporal information, multiple views, or adaptive spectral filtering.

\subsection{Near-depth limit without wavelength-selective filtering.}
A sufficiently close obstruction can be strongly defocused even without wavelength-selective filtering. This occurs when the sharp far-depth PSFs are spectrally shifted outside the detectable spectrum of the sensor. Using Equation~\eqref{eq:depth_wavelength_relationship}, this condition can be formulated as the following inequality:
\[
 f\cdot \frac{\min{\text{CRF}}}{z_\text{near}} + \min{\text{CRF}} > \max{\text{CRF}},
\]
where $\text{CRF}$ is the camera response function. Then, no detectable wavelength produces a sharp near-depth obstruction image, and wavelength-selective filtering is not necessary for near-depth suppression. 

\subsection{Additional validation across NAs.}
To further examine the applicability of our framework across a broad range of numerical apertures, with particular attention to the higher-NA regime, we additionally learned de-occluding broadband metalenses for NA values ranging from 0.1 to 0.9 in increments of 0.1 and analyzed their wavelength-dependent PSF behavior. Although the analytical depth--wavelength symmetry model in Equation~\eqref{eq:depth_wavelength_relationship} is derived using the paraxial approximation, its role in our framework is to provide a leading-order physical guide for the depth-wavelength coupling and for pass-band/stop-band selection, rather than to serve as the final forward model during optimization. The actual metalens optimization is performed using numerical propagation through the differentiable PSF and image simulators.

\begin{figure}[h]
	\centering
		\includegraphics[width=\textwidth]{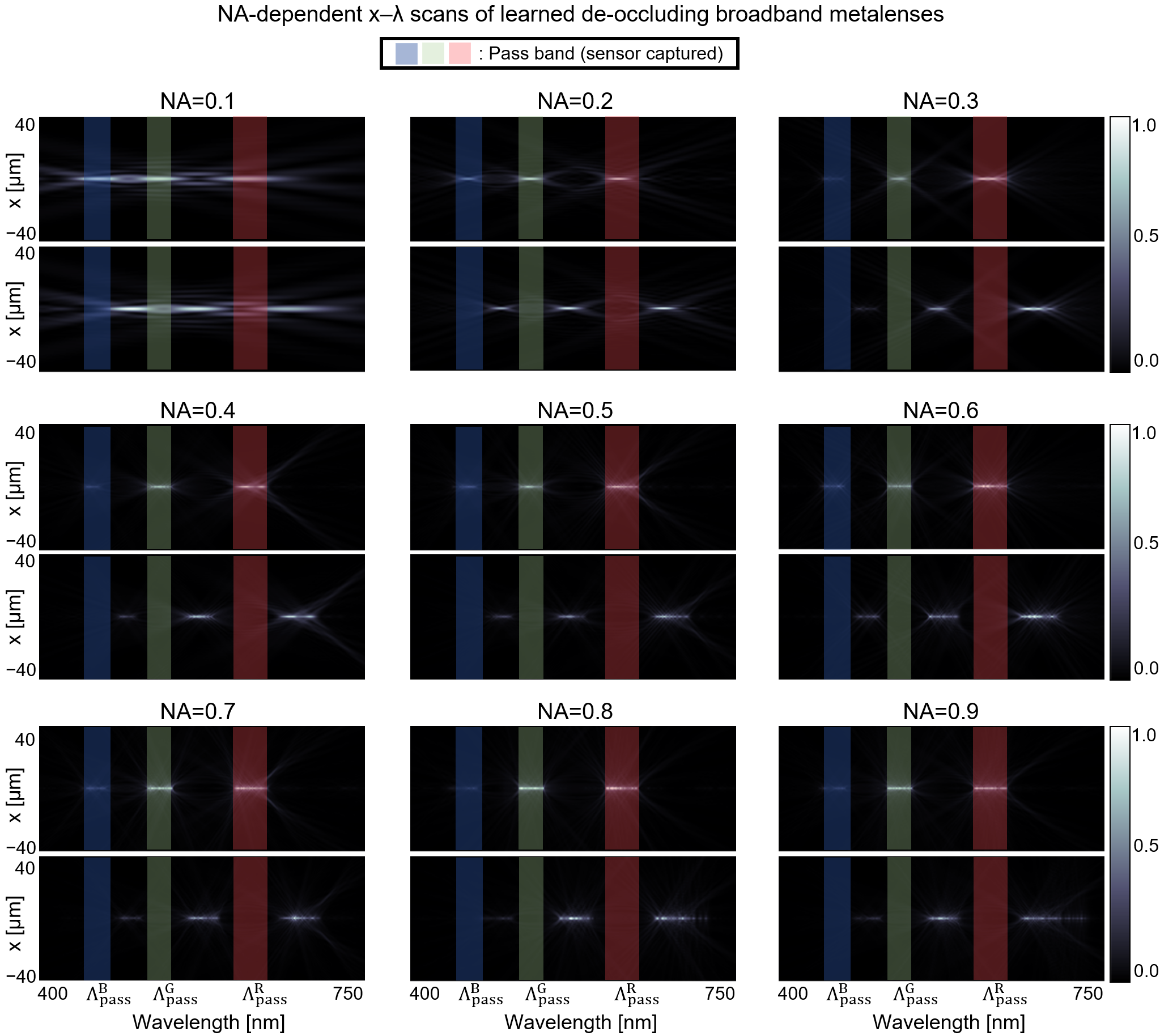}
		\caption{\textbf{NA-dependent wavelength scans of learned de-occluding broadband metalenses.} Each panel corresponds to a separately learned de-occluding broadband metalens with NA ranging from 0.1 to 0.9 in increments of 0.1. For each NA, the upper strip shows the far-depth $(z=\infty)$ $x$--$\lambda$ scan, and the lower strip shows the near-depth $(z=0.045\,\mathrm{m})$ scan. The colored vertical bands indicate the pass bands of the tri-band filter. Across the full NA range, the far-depth PSFs remain concentrated within the pass-band wavelengths, whereas the sharp near-depth spectral components are shifted into stop-band wavelengths or outside the sensor sensitivity range, leaving the near-depth response within the pass bands strongly suppressed and defocused. These results show that the learned de-occluding broadband metalens framework preserves the desired split-spectrum behavior across a wide NA range, including the higher-NA regime. For these additional simulations, a shorter focal length ($f = 1\,\mathrm{mm}$) was used to keep the optimization and numerical propagation tractable under GPU memory constraints.}
		\label{fig:supp_higher_NA_validation_of_the_learned_de-occluding_broadband_metalens_framework}
\end{figure}

Figure~\ref{fig:supp_higher_NA_validation_of_the_learned_de-occluding_broadband_metalens_framework} shows the far-depth and near-depth $x$--$\lambda$ scans of the nine learned de-occluding broadband metalenses across the tested NA range. Across all tested NAs, the sharp far-depth PSFs remain concentrated within the pass-band wavelengths. At near depth, the sharp spectral components are shifted into stop-band wavelengths or outside the sensor sensitivity range, leaving the near-depth response within the pass bands strongly suppressed and defocused. The persistence of this split-spectrum behavior from NA = 0.1 to 0.8, including the higher-NA cases, indicates that the proposed framework remains applicable beyond the moderate-NA regime, while the detailed high-NA optical behavior is naturally handled by numerical optimization and refinement rather than by the paraxial analytical model alone. For these additional NA-sweep simulations, we used a shorter focal length ($f = 1\,\mathrm{mm}$) to keep the optimization and propagation computationally tractable under GPU memory constraints while preserving the relevant NA-dependent behavior.

We also examined whether increasing NA alone could suppress near-depth obstructions without wavelength-selective filtering. A high-NA design reduces the monochromatic depth of field and can therefore increase ordinary defocus blur for a single wavelength. However, for chromatic diffractive lenses, the depth-wavelength symmetry still produces wavelength-shifted sharp responses for near-depth sources. To leading order by the depth-wavelength symmetry, these wavelength-shifted sharp responses are determined by $f$, $z_{\mathrm{near}}$, and $z_{\mathrm{far}}$, not by NA. Therefore, increasing NA does not eliminate the depth-wavelength symmetry itself.

\begin{figure}[h]
	\centering
		\includegraphics[width=\textwidth]{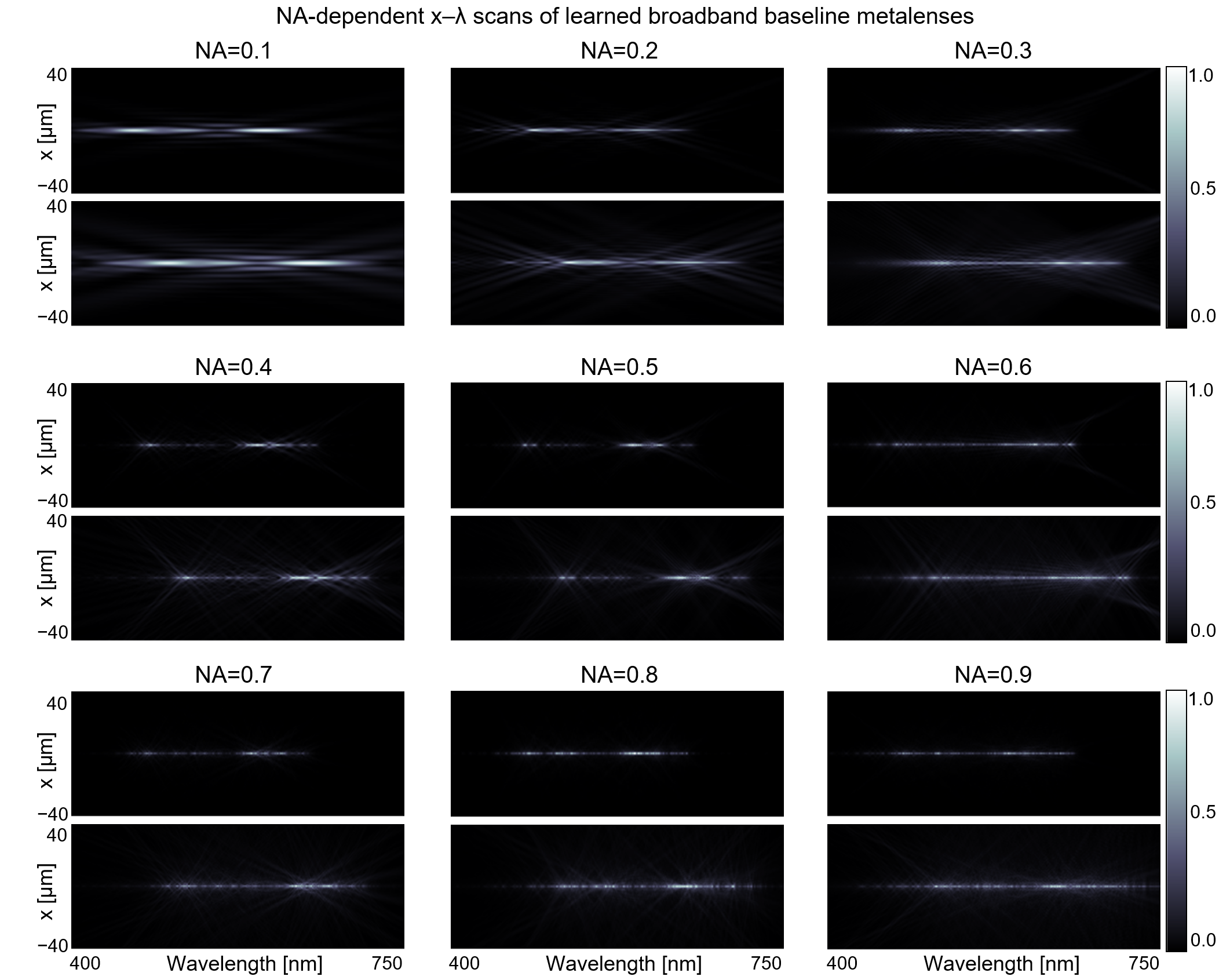}
		\caption{\textbf{NA-dependent wavelength scans of broadband baseline metalenses.} Each panel corresponds to a separately learned broadband baseline metalens with NA ranging from 0.1 to 0.9 in increments of 0.1. For each NA, the upper strip shows the far-depth $(z=\infty)$ $x$--$\lambda$ scan, and the lower strip shows the near-depth $(z=0.045\,\mathrm{m})$ scan.
        Increasing NA reduces the monochromatic depth of field and changes the PSF width, but the wavelength-shifted sharp-response branches predicted by the depth--wavelength symmetry model remain visible. Therefore, high NA alone does not eliminate near-depth sharp obstruction components when the shifted sharp wavelengths remain within the detectable spectral range, and wavelength-selective filtering is required for the de-occlusion.}
        \label{fig:supp_higher_NA_validation_of_the_broadband_baseline_metalens_framework}
\end{figure}

To verify this point, we trained Broadband baseline metalenses with different NAs and simulated their PSF wavelength scans. As shown in Figure~\ref{fig:supp_higher_NA_validation_of_the_broadband_baseline_metalens_framework}, the near-depth sharp-response remain present across the tested NA values. Increasing NA changes the PSF width and the monochromatic depth of field, but it does not remove the wavelength-shifted near-depth sharp responses. Therefore, high NA alone cannot reliably suppress near-depth obstructions over the broadband visible spectrum when the shifted sharp PSFs remain inside the detectable spectral range. Therefore, wavelength-selective filtering remains necessary to reject those shifted sharp components. 


\begin{table}
\centering
\setlength{\tabcolsep}{3pt}
\begin{footnotesize}
\makebox[\columnwidth][c]{%
\begin{tabular}{c c c c c}
\toprule
\textbf{Bandwidth} &
\textbf{Pass-band width} &
\textbf{Rejection window} &
\textbf{Window} &
\textbf{Relative} \\
\textbf{scale} \(s\) &
\textbf{B/G/R (nm)} &
\(z_{\mathrm{near}}\) \textbf{(mm)} &
\textbf{span (mm)} &
\textbf{throughput} \\
\midrule
2.00 & 44/40/56 & None & -- & 2.00 \\
1.50 & 33/30/42 & 45.6--53.4 & 7.8 & 1.50 \\
1.25 & 27.5/25/35 & 40.3--64.5 & 24.2 & 1.25 \\
\addlinespace
\textbf{1.00} & \textbf{22/20/28} & \textbf{36.0--81.1} & \textbf{45.1} & \textbf{1.00} \\
\addlinespace
0.75 & 16.5/15/21 & 32.5--108.8 & 76.3 & 0.75 \\
0.50 & 11/10/14 & 29.6--164.2 & 134.6 & 0.50 \\
0.25 & 5.5/5/7 & 27.1--330.4 & 303.2 & 0.25 \\
\bottomrule
\end{tabular}%
}
\end{footnotesize}
\caption{\textbf{Analytic near-depth rejection window as a function of pass-band bandwidth.}
The pass-band centers are fixed at 457, 530, and 628 nm, and the pass-band widths are scaled relative to the commercial tri-band filter used in this work. For a bandwidth scale \(s\), the B/G/R pass-band widths are \(22s/20s/28s\) nm, respectively. The near-depth rejection window is calculated from the band-edge inequalities derived from the depth--wavelength symmetry model with \(f=4~\mathrm{mm}\), under an ideal rectangular-band approximation. ``None'' indicates that the inequalities cannot be simultaneously satisfied and therefore no continuous near-depth rejection window exists. The relative throughput is approximated by the total pass-band width normalized to that of the current filter, assuming identical peak transmission and flat spectral response. The estimated depth range should be interpreted as an analytic spectral-rejection window rather than a full imaging-performance guarantee, as actual performance can be affected by residual PSF quality, sensor spectral response and meta-atom efficiency.}
\label{tab:supp_bandwidth_depth_range}
\end{table}

\clearpage 

\section*{Supplementary Note 2: Broadband metalens design in the visible}
\label{supp:note2}
Table~\ref{tab:supp_broadband_metalenses} provides a comparative benchmark of recent visible broadband diffractive lens studies based on their NA, aperture size, achromaticity, and reliance on computational backend.
Our approach, aligned with other high-performance computational imaging systems, exhibits depth-wavelength symmetry and leverages a computational backend to achieve a significantly larger aperture of $2500\,\mu$m and a high NA of 0.3.
The derivation provided in the Supplementary Note~1 assumes that the diffractive lens is chromatic, or the diffractive lens cannot satisfy the focusing phase profiles for wavelengths other than $\lambda_{\text{d}}$. The depth-wavelength symmetry model we propose is not applied to achromatic diffractive lenses. The achromatic diffractive lenses are typically achieved by dispersion-engineering that controls phase delays along with the group delay and group delay dispersion. However, the dispersion engineering strategy to achieve achromatic diffractive lenses is fundamentally constrained by low numerical aperture and small diameters\cite{shrestha2018broadband, presutti2020focusing}, limiting their application to real-world imaging systems. Therefore, numerous studies that aim for real-world metaoptics or diffractive optics imaging used chromatic diffractive lenses to achieve large-aperture and high NA, or fast lens, and were aided by a computational backend to correct aberrations. Therefore, diffractive lenses of the practical scale will exhibit the depth-wavelength symmetry we propose. 

\begin{table}
\centering
\setlength{\tabcolsep}{3pt}
\begin{footnotesize}
\makebox[\columnwidth][c]{%
\begin{tabularx}{\columnwidth}{c c c c c c}
\toprule
\textbf{Reference} & \textbf{NA} & \textbf{Aperture} & \centering\textbf{Dispersion engineering} & \textbf{Computational} & \textbf{Depth-wavelength} \\
 & & ($D, \mu$m) & \centering\ ($\Delta\lambda$, nm) & \textbf{backend} & \textbf{symmetry}\\
\midrule
Chen et al.\cite{chen2018broadband} & 0.02 & 220 & Yes (470--670) & No & No \\
Chen et al.\cite{chen2019broadband} & 0.2 & 26.4 & Yes (460--700) & No & No \\
Hu et al.\cite{hu2023asymptotic} & 0.164 & 50 & Yes (400--1000) & No & No \\
Chang et al.\cite{chang2024achromatic} & 0.02 & 400 & Yes (400--700) & No & No \\
Colburn et al.\cite{colburn2019optical} & 0.46 & 200 & No & Yes & Yes \\
Huang et al.\cite{huang2020design} & 0.45 & 200 & No & Yes & Yes \\
Bayati et al.\cite{bayati2022inverse} & 0.46 & 1000 & No & Yes & Yes \\
Fröch et al.\cite{froch2025beating} & 0.24 & 10000 & No & Yes & Yes \\
\addlinespace
\textbf{This work} & \textbf{0.30} & \textbf{2500} & \textbf{No} & \textbf{Yes} & \textbf{Yes} \\
\bottomrule
\end{tabularx}%
}
\end{footnotesize}
\caption{\textbf{Comparison of visible broadband diffractive optics imaging systems.} The table benchmarks their optical specifications, including numerical aperture (NA), aperture size, achromaticity, computational backend reliance, and depth-wavelength symmetry.}
\label{tab:supp_broadband_metalenses}
\end{table}

\clearpage 

\section*{Supplementary Note 3: Comparison with other obstruction-free imaging methods}
\label{supp:note3}
Table~\ref{tab:supp_obstructionfree_comparison} compares representative obstruction-free imaging paradigms using four practical criteria: module compactness, hallucination risk (whether the pipeline must synthesize occluded content rather than directly measure it), and whether the method is single-shot and single-view.

Single-image inpainting methods are attractive for compact deployment because they operate on a single captured image (single-shot, single-view). However, they inherently solve an ill-posed completion problem and may synthesize unseen structures, which introduces fidelity risks and typically incurs high computational cost. Multi-view (or burst) occlusion layer separation reduces hallucination risk by leveraging parallax or motion cues, but it generally requires multiple viewpoints and/or multiple frames, leading to capture overhead and sensitivity to layer/flow estimation errors. Crucially, such approaches are fundamentally inapplicable to obstructions fixed relative to the camera system--such as the cover-glass contaminants targeted in this work--as the absence of relative motion precludes the generation of necessary parallax cues.

Optical approaches based on synthetic apertures can physically suppress near-depth occluders via depth-of-field control and refocusing with low hallucination risk, but they usually require large baselines or camera arrays, resulting in bulky and costly setups that are difficult to integrate into compact platforms. Hybrid diffractive--compound refractive systems can remain single-shot and single-view, yet often rely on bulky compound optics to compensate for limited optical suppression, which negates the advantage of the compactness of the diffractive optics.

In contrast, our de-occluding broadband metalens is designed to achieve a favorable operating point: a single-layer metalens with a monolithic form factor that is both single-shot and single-view with low hallucination risk. 

\begin{table}
\centering
\setlength{\tabcolsep}{3pt}
\renewcommand{\arraystretch}{1.15}
\begin{footnotesize}
\begin{threeparttable}
\begin{tabularx}{\linewidth}{
>{\raggedright\arraybackslash}p{0.20\linewidth} %
c c c c
>{\raggedright\arraybackslash}X                 %
}
\toprule
\textbf{Approach} &
\textbf{Compact\tnote{a}} &
\textbf{Halluc.\tnote{b}} &
\textbf{Single-shot\tnote{c}} &
\textbf{Single-view\tnote{c}} &
\textbf{Short notes} \\
\midrule

Inpainting / generative removal\cite{hirohashi2020removal, zhang2021automatic, jonna2024deep} &
Yes &
High &
Yes &
Yes &
Synthesize unseen content (fidelity risk), high computational cost. \\

Multi-view (or burst) occlusion layer separation\cite{jonna2017stereo, zhu2023occlusion, liu2020learning, xue2015computational, tsogkas2023efficient, chugunov2024neural} &
Yes &
Moderate &
No &
No &
Relies on parallax/flow; capture overhead, layer/flow estimation errors, typically high computational cost, inapplicable to obstructions fixed relative to the camera system. \\

Synthetic aperture / camera array\cite{levoy2004synthetic, joshi2007synthetic, pei2016all, pei2019occluded} &
No &
Low &
Yes &
No &
Physically suppresses near obstructions via DoF/refocusing, but bulky/costly setups. \\

Compound--diffractive optics\cite{shi2022seeing} &
Moderate &
Low &
Yes &
Yes &
Single-view capture, but needs compound optics. \\

\addlinespace
\textbf{This work: de-occluding broadband metalens} &
\textbf{Yes} &
\textbf{Low} &
\textbf{Yes} &
\textbf{Yes} &
Single-layer metalens-enabled monolithic form-factor. Pass/stop-band splitting paradigm enables far focus while rejecting focused near-depth light in a single-layer chromatic metalens. \\
\\
\bottomrule
\end{tabularx}

\begin{tablenotes}[flushleft]
\footnotesize
\item[a] \textbf{Compact:} easy integration into space-constrained camera modules without large baselines/arrays or bulky compound optics.
\item[b] \textbf{Hallucination risk:} whether the pipeline must synthesize occluded content rather than directly measure it.
\item[c] \textbf{Single-shot / single-view:} one temporal capture / one viewpoint-sensor (no stereo/array/motion parallax dependence).
\end{tablenotes}

\caption{Comparison of obstruction-free imaging approaches using compactness, hallucination risk, and capture/viewpoint requirements.}
\label{tab:supp_obstructionfree_comparison}
\end{threeparttable}
\end{footnotesize}
\end{table}

\clearpage 

\section*{Supplementary Note 4: Details on image simulation and metalens learning}
\label{supp:note4}
This section describes the details of the image simulation, loss functions, and metalens learning. 

\subsection{Meta-atom phase.}
We choose a geometric phase\cite{berry1984quantal} for our metalens design. The geometric phase achieves phase delay solely through the rotation angle of the meta-atoms, regardless of the wavelength of the incident light, making them inherently broadband-working and providing efficient computational modeling.
The metalens phase delay, $\phi_\text{meta}(x,y)$ is directly mapped from the rotation angle of the meta-atom, $\theta_\text{meta}(x,y)$ as: 
\begin{equation}
    \phi_\text{meta}(x,y)=2\theta_\text{meta}(x,y).
\end{equation}

\subsection{PSF simulation.}
The PSF is computed for a given metalens design map $\theta_\text{meta}(x,y)$, depth of the spherical light source $z$, and color channel $c\in\{\text{R, G, B}\}$ as: \[
    P_z^c = f_{\text{psf}}(\theta_\text{meta}, z, c).
\]
The band-limited Angular Spectrum Method\cite{matsushima2009band} is used to propagate spherical light source $\phi_\text{S}$ modulated by the metalens phase to the sensor plane as:
\[
    f_{\text{psf}}(\theta_\text{meta}, z, c) = |f_\text{ASM}\left(\phi_\text{S}(x,y,z,\lambda)+\phi_\text{meta}(x,y)\right)|^2.
\]
The wavelength $\lambda$ is sampled from the spectral sensitivity function $S^{c}(\lambda)$:
\[
    S^{c}(\lambda) = \mathrm{CRF}^{c}(\lambda)
                \cdot T_{\mathrm{BPF}}(\lambda)
                \cdot \eta_{\mathrm{meta}}(\lambda),
\]
where $\mathrm{CRF}^{c}(\lambda)$ is the camera response function of the color channel $c$, $T_{\mathrm{BPF}}(\lambda)$ is the spectral filter transmission, and $\eta_{\mathrm{meta}}(\lambda)$ is the meta-atom conversion efficiency computed from rigorous coupled-wave analysis simulation. 

The broadband baseline metalens optimization omits the multi-band bandpass filter in the simulated response:
\[
    S_{\mathrm{broadband}}^{c}(\lambda) =
    \mathrm{CRF}^{c}(\lambda)
    \cdot \eta_{\mathrm{meta}}(\lambda).
\]

\begin{figure}[h]
	\centering
		\includegraphics[width=\textwidth]{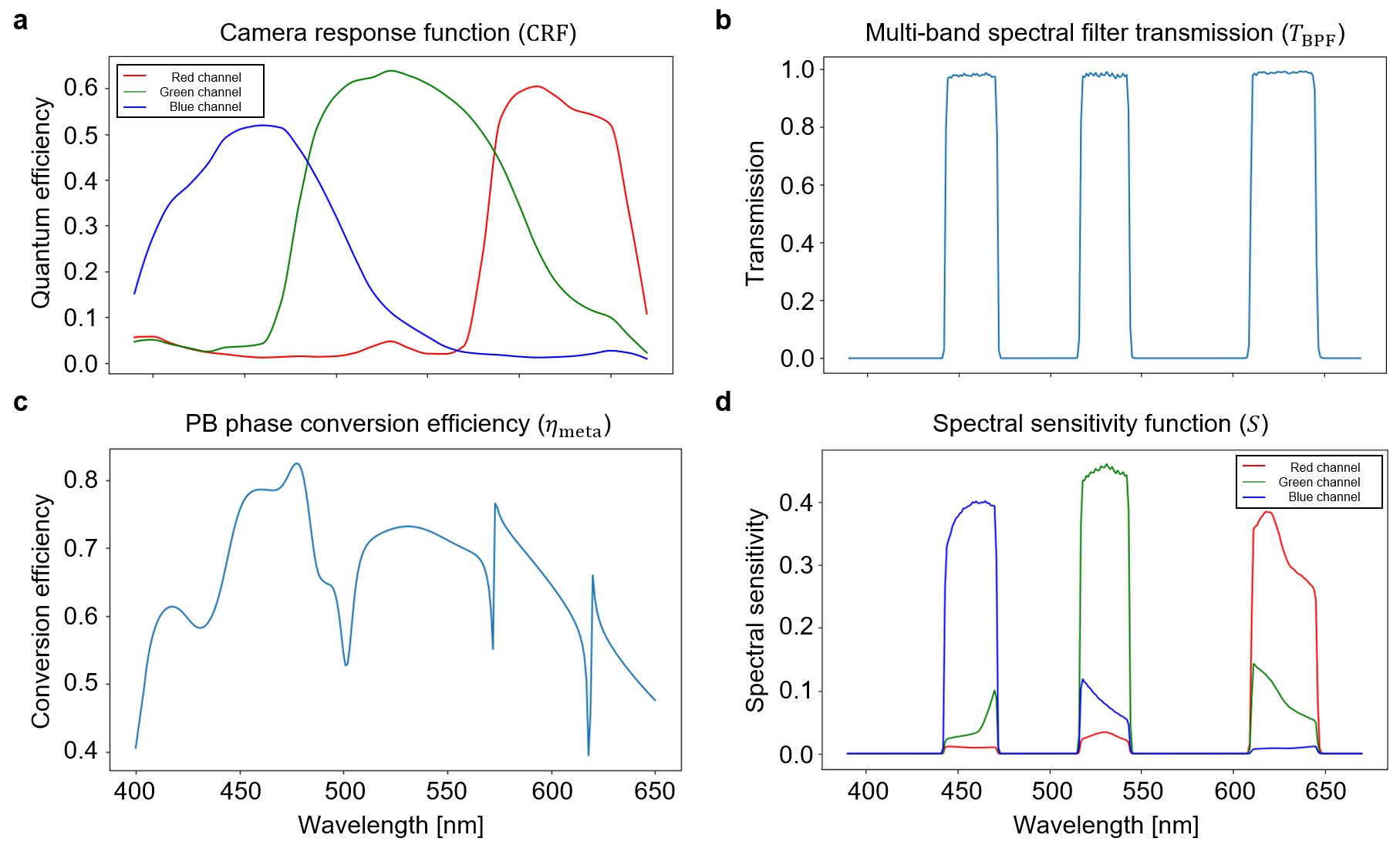}
		\caption{\textbf{Spectral characterization of the imaging system.} \textbf{a} Camera response functions for the RGB color channels provided by the camera vendor. \textbf{b} Multi-band spectral filter transmission graph provided by the vendor. \textbf{c} Metasurface conversion efficiency computed with rigorous coupled wave analysis simulation. \textbf{d} The resulting spectral sensitivity of the imaging system obtained as the product of \textbf{a}, \textbf{b}, and \textbf{c}.}
		\label{fig:supp_spectral_sensitivity}
\end{figure}

Each component of the spectral sensitivity function is presented in Figure~\ref{fig:supp_spectral_sensitivity}. PSFs calculated for each color channel are concatenated to form a three-channel PSF, $P_z$.

\subsection{Image simulation.}
The sensor-captured image $I_\text{captured}$ is simulated by an obstruction-aware alpha-blending model\cite{shi2022seeing} with the simulated PSF as:
\[
        I_\text{captured} = \alpha \odot (P_{z_{\text{near}}}*I_{\text{obs}}) + (1-\alpha) \odot (P_{z_{\text{far}}}*I_{\text{clean}}),
\]
\[
\alpha = \mathbf{1}\{I_\text{obs}(x,y) \neq0 \} * P_{z_{\text{near}}},
\]
where $*$ is the convolution operator, $\odot$ is the pixel-wise Hadamard product operator, and $\mathbf{1}\{\cdot\}$ is the indicator function. High-resolution three-channel image $I_{\text{clean}}$ is sampled from training set of DIV2K\cite{Agustsson_2017_CVPR_Workshops} images, and  $I_\text{obs}$ is randomly generated following the Perlin Noise-based dirt model\cite{perlin1985image, shi2022seeing}. $z_\text{near}$ is sampled from $[0.04\,\text{m}, 0.05\,\text{m}]$ and $z_\text{far}$ is sampled from $[1\,\text{m}, 20\,\text{m}]$. The convolution operation and the Hadamard product are done in a channel-wise manner. 

The obstruction-aware image simulation above is used for learning the proposed de-occluding broadband metalens. In contrast, the Broadband baseline metalens is trained without near-depth obstructions and without the alpha-blending obstruction model. Its simulated captured image is defined simply as
\[
        I_{\mathrm{captured}}
        =
        P_{z_{\mathrm{far}}} * I_{\mathrm{clean}}.
\]
Thus, the Broadband baseline objective encourages the lens to form sharp far-depth images over the broadband visible spectrum.

We further clarify that using RGB images from DIV2K does not mean that we samples only three RGB wavelengths throughout the optimization. In the PSF simulator, wavelengths are sampled from the broadband system spectral response function. The role of the DIV2K RGB image $I_{\text{clean}}$ is to provide high-resolution spatial content including natural textures and edges for the image-formation loss. Therefore, the RGB dataset acts as a spatial image prior, not as a hyperspectral model of the scene.

\subsection{Image similarity loss.}
The image similarity loss $\mathcal{L}_{\text{img}}$ encourages the simulated sensor-captured image $I_\text{captured}$ to be similar to the clean ground-truth image $I_\text{clean}$:
\[
\mathcal{L}_{\text{img}}\!\left(I_{\text{captured}}, I_{\text{clean}}\right) = \mathcal{L}_{\text{L1}}(I_{\text{captured}}, I_{\text{clean}}) + \mathcal{L}_{\text{SSIM}}(I_{\text{captured}}, I_{\text{clean}})+\mathcal{L}_{\text{DA}}(I_{\text{captured}}, I_{\text{clean}}),
\]
where $\mathcal{L}_{\text{L1}}$ is the pixel-wise L1 loss between the two images, $\mathcal{L}_{\text{SSIM}}$ is the structural similarity index measure loss defined by $1-\text{SSIM}(I_1, I_2)$, and $\mathcal{L}_{\text{DA}}$ is the Domain Adaptation loss\cite{chen2020adversarial}, which is a L1 loss computed between image features extracted by DeepLabv3+\cite{chen2018encoder}.

\subsection{PSF loss.}
In addition to the image similarity losses, the PSF-based loss $\mathcal{L}_{\text{psf}}$ is used as a regularizer to help the metalens learning:
\[
\mathcal{L}_{\text{psf}} = \mathcal{L}_{\text{bright}} + \mathcal{L}_{\text{sharp}},
\]
\[
\mathcal{L}_{\text{bright}} = \frac{3N}{\sum_x\sum_y P_{z_{\text{far}}}(x,y)} -1,
\]
\[
\mathcal{L}_{\text{sharp}} = \frac{N}{\sum_{i=-2}^{2} \sum_{j=-2}^{2} P_{z_{\text{far}}}(x_\text{c}+i, y_\text{c}+j)} -1,
\]
where $\mathcal{L}_{\text{bright}}$ encourages input energy is conserved at the sensor plane for $P_{z_{\text{far}}}$, and $\mathcal{L}_{\text{sharp}}$ prefers sharp $P_{z_{\text{far}}}$. N is the number of pixels computed by $\sum_{x,y}1$. Note that $\sum_x\sum_y P_{z_{\text{far}}}(x,y)=3N$ when light energy is conserved at the sensor plane, or the incident light energy does not leak out of the sensor, since PSF is three-channel. $(x_\text{c}, y_\text{c})$ is the center position of the PSF. We use a sensor of pitch 1.85\,$\mu $m and resolution 512$\times$512. We empirically found that the configurations used in $\mathcal{L}_{\text{sharp}}$ balances between the sharp far-PSFs and blurry near-PSFs; otherwise, far-PSFs and near-PSFs are both likely to be sharp.

\subsection{Metalens learning.}
The final objective function to learn the metalens is defined as follows:
\begin{equation}
    \underset{\theta}{\text{minimize}}\;
        \mathcal{L}_{\text{img}}\!\left(I_{\text{captured}}, I_{\text{clean}}\right)
        + \mathcal{L}_{\text{psf}}\!\left( P_{z_{\text{far}}}\right).
    \label{eq:metalens_optimization}
\end{equation}
We optimize this objective function with the first-order AdamW optimizer\cite{loshchilov2017decoupled}. We select a learning rate of 0.1 and use the cosine annealing\cite{loshchilov2016sgdr} for the learning rate scheduler, where $T_\text{max}$ is selected as 5000 steps. The training takes 26 hours on an NVIDIA A6000 GPU. 

\subsection{Rationale for separate metalens and reconstruction-network training.}
In this work, the metalens and the reconstruction network are optimized separately rather than in a fully end-to-end manner. The purpose of the metalens learning stage is to enforce the desired optical behavior at the sensor-captured image level: sharp far-depth PSFs and blurred near-depth PSFs in the selected pass bands. Including the reconstruction network in the metalens optimization could obscure this optical design objective. A sufficiently expressive neural network can restore blurred structures or remove obstruction-like patterns through learned priors. Therefore, an end-to-end loss computation may reduce the optimization pressure on the metalens to physically suppress near-depth obstruction at capture time. Such a solution results in less interpretable and less robust optics design, because the de-occluding effect would be shifted from the optical front-end to the computational backend.

In addition, an end-to-end optimization requires a highly accurate differentiable raw-image simulator that accounts for dense broadband wavelength integration, sensor noise, and angle-dependent aberrations. In practice, the memory and computation required for dense wavelength integration and angle-dependent aberration simulation are very high. In our metalens learning, only three wavelengths are sampled stochastically at each iteration due to memory constraints, which is sufficient for optimizing the desired optical behavior over the target spectrum, but is not intended to be a complete high-fidelity raw-image simulator for end-to-end training. Even with a more complex simulator, a synthetic-to-real gap would remain, and the reconstruction network still needs to be trained or fine-tuned using real captured data.

Therefore, the separated optimization strategy has advantages; it makes the physical role of the metalens explicit, reduces computational cost, and allows the reconstruction network to be trained as a modular backend using experimentally captured data, which can compensate for residual aberrations and noise without accurately calibrating and simulating the optical system.

\clearpage 

\section*{Supplementary Note 5: Details on image reconstruction neural network}
\label{supp:note5}

While dispersion engineering for broadband achromatic diffractive lenses is fundamentally limited by low NA and small diameter, the chromatic lens-based broadband diffractive imaging suffers from aberrations. Therefore, numerous studies co-design the optics with a computational backend or support the optics with neural networks to improve image quality. We use the lightweight neural network\cite{Kim_2024_CVPR} that can be integrated into edge devices for compact obstruction-free imaging. We train a separate neural network for each metalens. This section describes the image reconstruction neural network specifications. 

\subsection{Neural network architecture.}
\begin{figure}[h]
	\centering
		\includegraphics[width=\textwidth]{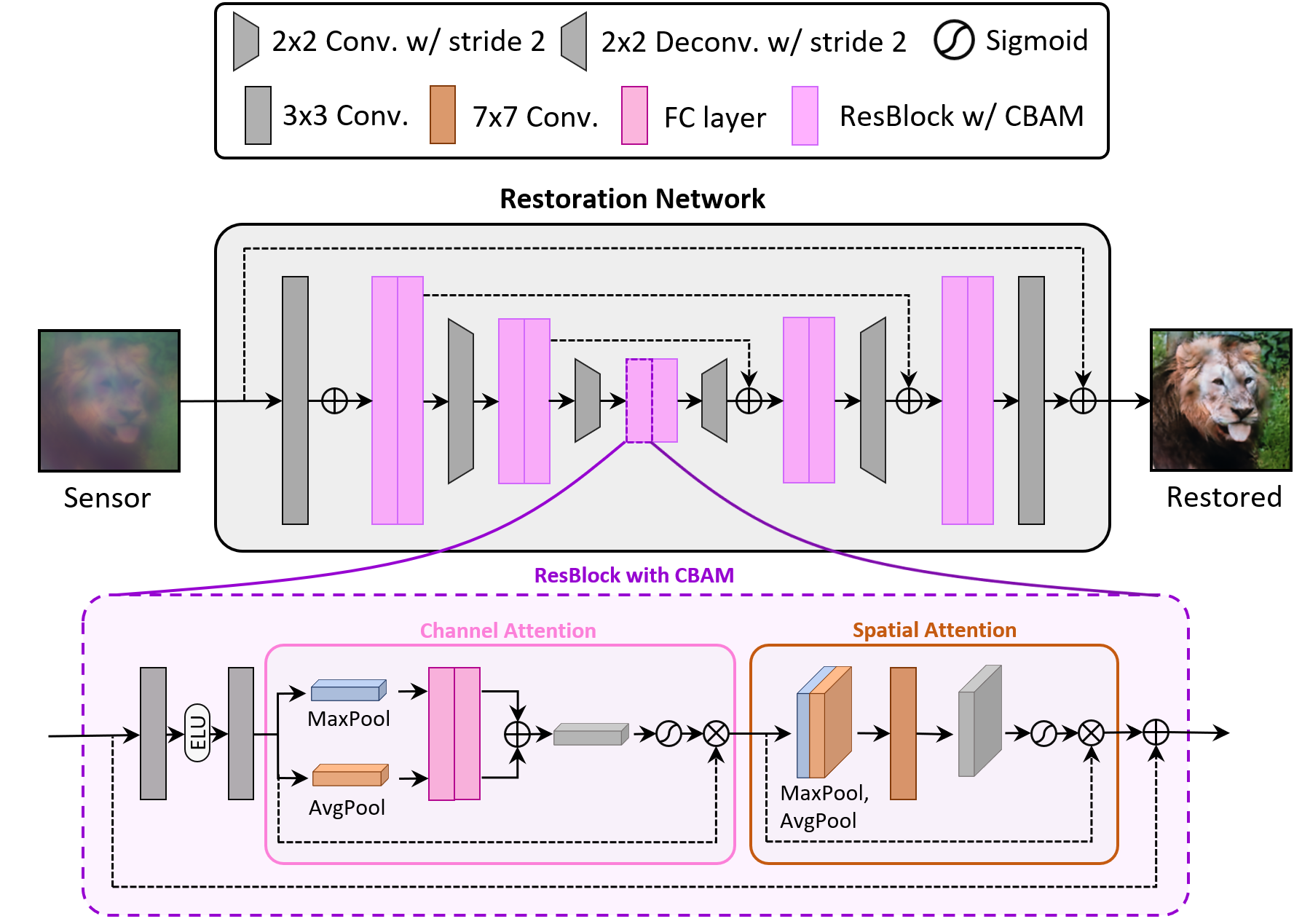}
		\caption{\textbf{Detailed architecture of the image reconstruction network.} Conv., convolution; Deconv., deconvolution; FC, fully connected; ResBlock, residual block; CBAM, convolutional block attention module; ELU, exponential linear unit; MaxPool, maximum polling; AvgPool, average pooling. The example source image is from DIV2K\cite{Agustsson_2017_CVPR_Workshops}.}
		\label{fig:supp_neural_network_architecture}
\end{figure}

Figure~\ref{fig:supp_neural_network_architecture} shows the detailed architecture of the image reconstruction network. The neural network is based on U-Net\cite{ronneberger2015u} architecture, and incorporates CBAM modules\cite{woo2018cbam}. The network predicts a residual and adds it to the input image for reconstruction. 
Additional input features, including gradient map, soft histogram, over-exposure mask, and positional encodings\cite{vaswani2017attention, mildenhall2021nerf}, are used to enhance the restoration quality. The positional encodings are especially helpful in addressing spatially varying aberrations of the metalens.

\subsection{Training details.}
The objective function for the neural network training is defined by as follows:
\begin{equation}
\mathcal{L}_{\text{total}} = \mathcal{L}_{\text{L1}} + \lambda_\text{SSIM}\mathcal{L}_{\text{SSIM}} + \lambda_\text{LPIPS}\mathcal{L}_{\text{LPIPS}},
\end{equation}
where $\mathcal{L}_{\text{LPIPS}}$ is the LPIPS loss\cite{zhang2018unreasonable}, and we set $\lambda_\text{SSIM}$ and $\lambda_\text{LPIPS}$ as 0.1. The neural network was trained with 4 NVIDIA RTX 6000 Ada GPUs for 60 hours. We employed the AdamW\cite{loshchilov2017decoupled} optimizer. We set the learning rate $\eta$ to $5\times10^{-6}$, which is adjusted following a cosine annealing schedule with warm-up and restart\cite{loshchilov2016sgdr}, where $T_\text{max}$ is set to 4000, and the linear warm-up from $0.1\eta$ is applied for the first 10 epochs. The training is performed with a batch size of 14 per GPU. Each image was randomly cropped to a patch size of $424\times424$ pixels. We applied random vertical and horizontal flips, along with random rotations. Random augmentations to brightness and contrast were applied, along with Gaussian noise with a standard deviation of $\sigma=5\times10^{-3}$. The Gaussian noise is designed to model sensor noise, preventing the network from overfitting to the ground-truth data. Meanwhile, the photometric augmentations (brightness and contrast) are crucial for compensating for potential inconsistencies between the metalens and the ground-truth compound lens cameras. Since factors such as auto-exposure and white balancing can vary independently between the two optical systems, these augmentations encourage the network to focus on structural recovery rather than global intensity mapping, ensuring robust reconstruction performance across diverse lighting conditions.

\subsection{Training dataset.}
\begin{figure}[h]
	\centering
		\includegraphics[width=0.8\textwidth]{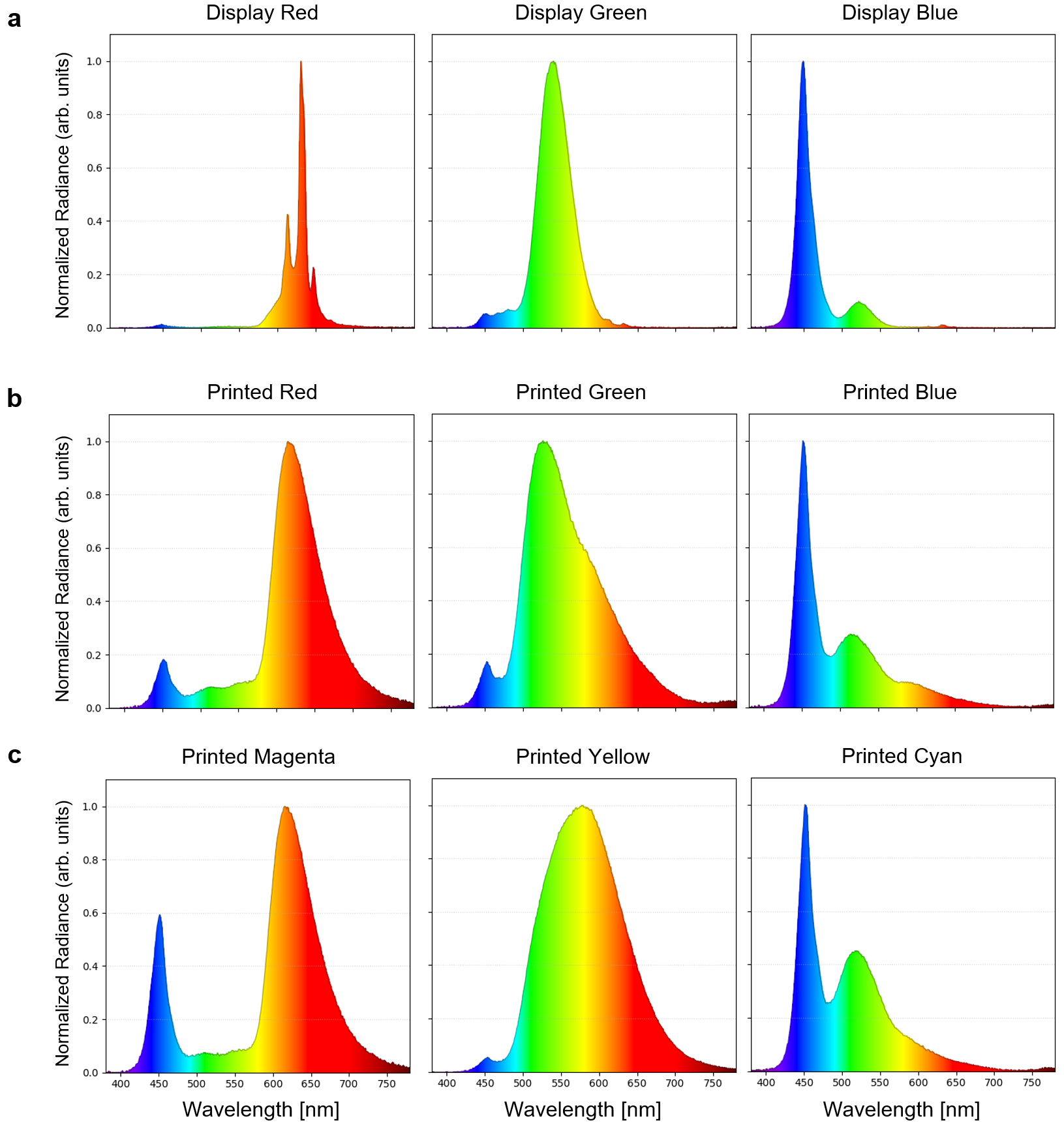}
		\caption{\textbf{Comparison of spectral radiance distribution between printed images and digital displays.} The spectral radiance of each primary color (Red, Green, and Blue) was measured with a spectroradiometer (JETI specbos 1211-2) to ensure spectral fidelity for broadband metalens characterization. \textbf{a} The digital display (LG 22-inch LCD monitor) shows narrow spectral peaks corresponding to its RGB channels, failing to represent the full continuous spectrum required for accurate broadband modeling. \textbf{b, c} Printed images under an LED light source exhibit a continuous and significantly broader spectrum, which is essential for preventing spectral undersampling during broadband imaging. We show red, green, and blue spectra along with magenta, yellow, and cyan, which are the basis colors for the printing. All curves are normalized to their respective peak intensities for clear comparison of the spectral width. The fill color indicates wavelength across the visible spectrum.}
		\label{fig:supp_spectral_curves}
\end{figure}

We use printed high-resolution images with rich structural and color content from the DIV2K dataset\cite{Agustsson_2017_CVPR_Workshops} as ground-truth scenes for the training dataset. Obstructed scenes were captured using the metalenses, and the ground-truth scenes were acquired with a compound lens with a focal length of $f=8$\,mm. The focal length of the compound lens is twice that of the metalenses to achieve better object space resolution\cite{engelberg2022generalized}. Note that the \textit{printed} images are used as training data, rather than digitally displayed images, to ensure spectral fidelity in broadband imaging. Displays typically exhibit narrow spectral peaks corresponding to their RGB channels. In contrast, printed images reflect a continuous and significantly broader spectrum. Figure~\ref{fig:supp_spectral_curves} compares the spectral radiance distribution of the printed image under a white LED light source against a display (LG 22-inch LCD monitor) for each color. This broader spectral content is essential for accurately characterizing broadband metalenses, as it prevents spectral undersampling.  

\begin{figure}[h]
	\centering
		\includegraphics[width=\textwidth]{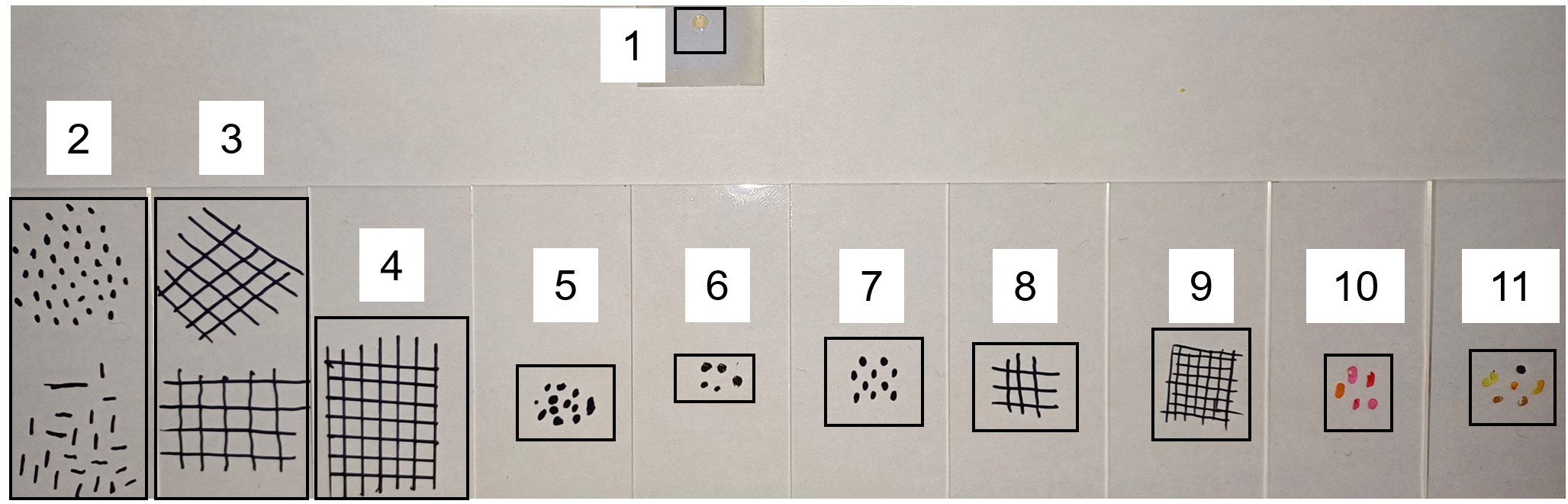}
		  \caption{\textbf{Obstructions used for the evaluations and the neural network training.} We show the images of obstructions used in this study. Our metalens is shown for the relative size comparison between the metalens and obstructions. \textbf{1} Our de-occluding broadband metalens for relative size comparison between the lens and obstructions. \textbf{2, 3} Obstructions used for the image reconstruction network training. \textbf{4, 5} Obstructions used for the image metric measurements. \textbf{6, 7, 8} Obstructions used for the in-the-wild scene imaging. \textbf{9, 10, 11} Obstructions used for the computer vision tasks imaging.}
		\label{fig:supp_obstructions}
\end{figure}

The dataset is composed of 200 training and 10 validation scenes from the DIV2K dataset for each lens. For each scene, we captured one clean image and an average of six obstructed counterparts, resulting in a total of 1,400 training and 70 validation images for each metalens.
Figure~\ref{fig:supp_obstructions} shows the obstructions used for the neural network training. Precise pixel-wise alignment between the obstructed metalens captures and the ground-truth compound lens images is essential for the supervised learning. To acquire such alignment, we use the stereo configuration for our dataset capture setup (Figure~\ref{fig:supp_imaging_setup}). We employ a homography-based warping strategy to compensate for the spatial offset and scaling differences between the two optical paths. Since our training scenes are planar, the depth-dependent parallax remains negligible, and the spatial relationship between the two lenses is accurately modeled by a homography transformation. The homography is estimated using a set of corresponding feature points from a learning-based feature extractor\cite{potje2024xfeat} and the RANSAC\cite{fischler1981random} algorithm.

\subsection{Inference time and memory consumption.}
We benchmarked the image reconstruction network's inference performance and memory footprint on an NVIDIA RTX 6000 Ada GPU to evaluate the practical viability of the imaging system for edge deployment. The benchmark was conducted at a resolution of $512\times512$ pixels, which is a standard input resolution scale for neural networks. As summarized in Table~\ref{tab:supp_inference_benchmark}, our model exhibits a throughput of 24.15\,FPS when optimized, showing its suitability for high-speed imaging. The high-performance efficiency, coupled with a manageable memory footprint of 946\,MiB, underscores the potential for integration into compact edge devices. Note that the throughput could be increased up to 75.44\,FPS when batch-processed, with the memory consumption of 4,118\,MiB.

\begin{table}
\centering
\setlength{\tabcolsep}{4pt}
\begin{footnotesize}
\begin{tabularx}{\columnwidth}{l @{\extracolsep{\fill}} ccccc}
\toprule
\textbf{Precision} & \textbf{Batch} & \textbf{Input size} & \textbf{Latency} & \textbf{Throughput} & \textbf{Memory} \\
 & \textbf{size} & (pixels) & (per image, ms) & (FPS) & (MiB) \\
\midrule
FP32 & 1 & $512 \times 512$ & 42.96 & 23.28 & 1,178 \\
FP32 & 8 & $512 \times 512$ & 21.13 & 47.33 & 7,136 \\
FP16 & 1 & $512 \times 512$ & 41.41 & 24.15 & 946 \\
FP16 & 8 & $512 \times 512$ & 13.26 & 75.44 & 4,118 \\
\bottomrule
\end{tabularx}
\end{footnotesize}
\caption{Inference performance and memory (VRAM) consumption of the image reconstruction network. Benchmarks were measured on an NVIDIA RTX 6000 Ada Generation GPU at a resolution of $512 \times 512$ pixels. Throughput and memory footprint highlight the lightweight nature of the proposed architecture for real-time edge-device applications.}
\label{tab:supp_inference_benchmark}
\end{table}

\clearpage 

\section*{Supplementary Note 6: Measured complex refractive index of the $\text{SiN}_x$ film}
\label{supp:note6}
\begin{figure}[h]
	\centering
		\includegraphics[width=0.5\textwidth]{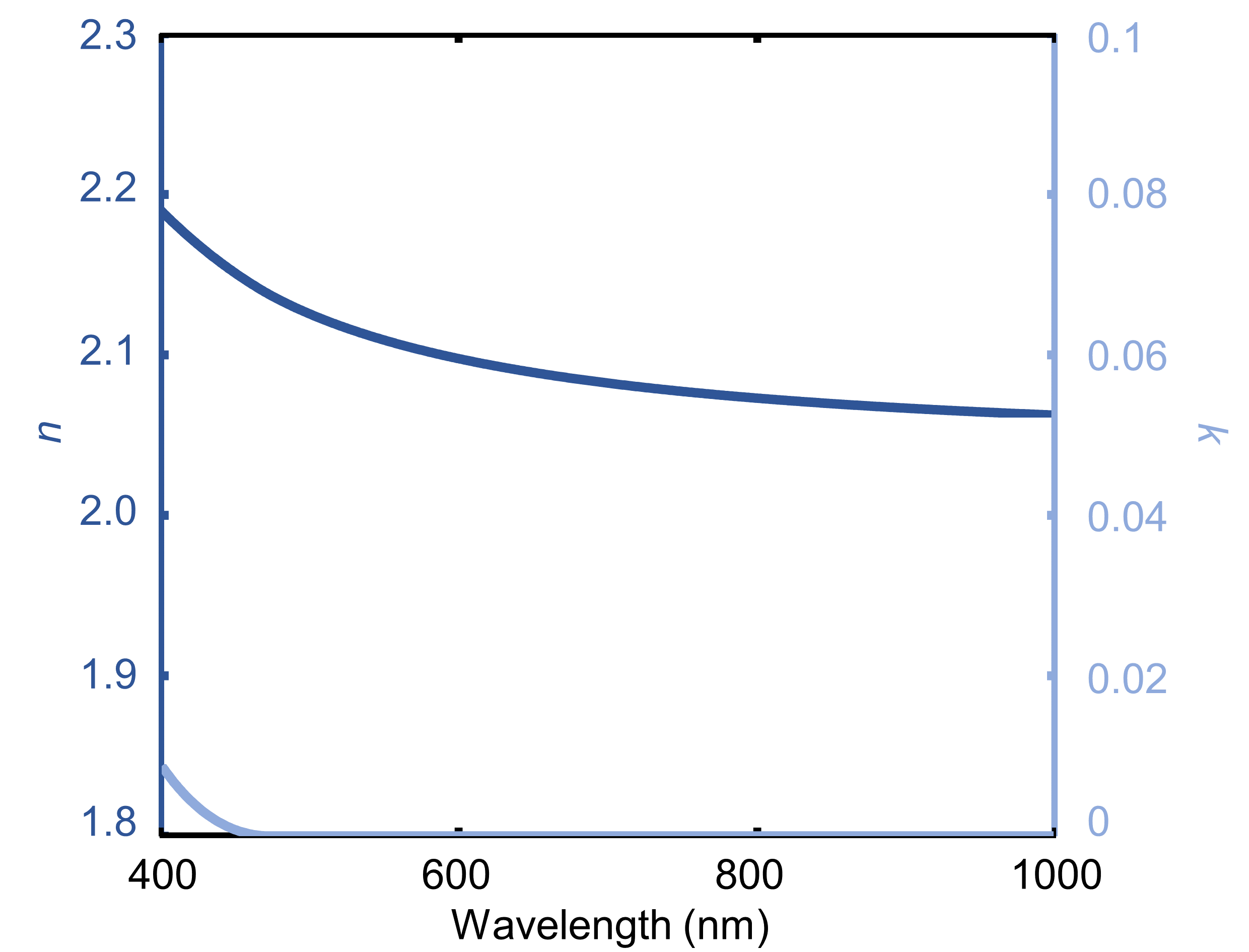}
		\caption{Measured refractive index and extinction coefficient of the $\text{SiN}_x$ film.}
		\label{fig:index}
\end{figure}

\clearpage 

\section*{Supplementary Note 7: Simulated conversion efficiency under variation of geometric parameters}
\label{supp:note7}

In this study, phase modulation is achieved using geometric phase, where each anisotropic meta-atom with an in-plane orientation $\theta$ imparts a spatially varying phase of $2\theta$ to the cross-polarized transmitted field.
These meta-atoms are designed to function as subwavelength half-wave plates to maximize polarization conversion efficiency, which requires a $\pi$ phase retardation between the two orthogonal eigenmodes.
We evaluated the conversion efficiency using rigorous coupled-wave analysis (RCWA) by sweeping the width $w$ from $210\,$nm to $360\,$nm and the length $l$ from $40\,$nm to $190\,$nm, while fixing the period at $p = 395\,$nm and the height at $h = 700\,$nm, consistent with fabrication constraints.
Although this period does not fully suppress higher-order diffraction on the reflection side, it was selected to ensure broadband operation and fabrication feasibility while maintaining a negligible impact on imaging performance.

\begin{figure}[h]
	\centering
		\includegraphics[width=1\textwidth]{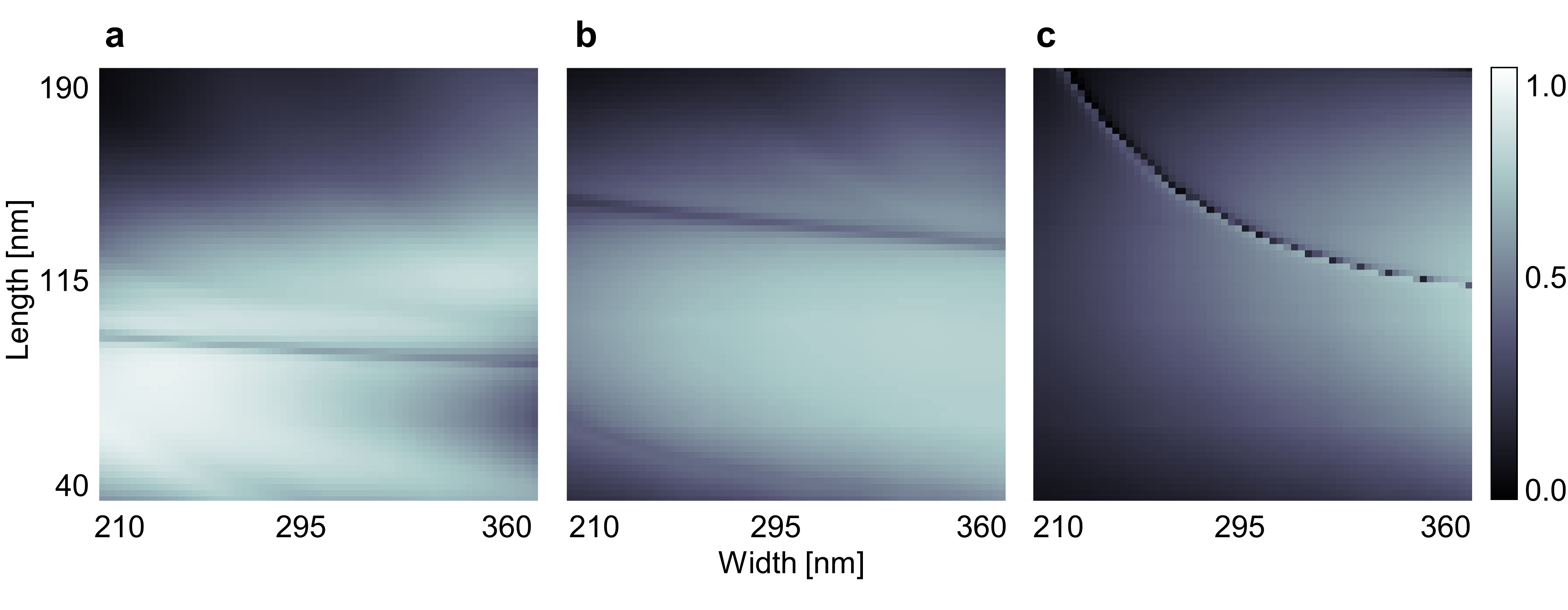}
		\caption{Calculated conversion efficiency of meta-atoms at wavelengths of $457\,$nm, $530\,$nm, and $628\,$nm, for a fixed period of $395\,$nm and a height of $700\,$nm, as function of the meta-atom width and length.}
		\label{fig:supp_CE}
\end{figure}

\begin{figure}[h]
	\centering
		\includegraphics[width=0.5\textwidth]{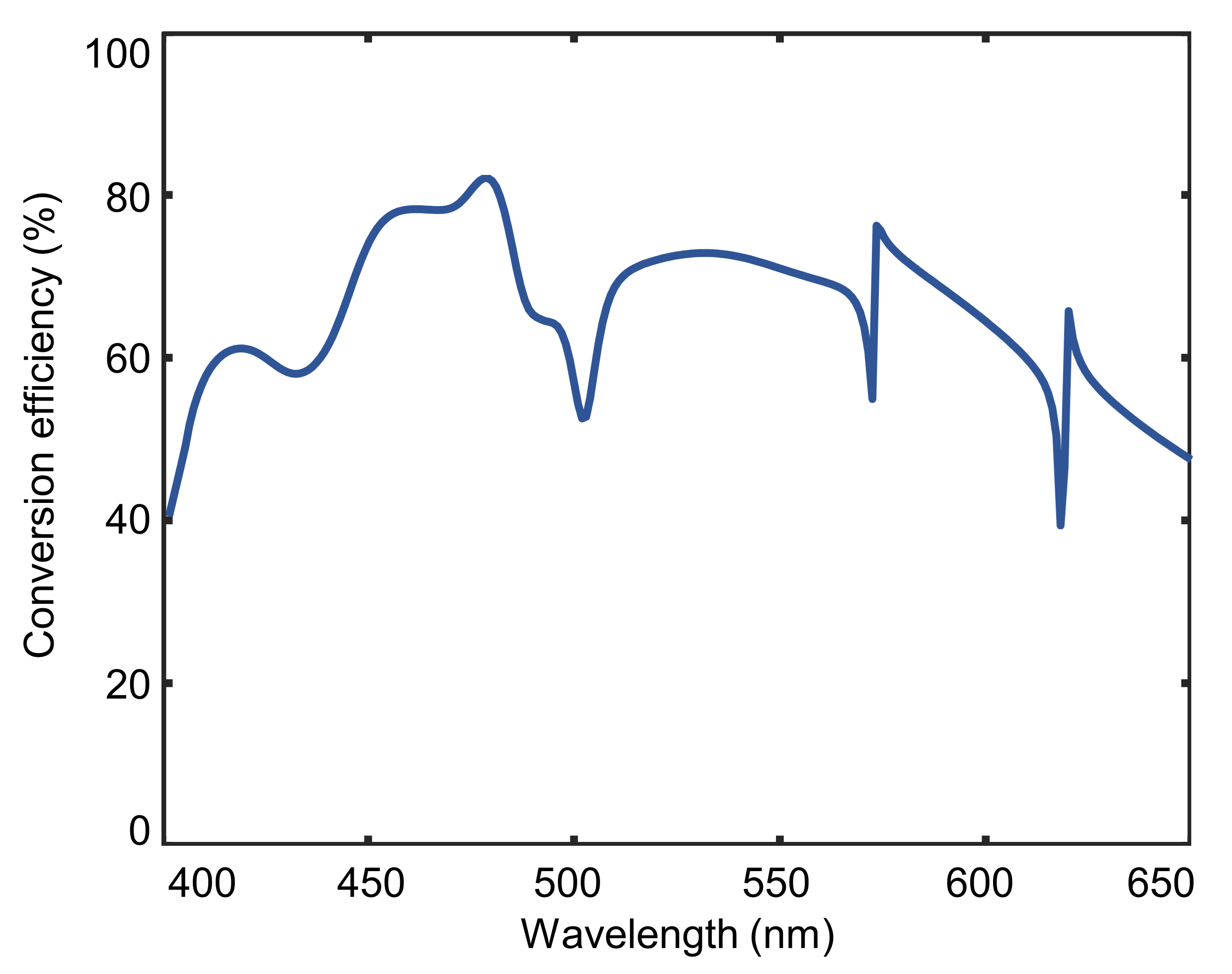}
		\caption{Calculated conversion efficiency of the designed meta-atom from $400\,$nm to $650\,$nm.}
		\label{fig:conversion_efficiency}
\end{figure}

Figure~\ref{fig:supp_CE} presents the calculated efficiency at the target wavelengths of $457\,$nm, $530\,$nm, and $628\,$nm. 
Based on these results, we selected a geometry with $w = 305\,$nm and $l = 125\,$nm, which ensures high efficiency while accommodating a fabrication tolerance of $10\,$nm.
This design achieves high average conversion efficiencies of $77.2\%$, $73.0\%$, and $54.4\%$ across the respective pass bands (Figure~\ref{fig:conversion_efficiency}).
In addition, this high efficiency is uniformly maintained, as evidenced by low standard deviations of the efficiencies across wavelengths within each band ($2.2\%$, $0.2\%$, and $4.9\%$, respectively; Table~\ref{tab:supp_passband_efficiency}). This indicates that the pass-band edge wavelengths are not suppressed and are therefore not expected to degrade the imaging quality. 

\begin{table}
\centering
\setlength{\tabcolsep}{4pt}
\renewcommand{\arraystretch}{1.15}
\begin{footnotesize}
\makebox[\columnwidth][c]{%
\begin{tabular}{c c c c c c}
\toprule
\textbf{Pass bands} &
\textbf{Wavelength} &
\textbf{Mean} &
\textbf{Min} &
\textbf{Max} &
\textbf{Std. Dev.} \\
\midrule
Blue  & 446 nm -- 468 nm & 77.2\% & 71.0\% & 78.6\% & 2.2\% \\
Green & 520 nm -- 540 nm & 73.0\% & 72.5\% & 73.2\% & 0.2\% \\
Red   & 614 nm -- 642 nm & 54.4\% & 39.5\% & 66.0\% & 4.9\% \\
\bottomrule
\end{tabular}%
}
\end{footnotesize}
\caption{\textbf{Statistical summary of conversion efficiencies across the target pass bands.}
The table reports the mean, minimum, maximum, and standard deviation of the conversion efficiency over each pass band, confirming broadband and spectrally uniform performance of the optimized geometric-phase meta-atoms.}
\label{tab:supp_passband_efficiency}
\end{table}

\clearpage 

\section*{Supplementary Note 8: Fabrication process of metalenses}
\label{supp:note8}
\begin{figure}[h]
	\centering
		\includegraphics[width=1\textwidth]{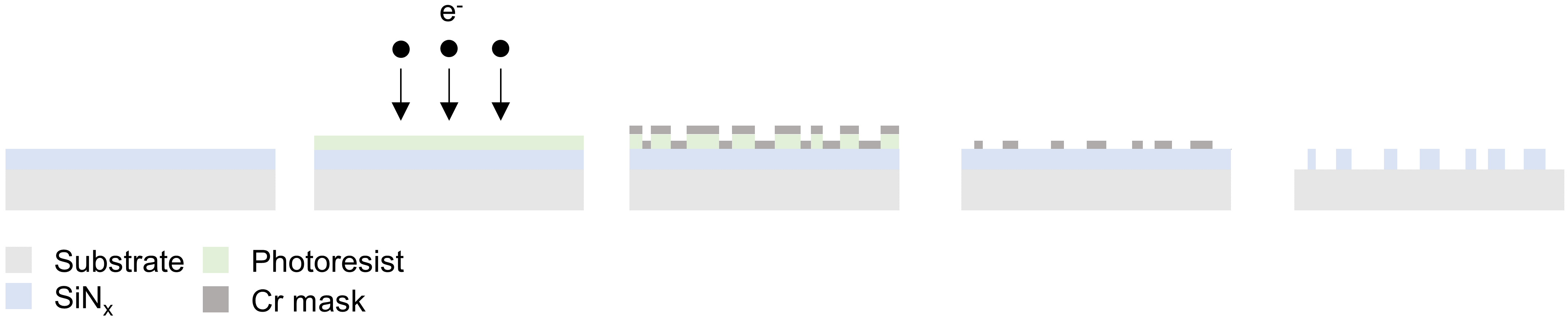}
		\caption{Fabrication process of metalenses.}
		\label{fig:fabrication}
\end{figure}

\clearpage 

\section*{Supplementary Note 9: Details on the experimental setup}
\label{supp:note9}
\input{sections_supplement/setup}
\clearpage 

\section*{Supplementary Note 10: Details on the image quality assessment}
\label{supp:note10}
We measure PSNR, SSIM, and LPIPS\cite{zhang2018unreasonable} for two representative obstructions: dirt (blob-like occluders) and fence (thin and structured occluders), which jointly span common near-depth obstruction geometries that provide a generalizable assessment of imaging fidelity. We consider (i) the compound lens GT image, (ii) the metalens unobstructed capture, (iii) the metalens obstructed capture, and (iv) the reconstructed output produced by the corresponding method. We compute metrics for each lens design: hyperbolic phase metalens (Hyperbolic), broadband metalens without a de-occluding strategy (Broadband), and the de-occluding broadband (Ours, De-occluding), and for each obstruction type (e.g., fence, dirt) using the same alignment protocol described below, ensuring fair comparisons across all conditions.
Due to the time-intensive nature of the optical acquisition process, we evaluate a subset of $N=10$ printed scenes per obstruction type and per lens design. For each scene, we capture both unobstructed and obstructed metalens images, and we additionally evaluate the corresponding reconstructed outputs. Qualitative examples are shown in Figures~\ref{fig:supp_additional_imaging_results_printed1} and~\ref{fig:supp_additional_imaging_results_printed2}, and quantitative results are summarized in Table~\ref{tab:supp_image_metrics}. The quantitative results demonstrate that our split-spectrum design achieves superior robustness against obstructions compared to baseline methods. Analyzing the performance gap ($\Delta$) between unobstructed and obstructed conditions, our method overall exhibits the minimal degradation across all metrics and conditions, validating the resilience of our method against obstructions.

\subsection{Precise alignment for image quality measurement.}
Because these metrics assume pixel-wise correspondence, accurate geometric alignment between the two imaging paths is critical. We employ a two-stage warping procedure: a global homography for coarse alignment (consistent with the neural network training pipeline in Supplementary Note~5), followed by a non-linear refinement for precise metric measurements. While a homography is sufficient for planar alignment in principle, we observed small residual misregistrations that can bias full-reference metrics, due to practical factors such as mild lens distortion, imperfect planarity of printed targets. Therefore, we apply an additional non-linear warping step based on a thin-plate spline (TPS) deformation field initialized from the homography-aligned pair. All reported metrics in Table~\ref{tab:supp_image_metrics} are computed after applying this identical alignment procedure across all lenses, obstructions, and image conditions.

\subsection{Photometric normalization for fair full-reference evaluation.}
Full-reference metrics such as PSNR, SSIM, and LPIPS assume not only pixel-wise geometric correspondence but also comparable photometric scaling.
In our setting, images are acquired through different optical paths and re-captured from printed targets, which can introduce global exposure and contrast shifts unrelated to imaging performance. To mitigate such global photometric offsets, we apply a temporary per-image photometric normalization prior to metric computation:
given an aligned reference image and a target image, we estimate an affine intensity transform (gain and bias) using the luminance mean and standard deviation of the reference, and apply the same gain/bias to the target image.
All reported PSNR/SSIM/LPIPS values are computed after this photometric matching, while the underlying geometry and spatial content remain unchanged.
We apply the same procedure consistently across all lens designs, obstruction types, and conditions (raw/reconstructed, unobstructed/obstructed) to ensure fair comparisons. This normalization only compensates for global brightness differences and does not correct local artifacts or structural misalignment; therefore, it does not artificially improve the metrics beyond removing trivial photometric bias.

\subsection{Qualitative observations.}
As shown in Figures~\ref{fig:supp_additional_imaging_results_printed1} and~\ref{fig:supp_additional_imaging_results_printed2}, obstruction patterns (dirt blobs and fence-like grids) strongly corrupt raw metalens captures and can propagate into reconstructions for baseline lens designs.
In contrast, the de-occluding broadband metalens maintains visually consistent reconstructions across unobstructed and obstructed inputs, exhibiting substantially reduced obstruction imprint and improved structural fidelity.
These qualitative trends align with the quantitative improvements in PSNR/SSIM and reductions in LPIPS reported in Table~\ref{tab:supp_image_metrics}, particularly under obstructed conditions.

\begin{table*}
\centering
\small
\setlength{\tabcolsep}{2pt}
\renewcommand{\arraystretch}{1.1}
\begin{tabular}{lcccccccc}
\toprule
\multirow{2}{*}{\textbf{Lens design}} & \multirow{2}{*}{\textbf{Capture}} & \multirow{2}{*}{\textbf{Condition}} &
\multicolumn{3}{c}{\textbf{Dirt Obstruction}} &
\multicolumn{3}{c}{\textbf{Fence Obstruction}} \\
\cmidrule(lr){4-6} \cmidrule(lr){7-9}
 & & &
\textbf{PSNR}~($\uparrow$) & \textbf{SSIM}~($\uparrow$) & \textbf{LPIPS}~($\downarrow$) &
\textbf{PSNR}~($\uparrow$) & \textbf{SSIM}~($\uparrow$) & \textbf{LPIPS}~($\downarrow$) \\
\midrule

\multirow{6}{*}{Hyperbolic} & \multirow{3}{*}{Raw} & Unobs. & 10.81 & 0.2840 & 0.8890 & 10.89 & 0.2873 & 0.8800 \\
 & & Obs. & 9.73 & 0.2678 & 0.9201 & 10.56 & 0.2756 & 0.9018 \\
 & & $\Delta$ (Drop) & \textit{-1.08} & \textit{-0.0162} & \textit{-0.0311} & \textit{-0.33} & \textit{-0.0117} & \textit{-0.0218} \\
\cmidrule{2-9}
 & \multirow{3}{*}{Recon.} & Unobs. & 18.77 & 0.7034 & 0.4640 & 19.01 & 0.7108 & 0.4531 \\
 & & Obs. & 15.98 & 0.6261 & 0.5419 & 15.67 & 0.6074 & 0.5611 \\
 & & $\Delta$ (Drop) & \textit{-2.79} & \textit{-0.0773} & \textit{-0.0779} & \textit{-3.34} & \textit{-0.1034} & \textit{-0.1080} \\
\midrule

\multirow{6}{*}{Broadband} & \multirow{3}{*}{Raw} & Unobs. & 15.00 & 0.3804 & 0.8478 & 14.98 & 0.3783 & 0.8501 \\
 & & Obs. & 14.14 & 0.3553 & 0.8799 & 14.36 & 0.3537 & 0.8841 \\
 & & $\Delta$ (Drop) & \textit{-0.86} & \textit{-0.0251} & \textit{-0.0321} & \textit{-0.62} & \textit{-0.0246} & \textit{-0.0340} \\
\cmidrule{2-9}
 & \multirow{3}{*}{Recon.} & Unobs. & 22.77 & 0.7521 & 0.4023 & 22.97 & 0.7526 & 0.4016 \\
 & & Obs. & 18.40 & 0.6725 & 0.4775 & 19.17 & 0.6498 & 0.5087 \\
 & & $\Delta$ (Drop) & \textit{-4.37} & \textit{-0.0796} & \textit{-0.0752} & \textit{-3.80} & \textit{-0.1028} & \textit{-0.1071} \\
\midrule

\multirow{6}{*}{\textbf{Ours}} & \multirow{3}{*}{\textbf{Raw}} & \textbf{Unobs.} & \textbf{15.58} & \textbf{0.4458} & \textbf{0.7221} & \textbf{15.30} & \textbf{0.4256} & \textbf{0.7463} \\
 & & \textbf{Obs.} & \textbf{15.15} & \textbf{0.4198} & \textbf{0.7559} & \textbf{14.72} & \textbf{0.4020} & \textbf{0.7937} \\
 & & \textbf{$\Delta$ \text{(Drop)}} & \textbf{\textit{-0.43}} & \textbf{\textit{-0.0260}} & \textbf{\textit{-0.0338}} & \textbf{\textit{-0.58}} & \textbf{\textit{-0.0236}} & \textbf{\textit{-0.0474}} \\
\cmidrule{2-9}
\textbf{(De-occluding)} & \multirow{3}{*}{\textbf{Recon.}} & \textbf{Unobs.} & \textbf{23.60} & \textbf{0.8282} & \textbf{0.2939} & \textbf{23.21} & \textbf{0.8129} & \textbf{0.3084} \\
 & & \textbf{Obs.} & \textbf{21.20} & \textbf{0.7803} & \textbf{0.3529} & \textbf{20.67} & \textbf{0.7316} & \textbf{0.4047} \\
 & & \textbf{$\Delta$ \text{(Drop)}} & \textbf{\textit{-2.40}} & \textbf{\textit{-0.0479}} & \textbf{\textit{-0.0590}} & \textbf{\textit{-2.54}} & \textbf{\textit{-0.0813}} & \textbf{\textit{-0.0963}} \\
\bottomrule
\end{tabular}

\caption{\textbf{Quantitative comparison of obstruction robustness.}
We consider (i) the compound lens GT image, (ii) the metalens unobstructed capture, (iii) the metalens obstructed capture, and (iv) the reconstructed output produced by the corresponding method. We compute metrics for each lens design: hyperbolic phase metalens (Hyperbolic), broadband metalens without a de-occluding strategy (Broadband), and the de-occluding broadband metalens (Ours, De-occluding), and for each obstruction type (e.g., fence, dirt). We define the metric gap as $\Delta=\mathrm{Obs}-\mathrm{Unobs}$ for PSNR/SSIM and $\Delta=\mathrm{Unobs}-\mathrm{Obs}$ for LPIPS, so that a more negative $\Delta$ indicates larger degradation under obstruction and values closer to zero indicate higher robustness.}
\label{tab:supp_image_metrics}
\end{table*}

\clearpage 

\section*{Supplementary Note 11: Details on the downstream vision tasks}
\label{supp:note11}
\input{sections_supplement/vision_tasks}
\clearpage 

\section*{Supplementary Note 12: Metalens point spread functions and optical MTFs}
\label{supp:note12}
\begin{figure}[h]
	\centering
		\includegraphics[width=1\textwidth]{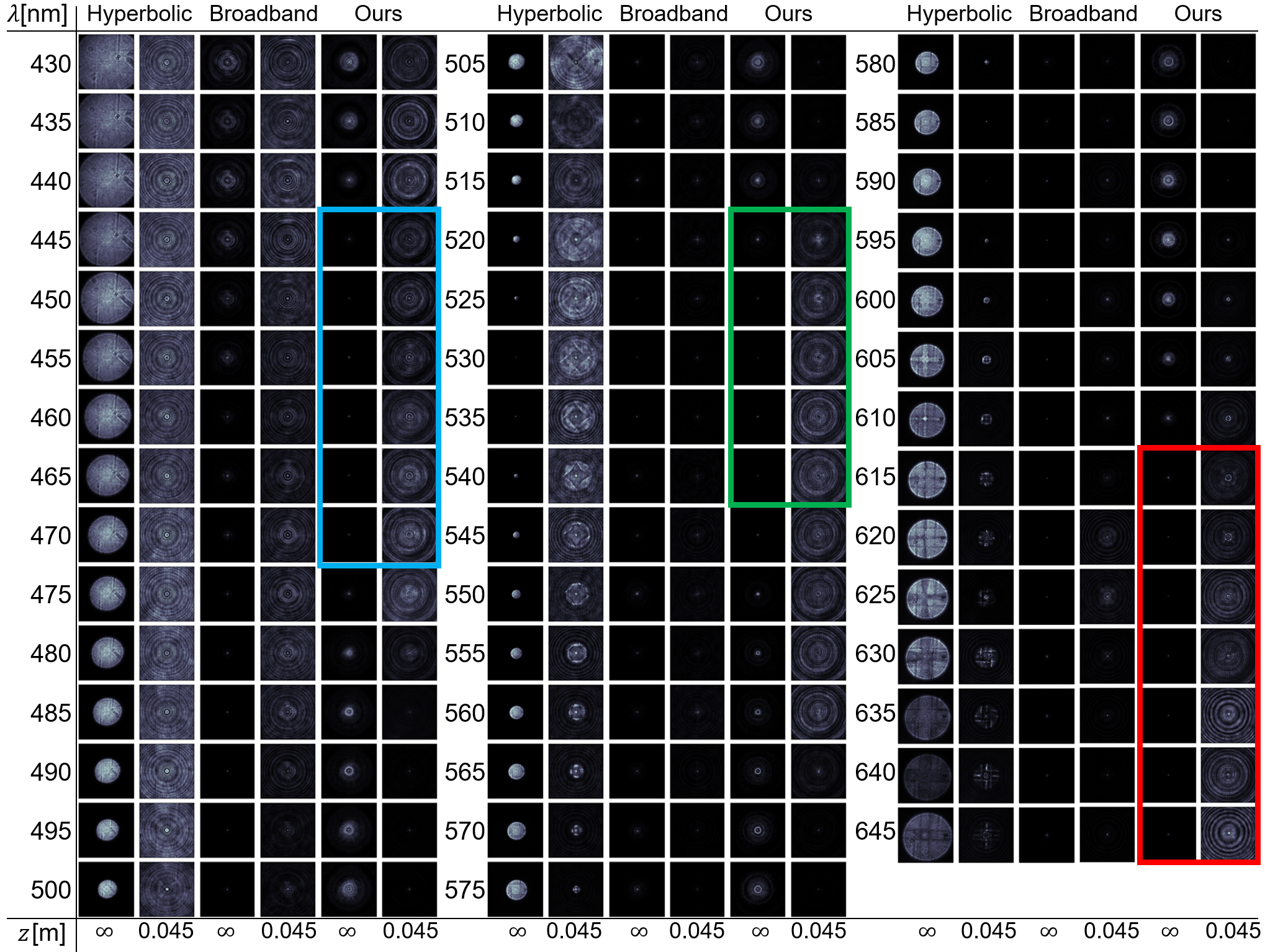}
		\caption{\textbf{Wavelength-dependent point spread functions (PSFs) at far and near depths.} We show 2D PSFs for far-depth and near-depth light sources for the three metalenses: hyperbolic phase metalens (Hyperbolic), broadband metalens without a de-occluding strategy (Broadband), and the de-occluding broadband metalens (Ours). The hyperbolic metalens shows the sharpest far-depth PSF near its nominal design wavelength, while the broadband metalens sustains sharp far-depth PSFs across a wider spectrum. Our far-depth PSFs remain sharp within the designated pass bands, highlighted by blue, green, and red boxes, enabling clear imaging of the background scene. Conversely, the near-depth PSFs within these same colored boxes are significantly blurred. This distinctive behavior--sharp far-depth PSFs and blurred near-depth PSFs within the pass bands--is central to our obstruction cloaking mechanism. We also observe a spectral shift in the focusing behavior as the depth of the incident light source changes; specifically, as the depth decreases (moving from far depth to near depth), the focused wavelengths shift toward the longer ones (red shift).}
		\label{fig:supp_full_psfs}
\end{figure}

We plot the wavelength-dependent point spread functions (PSFs) of metalenses: hyperbolic phase metalens (Hyperbolic), broadband metalens without a de-occluding strategy (Broadband), and the de-occluding broadband metalens (Ours), for the light sources at far depth and near depth in Figure~\ref{fig:supp_full_psfs}. The plot reveals the distinct spectro-optical characteristics of our design. The Hyperbolic metalens exhibits a standard diffractive behavior where a sharp PSF is confined strictly to its design wavelength. In contrast, the Broadband metalens effectively reduces chromatic aberration, maintaining consistently sharp far-depth PSFs across the majority of the visible spectrum. Our de-occluding broadband metalens, however, introduces a unique spectral modulation strategy. It is optimized to maintain sharp characteristic peaks only at specific pass bands for the far-depth sources. Crucially, at these same pass bands, the near-depth PSFs are strongly defocused (blurred). This contrast--sharp far-depth PSFs versus blurred near-depth PSFs--validates our spectral interleaving strategy, which allows the background information to be preserved while optically erasing the obstruction. Furthermore, consistent with claims in the main text and Supplementary Note~1, we observe a spectral shift in the focusing behavior as the depth of the incident light source changes; specifically, as the depth decreases (moving from far depth to near depth), the focused wavelengths shift toward the longer ones (red shift).

\subsection{PSF-based MTFs.}
\begin{figure}[h]
	\centering
		\includegraphics[width=1\textwidth]{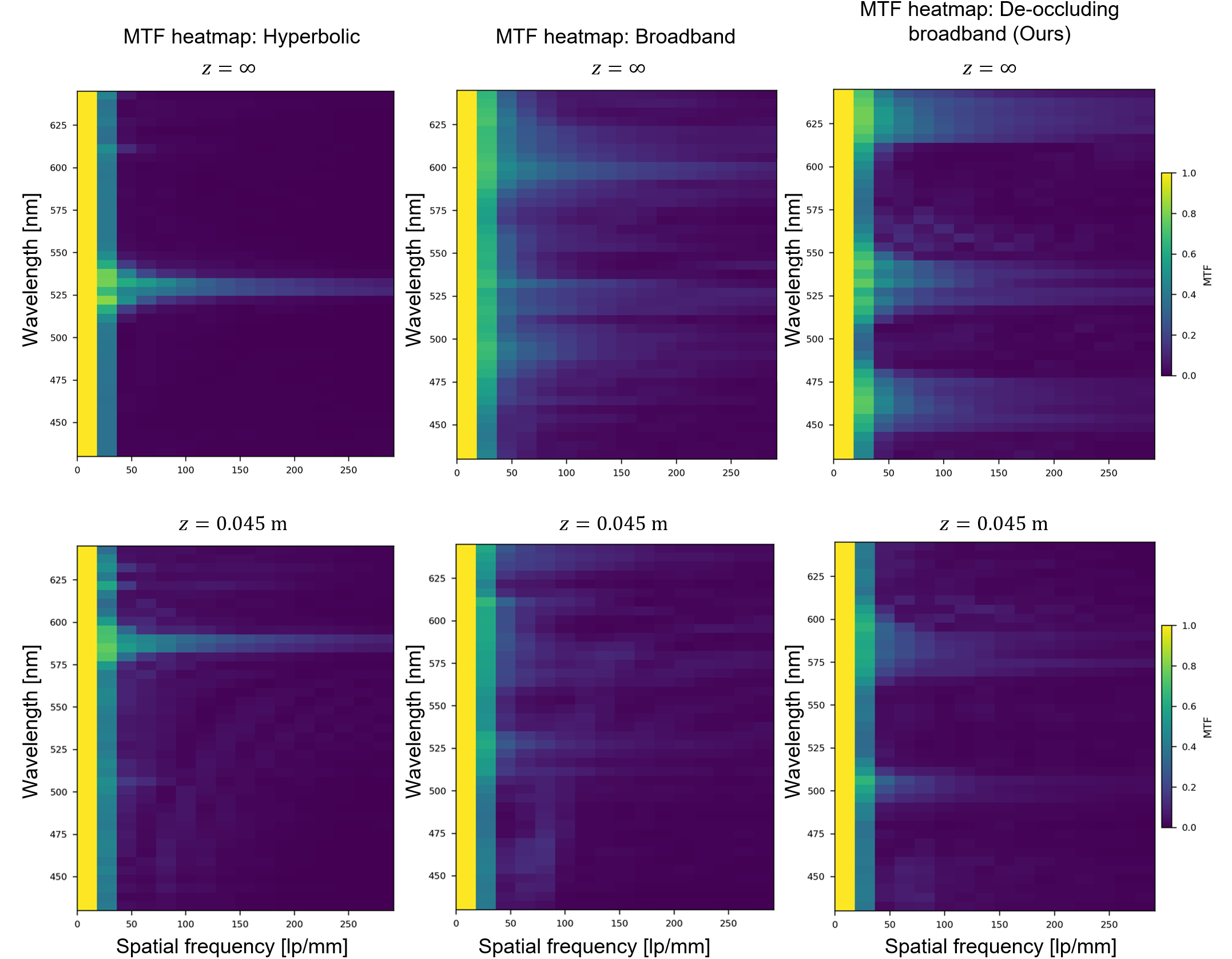}
		\caption{\textbf{PSF-based MTF heatmaps across wavelengths.} We show the MTF heatmaps of the near-depth PSFs and far-depth PSFs for three metalenses: hyperbolic phase metalens (Hyperbolic), learned broadband metalens without a de-occluding strategy (Broadband), and the de-occluding broadband metalens (Ours). The x-axis is spatial frequency (lp/mm) and the y-axis is wavelength (nm); color indicates MTF. The hyperbolic metalens shows the strongest response near its nominal design wavelength, while the broadband metalens sustains comparatively higher MTFs across a wider spectrum.
Our de-occluding broadband metalens exhibits enhanced MTFs at the multi-band spectral filter pass bands and suppressed response in the stop bands, consistent with the split-spectrum acquisition strategy. The high-MTF regions in the far depth ($z=\infty$) maps are spectrally shifted compared to the near depth ($z=0.045\,\text{m}$) maps.}
		\label{fig:supp_mtf_heatmaps}
\end{figure}

\begin{figure}[h]
	\centering
		\includegraphics[width=0.94\textwidth]{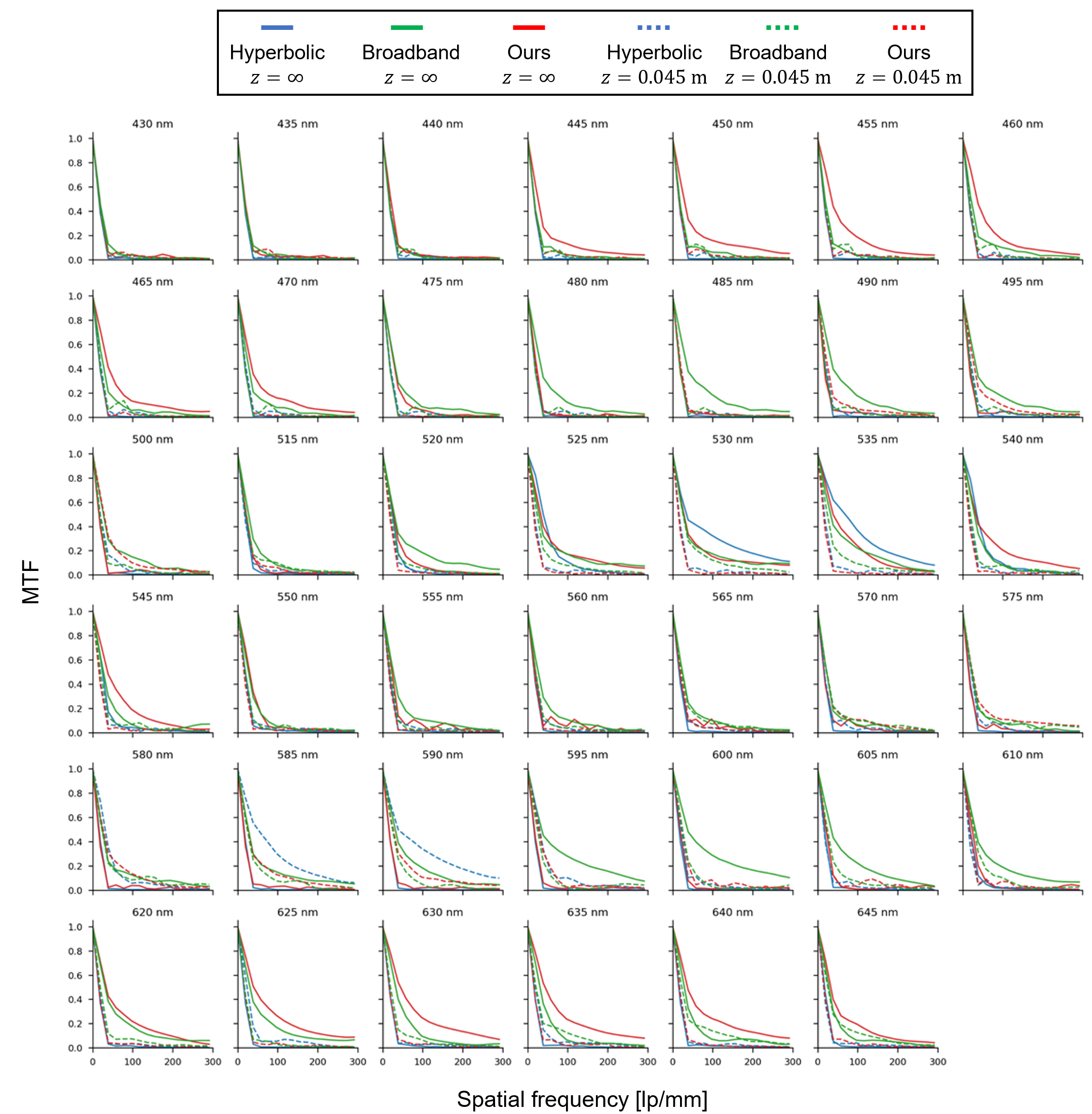}
		\caption{\textbf{Radially averaged MTF curves at each measured wavelength.}
Each panel plots MTF versus spatial frequency for one wavelength, comparing the three metalenses: hyperbolic phase metalens (Hyperbolic), broadband metalens without a de-occluding strategy (Broadband), and the de-occluding broadband metalens (Ours). Across wavelengths, the hyperbolic metalens peaks near its design wavelength, the broadband metalens remains relatively consistent over the spectrum, and our de-occluding broadband metalens provides higher MTFs in the pass bands and suppressed MTFs in the stop bands. The discrepancy between the solid lines ($z=\infty$) and dashed lines ($z=0.045\,\text{m}$) indicates the wavelength shift inherent to the near-depth MTF, which is consistent with the depth-wavelength symmetry.}
		\label{fig:supp_psf_based_mtfs}
\end{figure}

We computed the Modulation Transfer Function (MTF) from the measured point spread functions (PSFs) to quantitatively analyze the spectral focusing performance of the fabricated metalenses. Figures~\ref{fig:supp_mtf_heatmaps} and~\ref{fig:supp_psf_based_mtfs} show the characteristics of the Hyperbolic, Broadband, and de-occluding broadband (Ours) metalenses across the visible spectrum. The results highlight distinct spectral behaviors inherent to each design strategy: (i) the hyperbolic-phase metalens exhibits the strongest MTF near its nominal design wavelength (532\,nm) and substantially reduced response away from it, (ii) the broadband metalens maintains relatively elevated MTF over a wider spectral range, (iii) our de-occluding broadband metalens shows pronounced MTF enhancement at the filter pass bands while suppressing response in the stop bands. Furthermore, comparing the MTF profiles at $z=0.045\,\text{m}$ (near depth) with those at $z=\infty$ (far depth) reveals a spectral shift towards longer wavelengths. This observed wavelength shift experimentally validates the depth-wavelength symmetry.


\clearpage 

\section*{Supplementary Note 13: Slanted edge MTFs}
\label{supp:note13}
We conduct MTF measurements using the slanted-edge method based on ISO 12233 to provide a comprehensive evaluation of our imaging systems' performance. Our imaging systems consist of a joint hardware-software architecture with an image reconstruction neural network backend. Therefore, PSF-based MTF measurement does not fully reflect our system's imaging contrast fidelity, because it can typically only characterize the optical front-end. Therefore, we measure the holistic performance of the metalens and the restoration network by measuring the MTF directly from the reconstructed images of a slanted edge for each color channel. For a fair comparison and to demonstrate the generalization capability of our system, the checkerboard pattern used for this evaluation was strictly excluded from the image reconstruction network's training dataset.

\begin{figure}[h]
	\centering
		\includegraphics[width=0.9\textwidth]{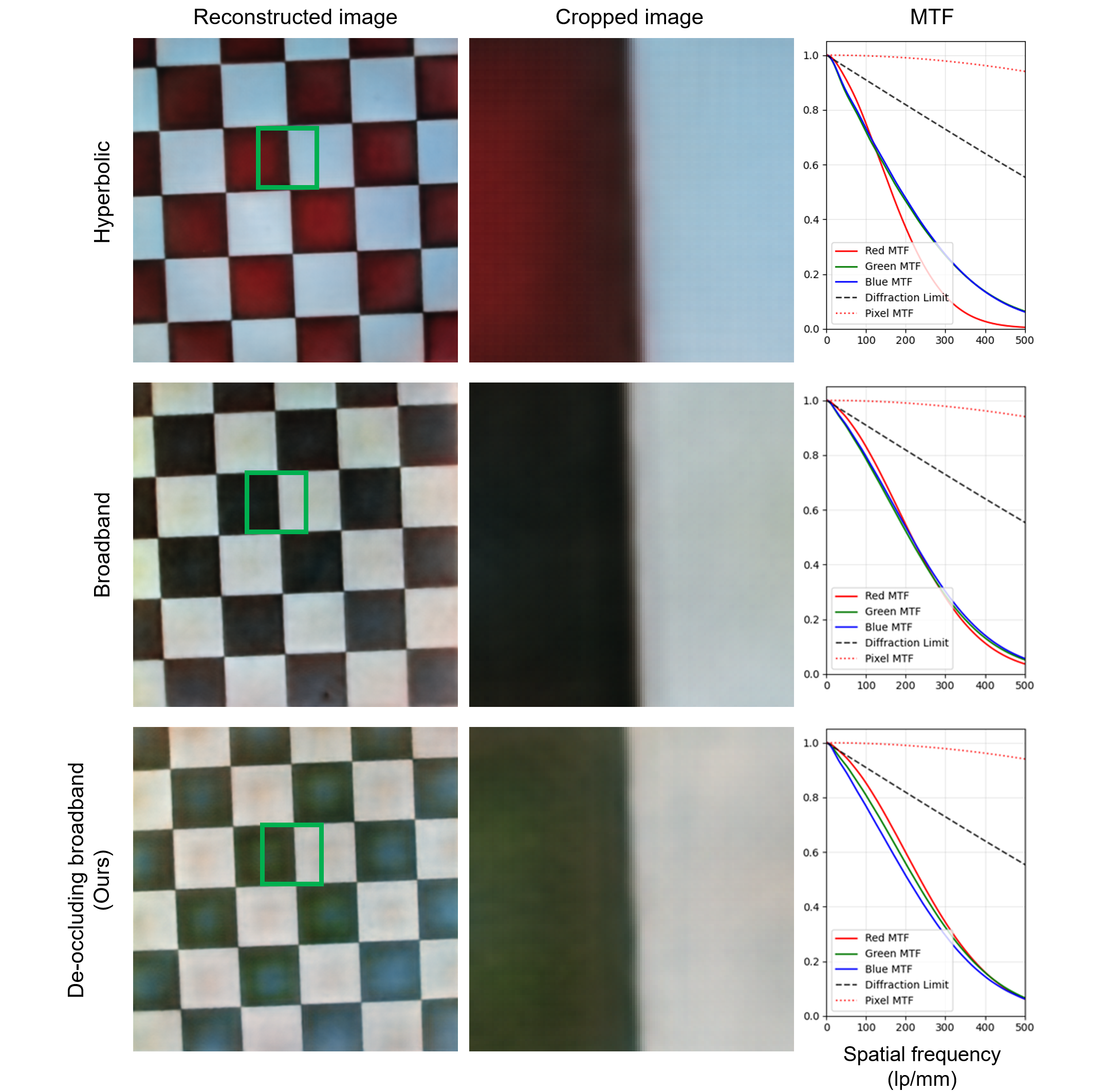}
		  \caption{\textbf{End-to-end MTF evaluation using the slanted-edge method.} We compare the imaging performance of the three metalenses: hyperbolic phase metalens (Hyperbolic), broadband metalens without a de-occluding strategy (Broadband), and the de-occluding broadband metalens (Ours), for each color channel of the sensor. The reconstructed checkerboard images (left) and the magnified cropped edges (middle) qualitatively demonstrate the superior sharpness achieved by the joint metalens-neural network systems. The quantitative MTF curves (right) show that our metalens and broadband metalens system maintain higher contrast up to high spatial frequencies, whereas the traditional hyperbolic metalens designs suffer from significant contrast loss for the red channel, reflecting the impact of residual chromatic aberrations. These results confirm that our system delivers high-quality, sharp imaging even for unobstructed scenes. The diffraction limit in the graph refers to the MTF of the diffraction-limited lens of our NA at wavelength 450\,nm. The pixel MTF is the Nyquist sampling limit of the imaging systems.} \label{fig:supp_slanted_edge_MTFs}
\end{figure}

As shown in Figure~\ref{fig:supp_slanted_edge_MTFs}, our approach is not only effective for obstruction-free imaging, but also excels in high-frequency imaging beyond occluded scenarios. Importantly, this performance is consistent across all color channels, confirming that the de-occluding broadband metalens effectively suppresses chromatic aberrations to enable robust broadband color imaging. In conclusion, the de-occluding broadband metalens can serve as a general-purpose compact camera.

\clearpage 

\section*{Supplementary Note 14: Additional experimental imaging results}
\label{supp:note14}
\paragraph{Static near-depth obstruction imaging on printed and in-the-wild scenes.}
We first provide additional qualitative results obtained under static near-depth obstructions, complementing the quantitative image-quality assessment in Supplementary Note~10. These results cover two scene categories: (i) high-resolution printed targets and (ii) real-world 3D scenes, hereafter referred to as in-the-wild scenes. 

\begin{figure}[h]
	\centering
		\includegraphics[width=0.94\textwidth]{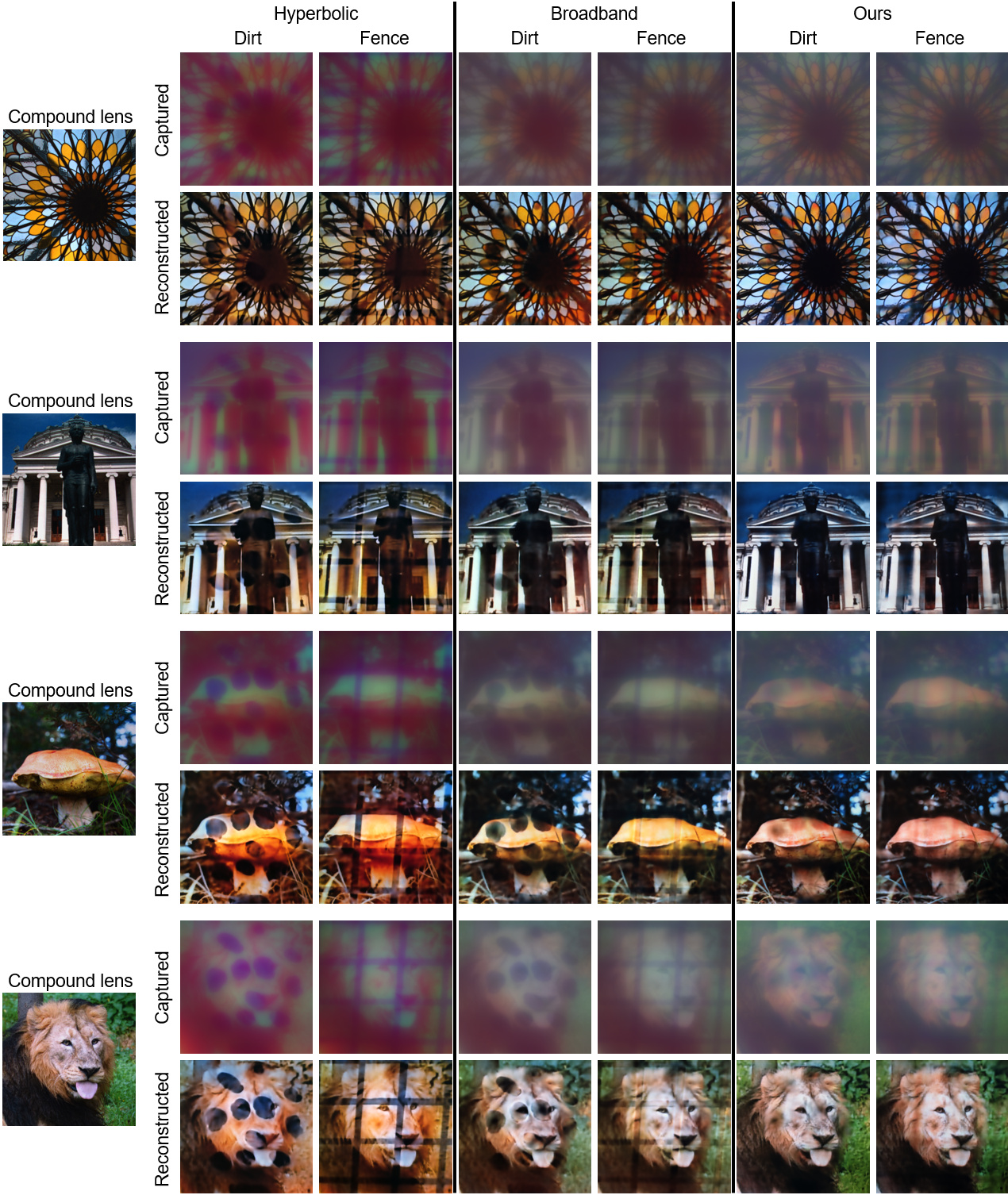}
		  \caption{\textbf{Qualitative comparisons under representative obstructions (dirt and fence).} We visualize the captured raw metalens image under dirt/fence and the reconstructed output produced by the corresponding method. The printed target images are from DIV2K\cite{Agustsson_2017_CVPR_Workshops}.}
		\label{fig:supp_additional_imaging_results_printed1}
\end{figure}

\begin{figure}[h]
	\centering
		\includegraphics[width=0.94\textwidth]{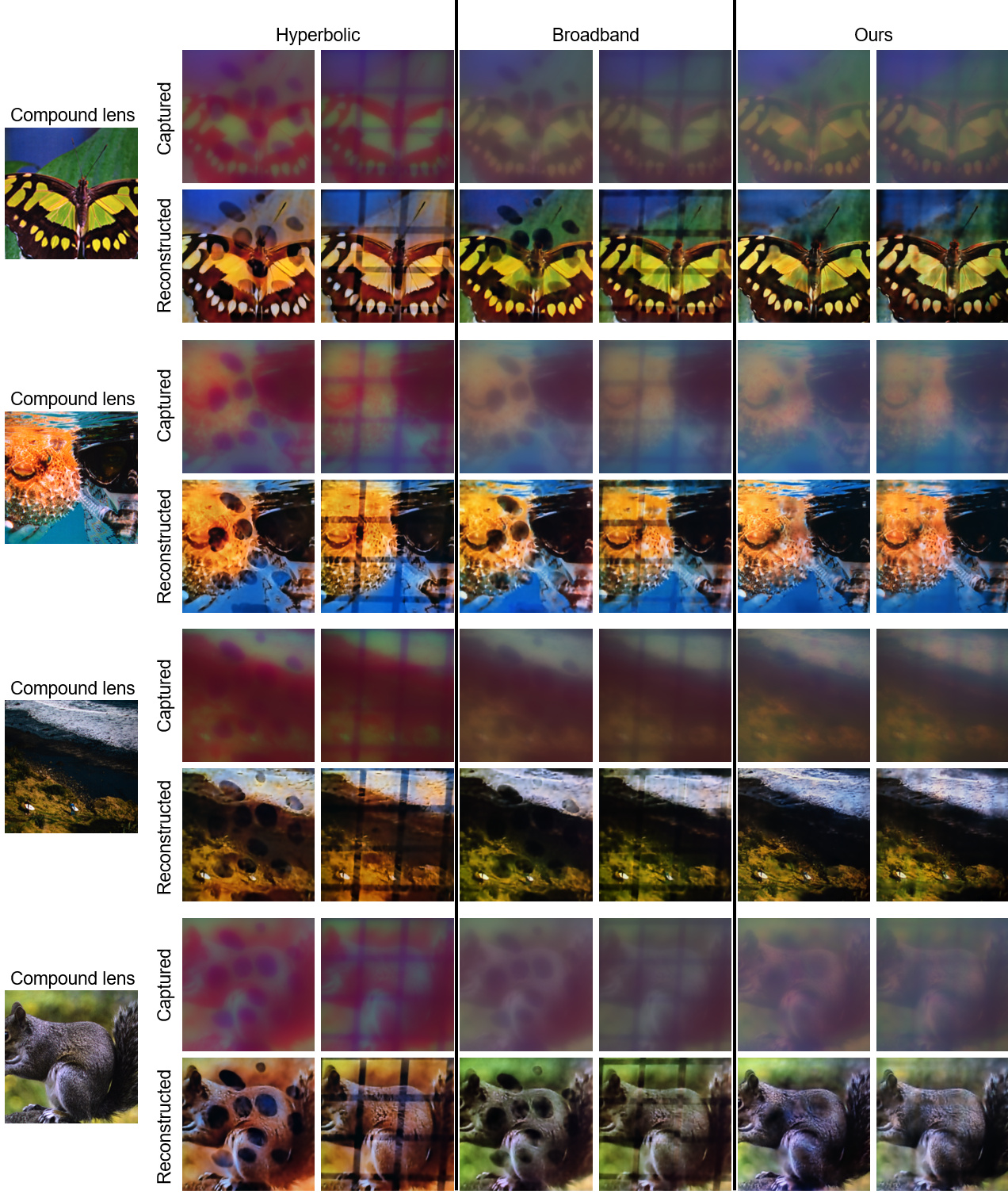}
		  \caption{\textbf{Additional qualitative results for dirt and fence obstructions.} We visualize the captured raw metalens image under dirt/fence and the reconstructed output produced by the corresponding method. The printed target images are from DIV2K\cite{Agustsson_2017_CVPR_Workshops}.}
		\label{fig:supp_additional_imaging_results_printed2}
\end{figure}

For the printed-target experiments, we used high-resolution images from the DIV2K dataset as planar scene targets. As discussed in the main text and Supplementary Note~5, printed targets are more suitable than digital display inputs for evaluating broadband imaging performance, because LCD or OLED displays typically exhibit narrow spectral peaks, whereas printed images under LED illumination reflect a continuous and broader spectrum (Figure~\ref{fig:supp_spectral_curves}). Figures~\ref{fig:supp_additional_imaging_results_printed1} and~\ref{fig:supp_additional_imaging_results_printed2} show qualitative comparisons of the printed scene captures under representative static near-depth obstructions, including dirt-like and fence-like patterns.

\begin{figure}[h]
	\centering
		\includegraphics[width=\textwidth]{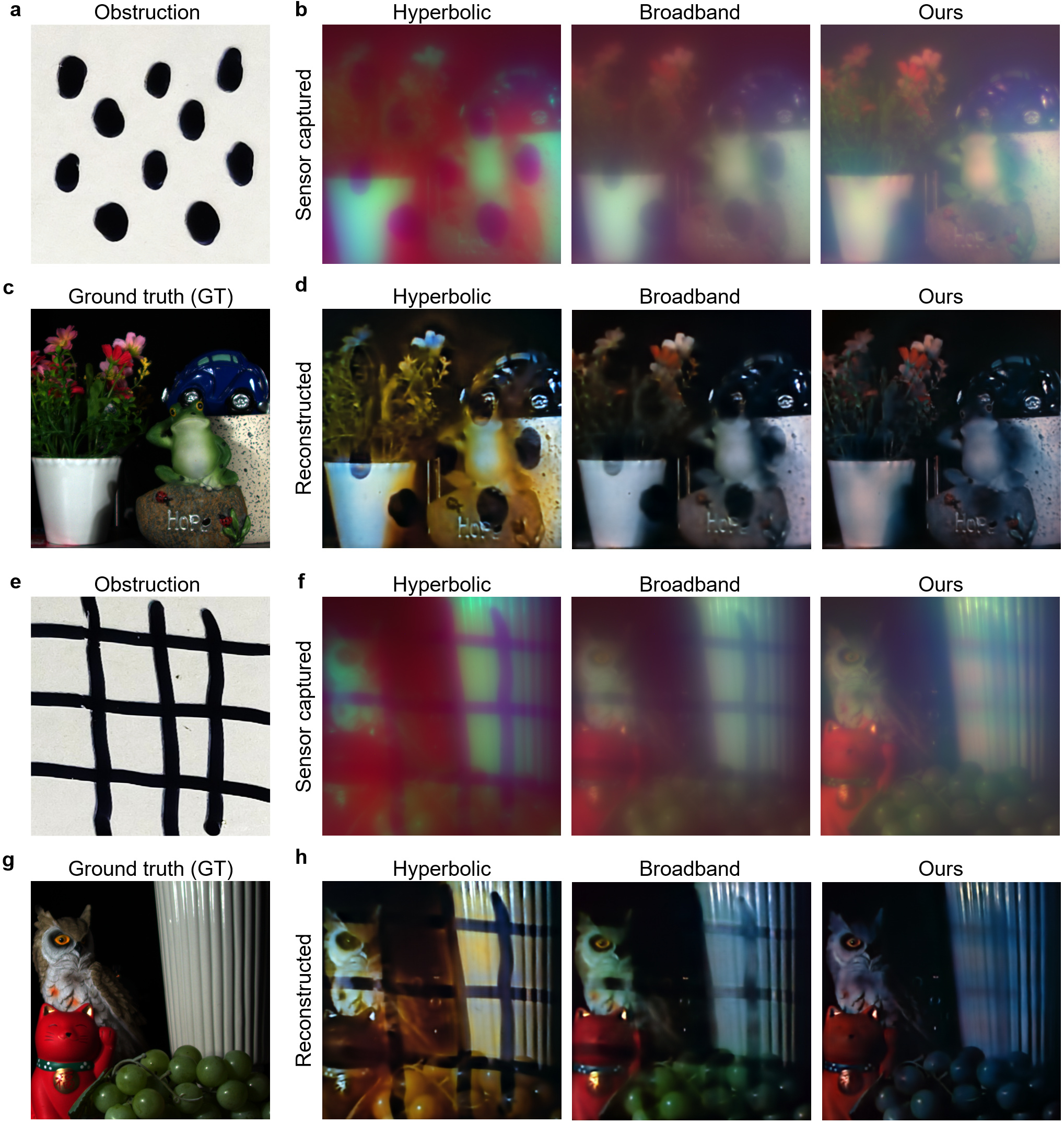}
		  \caption{\textbf{Additional in-the-wild qualitative results.} \textbf{a, e} Near-depth obstruction patterns. \textbf{b, f} Raw sensor images captured under the obstruction for the three metalens designs. \textbf{c, g} Ground-truth (GT) unobstructed reference images captured with a compound lens with $f=8\,$mm (two times longer than metalenses) for better object plane resolution. \textbf{d, h} Reconstructed images using the neural network trained for each metalens. \textbf{b, d, f, h} Our metalens achieves superior performance across all conditions.}
		\label{fig:supp_additional_imaging_results_in-the-wild}
\end{figure}

To demonstrate the imaging capability of our system beyond planar printed targets, we further captured real-world 3D objects under ambient indoor lighting conditions. Figure~\ref{fig:supp_additional_imaging_results_in-the-wild} shows in-the-wild imaging results with near-depth obstructions. These captures illustrate the applicability of the proposed de-occluding broadband metalens to non-planar scenes.

As shown in these figures, the images captured by the Hyperbolic and Broadband metalenses exhibit distinct obstruction artifacts, even after image reconstruction. In contrast, our de-occluding broadband metalens suppresses the near-depth obstruction imprint in the raw capture, enabling the reconstruction network to recover high-fidelity details and accurate colors that closely match the ground truth. 

\paragraph{Dynamic near-depth obstruction imaging.}
The experiments in the main text primarily use static near-depth obstruction patterns to enable repeatable image-quality measurements. To examine whether the proposed method is applicable to time-varying occlusions, we additionally captured videos with moving near-depth obstructions placed around \(z_{\mathrm{near}}\simeq0.045~\mathrm{m}\), while the far-depth target was kept fixed at \(z_{\mathrm{far}}\simeq0.6~\mathrm{m}\). Representative video frames are shown in Figure~\ref{fig:supp_dynamic_obstructions}, and the corresponding videos are provided in Supplementary Video~1.

The proposed method is inherently compatible with dynamic near-depth obstructions because it is a single-shot, single-view imaging method. Each video frame is captured and reconstructed independently, without requiring temporal parallax, optical flow, or multi-frame fusion. Therefore, unlike multi-view or burst-based occlusion separation methods, the method does not require the obstruction to remain static across frames. The split-spectrum metalens suppresses near-depth obstruction components optically at the time of capture, and the same reconstruction network is then applied frame-by-frame.

Furthermore, when the near-depth obstruction moves during the exposure time, the obstruction contribution is additionally reduced by ordinary motion blur. Because our optical design already defocuses near-depth obstruction components, this motion blur can further reduce obstruction artifacts.

For the dynamic video acquisition, we used the same benchtop 4f imaging setup described in the Supplementary Note~9, but with a higher-frame-rate capture setting than that used for the static image-quality measurements. Specifically, the exposure time is reduced and pixel binning is enabled to increase the video frame rate to 10\,FPS (the maximum the sensor can achieve). These acquisition changes allow us to capture time-varying obstruction motion more clearly. However, the shorter exposure and pixel binning can lead to increased sensor noise and reduced high-frequency detail compared with the longer-exposure static captures.

To quantify the dynamic-obstruction behavior, we sampled 21 frames from the dynamic sequences of the Supplementary Video~1 and computed PSNR, SSIM, and LPIPS against the unobstructed compound-lens ground truth using the same alignment and photometric normalization protocol as in Supplementary Note~10. Table~\ref{tab:supp_dynamic_obstruction} summarizes the frame-wise mean and standard deviation. 

The imaging system for dynamic obstructions can be further optimized by training the reconstruction network with dynamic-obstruction augmentation, including randomized obstruction motion, motion blur, and temporal intensity variation. Optional temporal consistency or recurrent reconstruction can also be incorporated, although the proposed optical de-occlusion itself is inherently snapshot and does not rely on multi-frame processing.

\paragraph{Color differences in the experimental comparisons.} The Hyperbolic, Broadband, and Ours cases were captured using different effective spectral responses. The Broadband and Hyperbolic baselines do not use the multi-band bandpass filter, whereas the proposed system uses the filter. 

In the experiment, we adjusted the acquisition and display normalization to match the overall brightness across the three metalens cases. However, the commercial tri-band filter used in the proposed system has different pass-bandwidths across the RGB bands, with the red pass band being wider than the blue and green pass bands. As a result, red-dominant regions can receive relatively stronger signal in the proposed system. 

This saturation effect can be mitigated by exposure bracketing, per-channel white balancing, HDR acquisition, saturation-aware training, or by using a custom multi-band filter with balanced band transmissions. 

\begin{figure}[h]
	\centering
		\includegraphics[width=\textwidth]{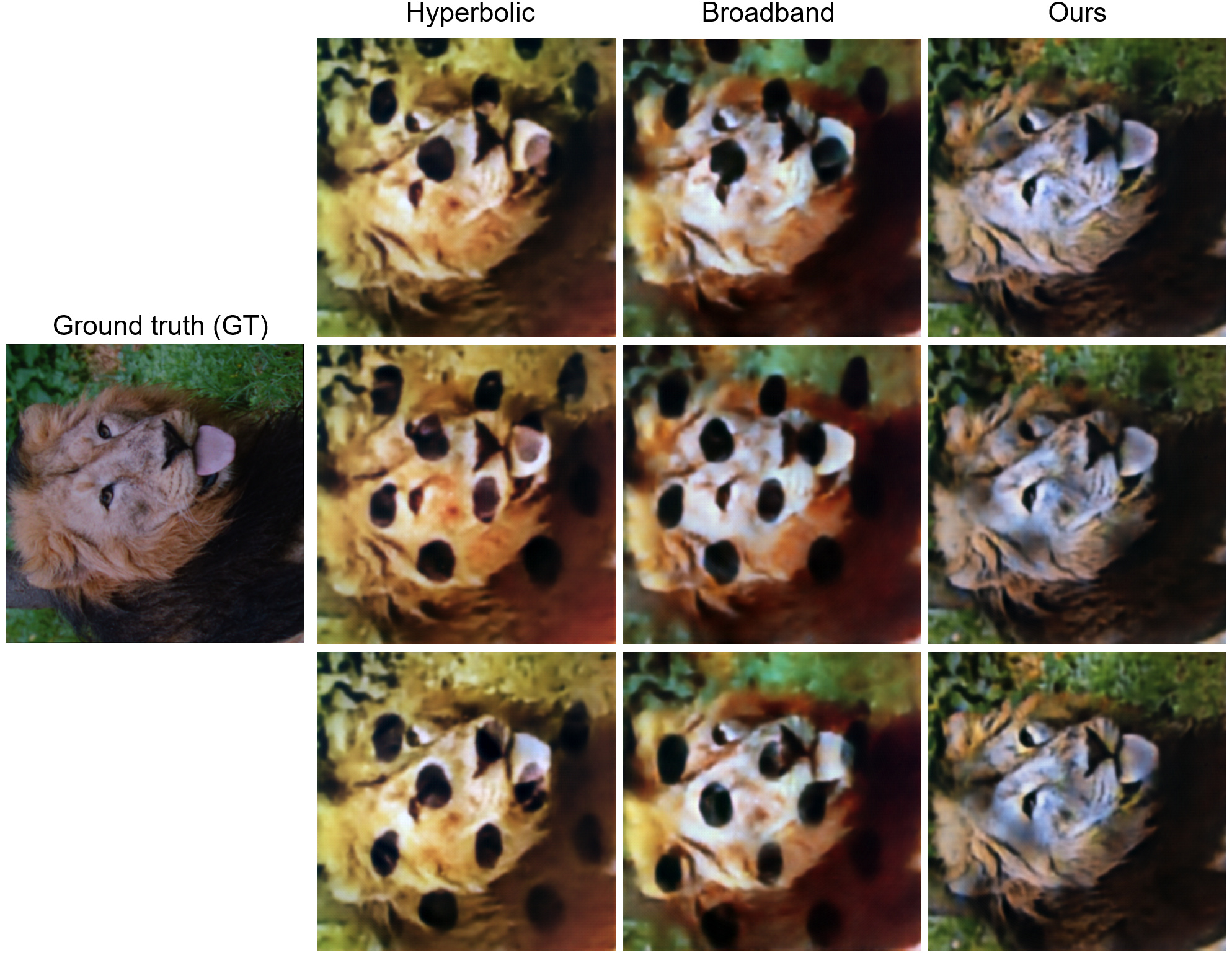}
		  \caption{\textbf{Imaging under dynamic near-depth obstructions.} Representative frames from Supplementary Video~1 showing imaging results under moving near-depth obstructions. The Hyperbolic and Broadband baselines exhibit stronger obstruction imprints after reconstruction, whereas the proposed metalens suppresses the moving near-depth obstruction optically and maintains more stable far-depth scene structures. Because the method is single-shot and single-view, each video frame is reconstructed independently without temporal fusion or optical-flow estimation. Each row corresponds to consecutive frames. The printed target image is from DIV2K\cite{Agustsson_2017_CVPR_Workshops}.}
		\label{fig:supp_dynamic_obstructions}
\end{figure}

\begin{figure}[h]
	\centering
		\includegraphics[width=1\textwidth]{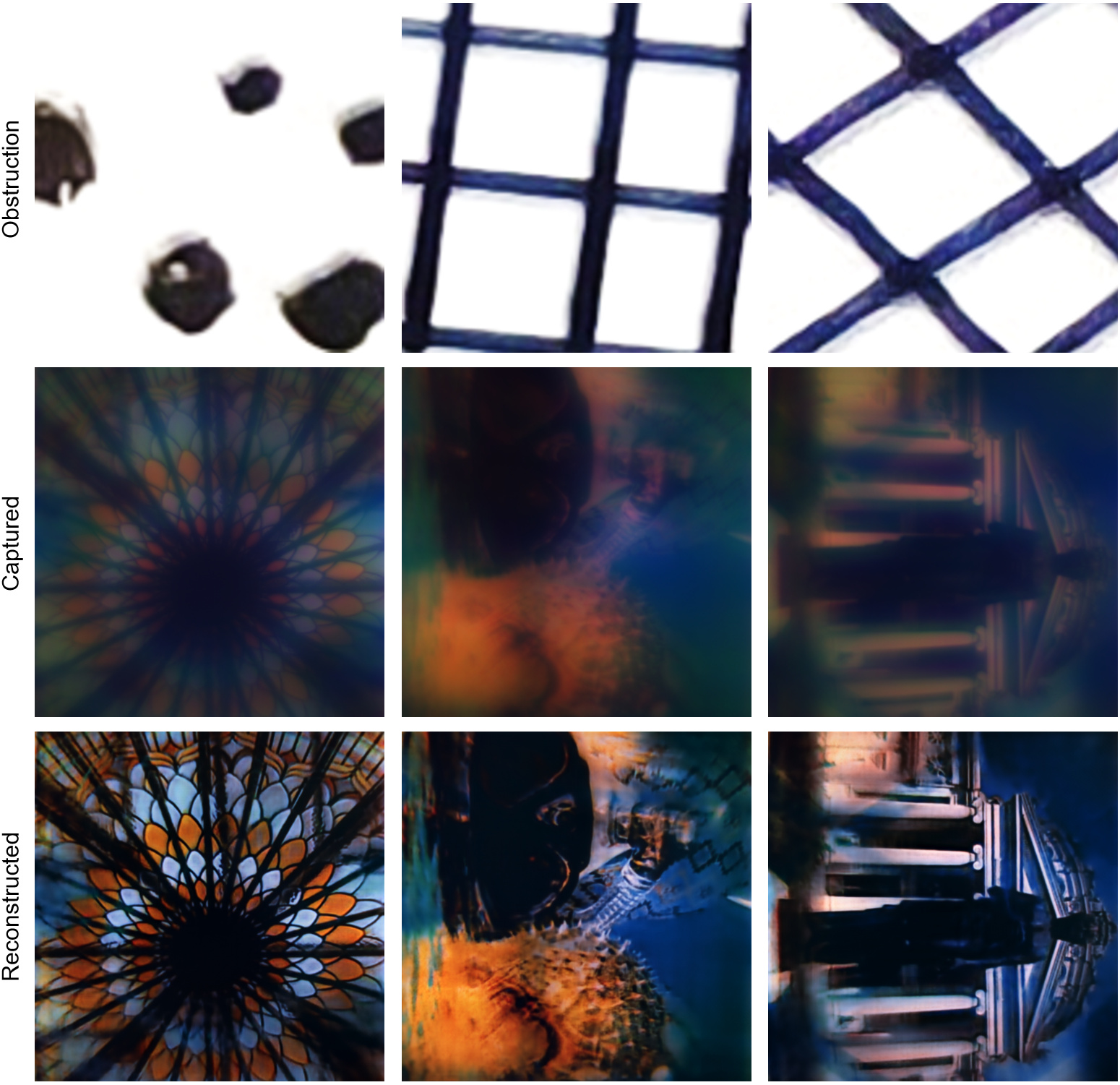}
		\caption{\textbf{Results using the compact prototype on the printed images.} The printed target images are from DIV2K\cite{Agustsson_2017_CVPR_Workshops}.}
		\label{fig:supp_compactcamera_printedcapture}
\end{figure}

\begin{table}
\centering
\setlength{\tabcolsep}{3pt}
\renewcommand{\arraystretch}{1.12}
\begin{footnotesize}
\begin{tabular}{l c c c c}
\toprule
\textbf{Lens design} &
\textbf{\# of frames} &
\textbf{PSNR (dB) $\uparrow$} &
\textbf{SSIM $\uparrow$} &
\textbf{LPIPS $\downarrow$} \\
\midrule
Hyperbolic &
21 &
20.72 &
0.6640 &
0.5402  \\
Broadband &
21 &
22.16 &
0.6938 &
0.5203 \\
\textbf{Ours} &
\textbf{21} &
\textbf{25.39} &
\textbf{0.8016} &
\textbf{0.4052} \\
\bottomrule
\end{tabular}
\end{footnotesize}
\caption{\textbf{Frame-wise quantitative comparison under dynamic near-depth obstruction.}
Metrics were computed from 21 sampled frames of the dynamic-obstruction video using the unobstructed compound-lens image as the ground truth. The same geometric alignment and photometric normalization protocol as in Supplementary Note~10 was used. The dynamic sequences were captured using a shortened exposure time and pixel-binned readout to increase the frame rate (10\,FPS, the maximum the sensor can achieve).}
\label{tab:supp_dynamic_obstruction}
\end{table}

\clearpage 

\section*{Supplementary Note 15: Robustness to illumination spectrum, polarization, and intensity.}
\label{supp:note15}

\subsection{Robustness to illumination spectrum of the proposed system.}
The proposed system uses a fixed multi-band bandpass filter with pass bands around 457, 530, and 628 nm, and a practical question is whether real-world illumination spectra and object reflectance spectra provide sufficient optical power within these pass bands. Prior measurements show that many natural and man-made object reflectance spectra are broad, smooth, and low-dimensional\cite{vrhel1994measurement,westland2000colour}, and practical scenes are therefore expected to rarely contain zero spectral energy across all selected pass bands. To evaluate this, we performed a hyperspectral-dataset-based throughput analysis using real-world spectral data\cite{yasuma2010generalized, chakrabarti2011statistics}.

\subsection{Dataset-driven spectral throughput analysis.}
For each hyperspectral scene, we computed the scene-averaged filtered intensity of the proposed de-occluding imaging system relative to the conventional RGB camera as follows:
\[
\tau_{\mathrm{spec}} = \frac{1}{|\Omega|} 
\int_{\Omega}
\int_{\Lambda}
L(\mathbf{x},\lambda)T_{\mathrm{BPF}}(\lambda)\eta_{\mathrm{meta}}(\lambda)
d\lambda d\mathbf{x},
\]
where $T_{\mathrm{BPF}}(\lambda)$ is the multi-band bandpass filter transmission, \(\eta_{\mathrm{meta}}\) is the meta-atom polarization-conversion efficiency, and $L(\textbf{x},\lambda)$ is the hyperspectral intensity.
To isolate spectral-shape mismatch from overall brightness, each hyperspectral scene was normalized to have the same total visible intensity before applying $S(\lambda)$. Table~\ref{tab:supp_spectral_throughput} summarizes the resulting filtered-intensity distribution across the dataset. The maximum absolute difference between the brightest and darkest scenes after system spectral sensitivity filtering was 7.44\,\%, indicating that realistic spectral variations do not strongly suppress the selected pass-band throughput. This is consistent with the broad and smooth spectral statistics of natural reflectance spectra. The main consequence of bandpass filtering is reduced throughput rather than complete loss of object information, and our system is expected to retain color vividness.

\subsection{Robustness to polarization of the PB-phase implementation.}
The present experimental implementation uses a Pancharatnam--Berry (PB) phase metalens, whose geometric phase is imparted to the cross-polarized circular component. Therefore, the current optical stack includes an LCP generator before the metalens and an RCP analyzer after the metalens, implemented using LP and QWP elements, as shown in Figure~\ref{fig:supp_real_imaging_setup}. These polarization optics define the input and detected polarization states required for the PB-phase response.

Importantly, the method does not require the scene illumination itself to be circularly polarized. The imaging experiments in this work were conducted under LED or ambient indoor illumination without assuming pre-polarized scene light. The input LP selects one linear component, and the QWP converts it into the desired circular polarization. Thus, the PB-phase metalens receives a well-defined circular polarization state, and the primary consequence is a throughput penalty, typically a factor of two at the input polarizer, rather than a failure of the phase response. 

This behavior can be expressed using the Stokes vector of the light incident on the input LP,
\[
\mathbf{S}=[S_0,S_1,S_2,S_3]^\top .
\]
For an input linear polarizer with transmission axis \(\psi\), the transmitted intensity is
\[
I_{\mathrm{LP}} = \frac{1}{2} \left(S_0 + S_1\cos 2\psi + S_2\sin 2\psi \right).
\]
Equivalently, for partially linearly polarized light with degree of linear polarization \(\rho_L\) and polarization angle \(\phi\),
\[
I_{\mathrm{LP}} = \frac{S_0}{2} \left[1+\rho_L\cos 2(\phi-\psi)\right].
\]
Therefore, unpolarized light \((\rho_L=0)\) transmits \(S_0/2\), while fully linearly polarized light can vary between full transmission and nearly zero transmission depending on its angle relative to the LP axis. 

The most challenging case occurs when the scene radiance reaching the camera is dominated by a fully linearly polarized component nearly orthogonal to the input LP axis. This can occur for strongly specular reflections or reflections near the Brewster angle under linearly polarized light illumination. In this case, the selected polarization channel may receive little light, leading to reduced SNR or missing local scene information. In contrast, many practical scenes contain mixtures of diffuse and specular reflection, multiple surface orientations, multiple scattering paths, and are illuminated by unpolarized light. Therefore, the scene radiance is generally partially polarized rather than a single fully polarized state. Recent real-world spectro-polarimetric datasets further show diversity of polarization states in natural scenes\cite{Jeon_2024_CVPR}. Therefore, the present PB-phase implementation is compatible with unpolarized and moderately partially polarized illumination, but it can suffer throughput loss only in strongly polarized edge cases.

Note that the proposed split-spectrum de-occluding principle is not restricted to PB-phase metalenses. As shown in Supplementary Note~1, the depth-wavelength symmetry model is also applicable to propagation-phase metalenses. A polarization-insensitive propagation-phase implementation using isotropic or dual-polarization meta-atoms would remove the need for the LCP generator and RCP analyzer, reducing both polarization sensitivity and optical-stack thickness. This provides a route toward more polarization-robust and integration-friendly future implementations.

\subsection{Dataset-driven polarization throughput analysis.}
To quantify the effect of reduced light throughput under realistic polarization statistics, we performed a dataset-driven polarization-throughput analysis using the real-world spectro-polarimetric dataset\cite{Jeon_2024_CVPR}, which provides full-Stokes measurements for diverse scenes.

The scene-level polarization throughput is defined as
\[
\tau_{\mathrm{pol}}(\psi)
=
\frac{
\sum_{\mathbf{x}} I_{\mathrm{LP}}(\mathbf{x};\psi)
}{
\sum_{\mathbf{x}} S_0(\mathbf{x})
}.
\]
For unpolarized light, \(\tau_{\mathrm{pol}}=0.5\), corresponding to the ideal factor-of-two intensity loss at the input polarizer. For partially polarized real-world scenes, \(\tau_{\mathrm{pol}}\) can deviate from 0.5 depending on the degree and angle of linear polarization. 

We report the average throughput for the input-LP axis of $\psi$ from $0^\circ$ to $90^\circ$ for $5^\circ$ steps, the throughput of the worst-case value, and the best-case value. Table~\ref{tab:supp_polarization_throughput} summarizes the distribution of \(\tau_{\mathrm{pol}}\) over the evaluated Stokes images. The results show that the selected-channel throughput remains close to the unpolarized value for most real-world scenes, indicating that practical scene radiance is rarely dominated by a single fully linearly polarized component orthogonal to the input LP. The main effect of polarization selection is therefore a predictable throughput reduction and SNR penalty, rather than a failure of the PB-phase response. Strong specular reflections or Brewster-angle reflections can still create highly polarized local regions, and these cases remain a limitation of the current PB-phase implementation. Again, since the depth-wavelength symmetry model is not restricted to the PB-phase case, a polarization-insensitive propagation-phase implementation can remove the need for the input/output polarizers.

\subsection{Throughput, focusing efficiency, and far-PSF MTF analysis of the complete optical stack.}

We combined the spectral-throughput and polarization-throughput analyses to estimate the detected photon throughput of the complete optical stack. The relative detected throughput was computed as
$
\tau_{\mathrm{det}}
=
\tau_{\mathrm{spec}}
\cdot
\tau_{\mathrm{pol}}
$
where \(\tau_{\mathrm{spec}}\) is the mean spectral throughput obtained from the hyperspectral analysis and \(\tau_{\mathrm{pol}}\) is the polarization throughput obtained from the Stokes-image analysis. For the Hyperbolic and Broadband metalens baselines, the same PB-phase polarization optics are included, but no multi-band bandpass filter is used. For our de-occluding broadband metalens, both the PB-phase polarization optics and the multi-band filter are included. Table~\ref{tab:supp_detected_throughput_and_focusing} summarizes the estimated detected photon throughput result.

However, raw detected photon throughput alone does not fully determine imaging quality. The useful signal for reconstruction also depends on how much of the detected light contributes to a high-frequency-preserving far-depth PSF. Therefore, we additionally evaluated both the focused-signal efficiency and the far-PSF sharpness. For each lens design, we first computed the focusing efficiency as the fraction of optical power concentrated within a
10-pixel-radius circular window around the far-depth PSF peak,
\[
\eta_{\mathrm{focus}}
=
\frac{
\int_{\Lambda} S(\lambda)
\left[
\sum_{\mathbf{x}\in \mathcal{C}_{10}(\lambda)}
P_{\mathrm{far}}(\mathbf{x},\lambda)
\right]
d\lambda
}{
\left[
\sum_{\mathbf{x}\in \Omega_{\mathrm{in}}}
P_{\mathrm{in}}(\mathbf{x})
\right]
d\lambda
},
\]
where \(S(\lambda)\) is the system spectral sensitivity used for each lens
design, \(P_{\mathrm{far}}(\mathbf{x},\lambda)\) is the far-depth PSF intensity
on the sensor plane, \(P_{\mathrm{in}}(\mathbf{x},\lambda)\) is the incident
intensity on the metalens aperture, and \(\mathcal{C}_{10}(\lambda)\) denotes
a circular region with a radius of 10 pixels centered at the peak of
\(P_{\mathrm{far}}(\mathbf{x},\lambda)\). The region \(\Omega_{\mathrm{in}}\)
denotes the incident aperture region used to normalize the focused power.

To quantify far-PSF sharpness, we computed the averaged modulation transfer function weighted by the system spectral sensitivity and computed the area under the MTF curve $A_\text{MTF}$. A higher $A_\text{MTF}$ indicates stronger preservation of spatial frequencies and sharper far-depth imaging.

Table~\ref{tab:supp_detected_throughput_and_focusing} summarizes the estimated detected photon throughput, focused-signal efficiency, and far-depth MTF sharpness for the three PB-phase metalens systems. The proposed system has a lower raw detected throughput because of the additional multi-band filtering. However, the proposed system maintains comparable focused-signal efficiency and achieves the highest \(A_{\mathrm{MTF}}\), indicating that the filter reduces raw photon throughput while concentrating useful high-frequency information into the selected pass-band regions.

\subsection{Sensitivity under reduced photon budget.}
We further evaluated the robustness of the proposed de-occluding metalens to reduced photon budget by capturing an additional 10 scenes without obstructions while decreasing the exposure time and increasing the sensor gain. This experiment emulates practical low-throughput conditions caused by any kind of spectral or polarization mismatch, or simply low-light conditions. Table~\ref{tab:supp_exposure_gain} shows that the image metric degrades only moderately as the exposure time is reduced, because the reconstruction network also provides a denoising effect. 

Nevertheless, we note that severe low-light conditions or extremely narrow-band illuminations can still reduce SNR, and backlight or overexposure can cause irreversible sensor saturation. These cases can be mitigated by exposure control, auto-gain, HDR acquisition, illumination augmentation during training, or application-specific co-design of the filter and metalens.

\begin{table}
\centering
\setlength{\tabcolsep}{3pt}
\renewcommand{\arraystretch}{1.12}
\begin{footnotesize}
\makebox[\columnwidth][c]{%
\begin{tabular}{l c c}
\toprule
\textbf{Dataset / spectra} &
\textbf{Mean $\tau_\text{spec}$ (\%)} &
\textbf{Min.--max. $\tau_\text{spec}$ (\%)} \\
\midrule
Indoor \& outdoor scenes under daylight\cite{chakrabarti2011statistics} &
35.77 &
35.11--37.48 \\
Indoor scenes under artificial \& mixed illumination\cite{chakrabarti2011statistics} &
37.87 &
34.56--42.01 \\
Natural \& man-made object spectra\cite{yasuma2010generalized} &
37.49 &
35.62--42.57 \\
\bottomrule
\end{tabular}%
}
\end{footnotesize}
\caption{\textbf{Dataset-driven spectral-throughput analysis using hyperspectral image dataset}
Each hyperspectral scene or reflectance spectrum was normalized to have the same total visible intensity before applying the effective system spectral sensitivity \(S(\lambda) = \sum_{c\in\{\text{R,G,B}\}}\mathrm{CRF}^{c}(\lambda)\cdot T_{\mathrm{BPF}}(\lambda)\cdot \eta_{\mathrm{meta}}(\lambda)\). The table reports the scene-averaged filtered intensity of the proposed imaging system relative to the conventional RGB camera. The small relative variation indicates that realistic spectral radiance/reflectance variations do not strongly suppress the selected pass-band throughput.}
\label{tab:supp_spectral_throughput}
\end{table}

\begin{table}
\centering
\setlength{\tabcolsep}{3pt}
\renewcommand{\arraystretch}{1.12}
\begin{footnotesize}
\makebox[\columnwidth][c]{%
\begin{tabular}{l c c c}
\toprule
\textbf{Dataset} &
\textbf{Mean \(\boldsymbol{\tau_{\mathrm{pol}}}\) (\%)} &
\textbf{Min.-max. \(\boldsymbol{\tau_{\mathrm{pol}}}\) (\%)}  \\
\midrule
Trichromatic Stokes scenes\cite{Jeon_2024_CVPR} &
49.81 &
40.48--58.10 \\
\bottomrule
\end{tabular}%
}
\end{footnotesize}
\caption{\textbf{Dataset-driven polarization-throughput analysis using real-world trichromatic Stokes images.}
The polarization throughput \(\tau_{\mathrm{pol}}\) is the fraction of incident scene intensity transmitted by the input linear polarizer before QWP conversion. For unpolarized light, \(\tau_{\mathrm{pol}}=0.5\). We report the average throughput for the input-LP axis from $0^\circ$ to $90^\circ$ for $5^\circ$ steps, the throughput of the brightest case, and the darkest case. Values close to 0.5 indicate that the selected-channel throughput is not dominated by the severe polarization mismatch.}
\label{tab:supp_polarization_throughput}
\end{table}

\begin{table}
\centering
\setlength{\tabcolsep}{3pt}
\renewcommand{\arraystretch}{1.12}
\begin{footnotesize}
\makebox[\columnwidth][c]{%
\begin{tabular}{l c c c c c}
\toprule
\textbf{Lens design} &
\begin{tabular}[c]{@{}c@{}}\(\boldsymbol{\tau_{\mathrm{spec}}}\)\\\textbf{(\%)}\end{tabular} &
\begin{tabular}[c]{@{}c@{}}\(\boldsymbol{\tau_{\mathrm{pol}}}\)\\\textbf{(\%)}\end{tabular} &
\begin{tabular}[c]{@{}c@{}}\(\boldsymbol{\tau_{\mathrm{det}}}\)\\\textbf{(\%)}\end{tabular} &
\begin{tabular}[c]{@{}c@{}}\(\boldsymbol{\eta_{\mathrm{focus}}}\)\\\textbf{(\%)}\end{tabular} &
\begin{tabular}[c]{@{}c@{}}\(\boldsymbol{A_{\mathrm{MTF}}}\)\\\textbf{(lp/mm)}\end{tabular} \\
\midrule
Hyperbolic &
100 &
49.81 &
49.81 &
7.03 &
12.671 \\
Broadband &
100 &
49.81 &
49.81 &
5.00 &
7.124 \\
\textbf{Ours} &
\textbf{37.04} &
\textbf{49.81} &
\textbf{18.45} &
\textbf{6.36} &
\textbf{13.410} \\
\bottomrule
\end{tabular}%
}
\end{footnotesize}
\caption{\textbf{Estimated detected photon throughput, focused-signal efficiency, and far-depth MTF sharpness.}
All three metalenses in this comparison are PB-phase implementations and share the same polarization-throughput factor $\tau_{\mathrm{pol}}$. Note that the lenses can be implemented with the polarization insensitive propagation phase (Supplementary Note~1), with $\tau_{\mathrm{pol}}=1$. The spectral-throughput factor $\tau_{\mathrm{spec}}$ is estimated from the system spectral response and hyperspectral scene statistics. The polarization-throughput factor $\tau_{\mathrm{pol}}$ is estimated from real-world Stokes images and represents the selected-channel throughput of the input polarization optics. The detected-throughput factor is computed as $\tau_{\mathrm{det}}=\tau_{\mathrm{spec}}\tau_{\mathrm{pol}}$. The proposed system has lower raw detected throughput because of the additional multi-band filtering, but maintains comparable focused-signal efficiency and achieves the highest far-depth MTF-AUC. The focused-signal efficiency $\eta_{\mathrm{focus}}$ is the fraction of optical power concentrated within a 10-pixel-radius circular window around the far-depth PSF peak. The far-PSF sharpness is quantified by $A_{\mathrm{MTF}}$, the area under the radially averaged and zero-frequency-normalized MTF curve.}
\label{tab:supp_detected_throughput_and_focusing}
\end{table}

\begin{table}
\centering
\setlength{\tabcolsep}{3pt}
\renewcommand{\arraystretch}{1.12}
\begin{footnotesize}
\makebox[\columnwidth][c]{%
\begin{tabular}{l c c c c}
\toprule
\textbf{Condition} &
\textbf{Exposure / gain} &
\textbf{PSNR (dB) $\uparrow$} &
\textbf{SSIM $\uparrow$} &
\textbf{LPIPS $\downarrow$} \\
\midrule
Nominal &
500 ms / 0 dB &
24.80 &
0.856 &
0.257 \\
Reduced exposure 1 &
250 ms / 6.03 dB &
24.42 &
0.849 &
0.264 \\
Reduced exposure 2 &
175 ms / 9.12 dB &
24.52 &
0.846 &
0.261 \\
Reduced exposure 3 &
92.5 ms / 14.66 dB &
24.37 &
0.838 &
0.263 \\
\bottomrule
\end{tabular}%
}
\end{footnotesize}
\caption{\textbf{Image metric robustness under reduced exposure and increased gain.}
The exposure time was decreased and sensor gain was increased to emulate reduced photon-budget conditions. We further captured 10 scenes, and metrics were computed using the same ground-truth, alignment, and photometric normalization protocol as in Supplementary Note~10. As mentioned in the Supplementary Note~9, the exposure time can be further reduced if the 4f-relay system is removed. The reconstruction network suppresses noise, and the image metric degrades only moderately as the exposure time is reduced.}
\label{tab:supp_exposure_gain}
\end{table}

\clearpage 

\section*{Supplementary Note 16: Comparison with computational inpainting baselines}
\label{supp:note16}
To further compare the proposed optical de-occlusion approach with computational de-occluding methods, we evaluated representative image inpainting baselines on the same obstructed captures used for the image-quality assessment. We considered two categories of computational baselines: a mask-free instruction-guided object-removal model and mask-guided inpainting pipeline. For the mask-free baseline, we used GPT-5.5 Pro\cite{openai2026systemcard}, which can remove obstructions based on a text instruction without an externally provided binary mask. For the mask-guided baseline, we used Inpaint Anything\cite{yu2023inpaint}, which combines segmentation and inpainting models to remove user-specified image regions. 

\begin{figure}[h]
	\centering
		\includegraphics[width=\textwidth]{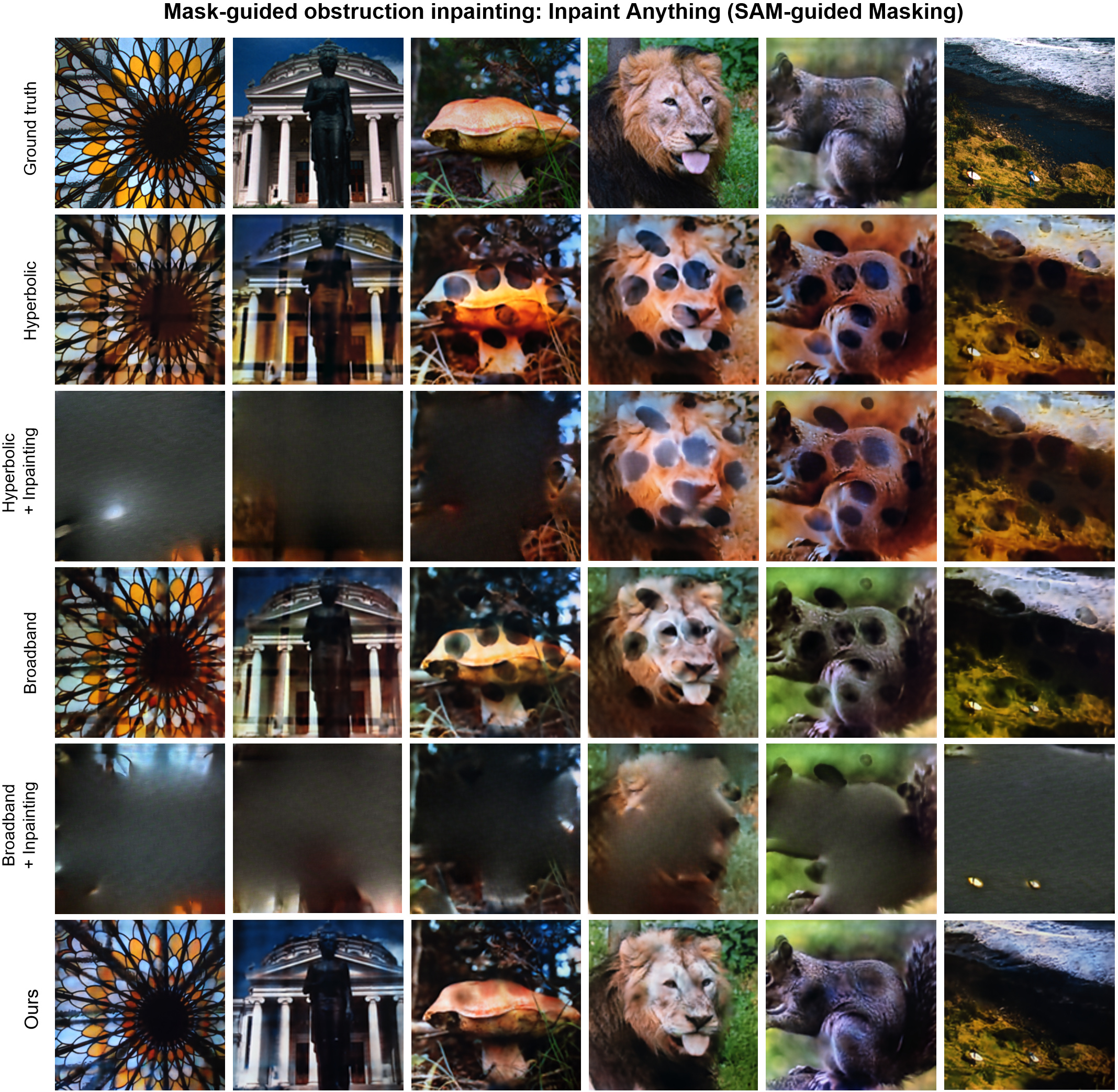}
		\caption{\textbf{Qualitative comparison with mask-guided Inpaint Anything with Segment Anything Model (SAM)-guided mask generation.}
Representative examples comparing the proposed optical de-occlusion approach with the mask-guided Inpaint Anything\cite{yu2023inpaint} pipeline with SAM-guided mask generation. The printed target images are from DIV2K\cite{Agustsson_2017_CVPR_Workshops}.}
\label{fig:supp_inpaint_sam_mask}
\end{figure}

\begin{figure}[h]
	\centering
		\includegraphics[width=\textwidth]{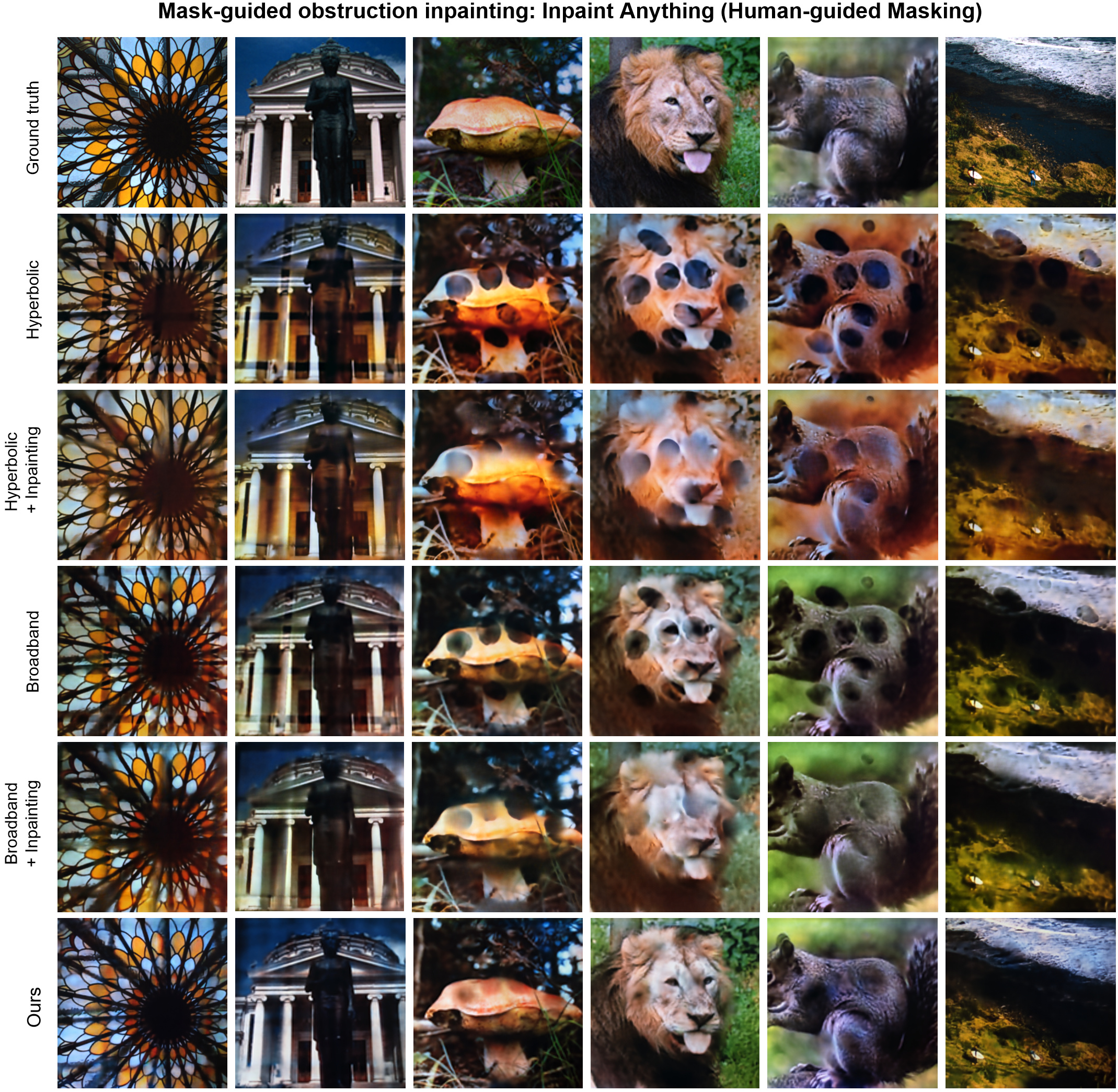}
		\caption{\textbf{Qualitative comparison with mask-guided Inpaint Anything with human-guided mask generation.}
Representative examples comparing the proposed optical de-occlusion approach with the mask-guided Inpaint Anything\cite{yu2023inpaint} pipeline with human-guided mask generation. The printed target images are from DIV2K\cite{Agustsson_2017_CVPR_Workshops}.}
\label{fig:supp_inpaint_human_mask}
\end{figure}

The inpainting baselines were applied to the reconstructed RGB outputs of the Hyperbolic and Broadband metalens captures. For the mask-guided baseline, the obstruction region was provided using human-prompted SAM-generated mask (Figure~\ref{fig:supp_inpaint_sam_mask}) or human-generated mask (Figure~\ref{fig:supp_inpaint_human_mask}). This mask-provided setting favors the inpainting baseline because practical mask-detection errors and mask-generation costs are not included in the image-quality metrics. For the mask-free baseline, we used fixed text instructions for each obstruction type, such as ``Remove the dirt obstruction from the image and keep the background unchanged'' and ``Remove the fence obstruction from the image and keep the background unchanged''.

We used the same full-reference evaluation protocol as in Supplementary Note~10. The compound-lens image without obstruction was used as the ground truth, and PSNR, SSIM, and LPIPS were computed after the same geometric alignment and photometric normalization procedure. Table~\ref{tab:supp_inpainting_comparison} summarizes the quantitative comparison. We also report inference latency for the inpainting baselines to quantify the additional computational cost beyond the lens-specific reconstruction network.

\begin{figure}[h]
	\centering
		\includegraphics[width=\textwidth]{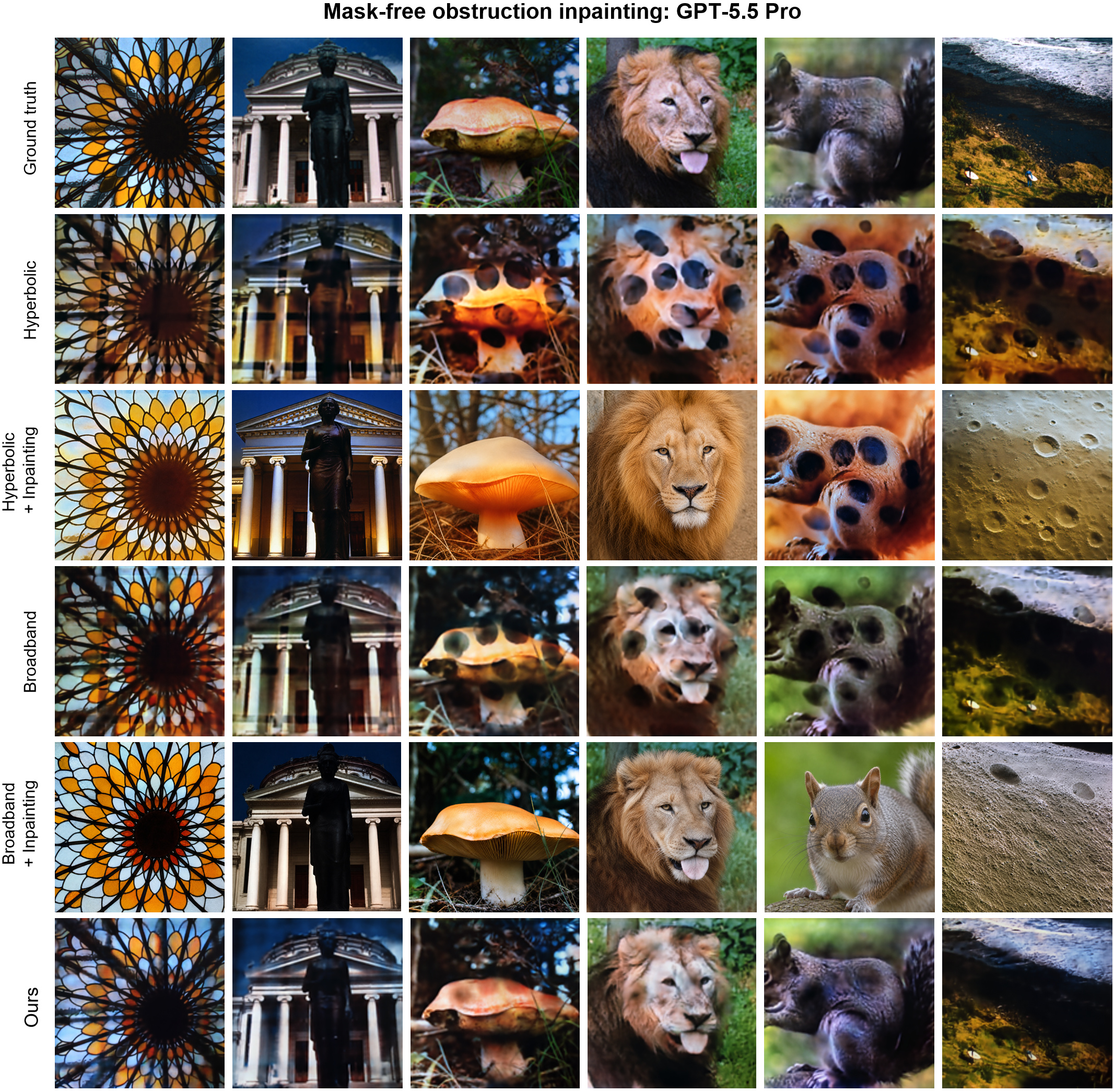}
		\caption{\textbf{Qualitative comparison with mask-free instruction-guided inpainting.}
Representative examples comparing the proposed optical de-occlusion approach with a mask-free instruction-guided inpainting baseline (GPT-5.5 Pro\cite{openai2026systemcard}). The printed target images are from DIV2K\cite{Agustsson_2017_CVPR_Workshops}.}
		\label{fig:supp_inpaint_mask_free}
\end{figure}

Figures~\ref{fig:supp_inpaint_mask_free} and~\ref{fig:supp_inpaint_human_mask} show representative qualitative results. The computational inpainting baselines can attenuate some visible obstruction artifacts, but they often alter scene structures inside or around the removed regions. These artifacts are consistent with the ill-posed nature of inpainting; the method must synthesize information in the masked or edited region when the underlying scene content is not physically measured. In contrast, the proposed de-occluding broadband metalens optically suppresses the near-depth obstruction before reconstruction. This allows the restoration network to recover the underlying scene from the measured optical signal with weaker obstruction imprint, rather than relying on a separate generative completion step. 

\begin{table}
\centering
\setlength{\tabcolsep}{3pt}
\begin{footnotesize}
\begin{tabular}{l c c c c c}
\toprule
\textbf{Input / method} &
\textbf{Mask} &
\textbf{PSNR} &
\textbf{SSIM} &
\textbf{LPIPS} &
\textbf{Latency} \\
 & & \textbf{(dB) $\uparrow$} & \textbf{$\uparrow$} & \textbf{$\downarrow$} & \textbf{(ms) $\downarrow$} \\
\midrule
Hyperbolic + Recon. + GPT-5.5 Pro & No & 13.47 & 0.5210 & 0.5901 & 8675$^\dagger$ \\
Hyperbolic + Recon. + Inpaint Anything (SAM-guided mask) & Yes & 11.50 & 0.5084 & 0.8146 & 189$^\ddagger$ \\
Hyperbolic + Recon. + Inpaint Anything (Human-guided mask) & Yes & 16.70 & 0.6784 & 0.5697 & 152$^\ddagger$ \\
\midrule
Broadband + Recon. + GPT-5.5 Pro & No & 15.76 & 0.4984 & 0.5102 & 8675$^\dagger$ \\
Broadband + Recon. + Inpaint Anything (SAM-guided mask) & Yes & 11.76 & 0.4902 & 0.8094 & 189$^\ddagger$ \\
Broadband + Recon. + Inpaint Anything (Human-guided mask) & Yes & 18.50 & 0.6953 & 0.5151 & 152$^\ddagger$ \\
\midrule
\textbf{Ours + Recon.} & \textbf{No} & \textbf{20.94} & \textbf{0.7560} & \textbf{0.3788} & \textbf{41$^*$} \\
\bottomrule
\end{tabular}
\end{footnotesize}
\caption{\textbf{Quantitative comparison with computational inpainting baselines.}
The inpainting baselines were applied to the reconstructed RGB outputs of the Hyperbolic and Broadband metalens captures. GPT-5.5 Pro\cite{openai2026systemcard} was evaluated as a mask-free instruction-guided baseline using fixed text prompts, while Inpaint Anything\cite{yu2023inpaint} was evaluated as a mask-guided baseline using human-prompted SAM-generated mask obstruction masks. PSNR, SSIM, and LPIPS were computed against the compound-lens ground truth using the same geometric alignment and photometric normalization protocol as in Supplementary Note~10.  *Latency was measured at $512\times512$ resolution on an NVIDIA RTX 6000 Ada GPU, $\dagger$ measured at $848\times848$ resolution on using ChatGPT\cite{openai2026systemcard} with 10 image batched query, and $\ddagger$ measured at $848\times848$ resolution on an NVIDIA RTX 5070 GPU. The mask-guided setting favors the inpainting baseline because practical mask-detection errors and mask-generation costs are not included in the image-quality metrics.}
\label{tab:supp_inpainting_comparison}
\end{table}

\clearpage 

\section*{Supplementary Note 17: Additional real-scene downstream vision demonstrations}
\label{supp:note17}
To complement the printed-benchmark protocol used in Supplementary Note~11, we additionally performed outdoor real-scene acquisitions. The printed-benchmark protocol remains important because it enables controlled optical re-capture of complex application-specific scenes while preserving standard benchmark annotations, which are difficult to obtain in matched form for practical scenarios such as urban, aerial, or endoscopic imaging. In addition, printed targets under broadband illumination provide spectrally suitable scene content for broadband metalens evaluation. However, printed targets do not fully reproduce real 3D scene geometry, environmental clutter, and uncontrolled illumination variations. Therefore, we further included outdoor real-scene captures as a qualitative validation beyond the controlled printed-benchmark setting.

For this additional validation, we captured outdoor scenes containing people, vehicles, and vegetation using the three fabricated metalenses: the Hyperbolic metalens, the Broadband metalens without the de-occluding strategy, and the de-occluding broadband metalens. Near-depth obstructions were placed in front of the metalens path to emulate practical obstruction conditions consistent with the main experiments. We then applied the same image reconstruction network and off-the-shelf semantic segmentation model used in the main text.

\begin{figure}[h]
	\centering
		\includegraphics[width=0.75\textwidth]{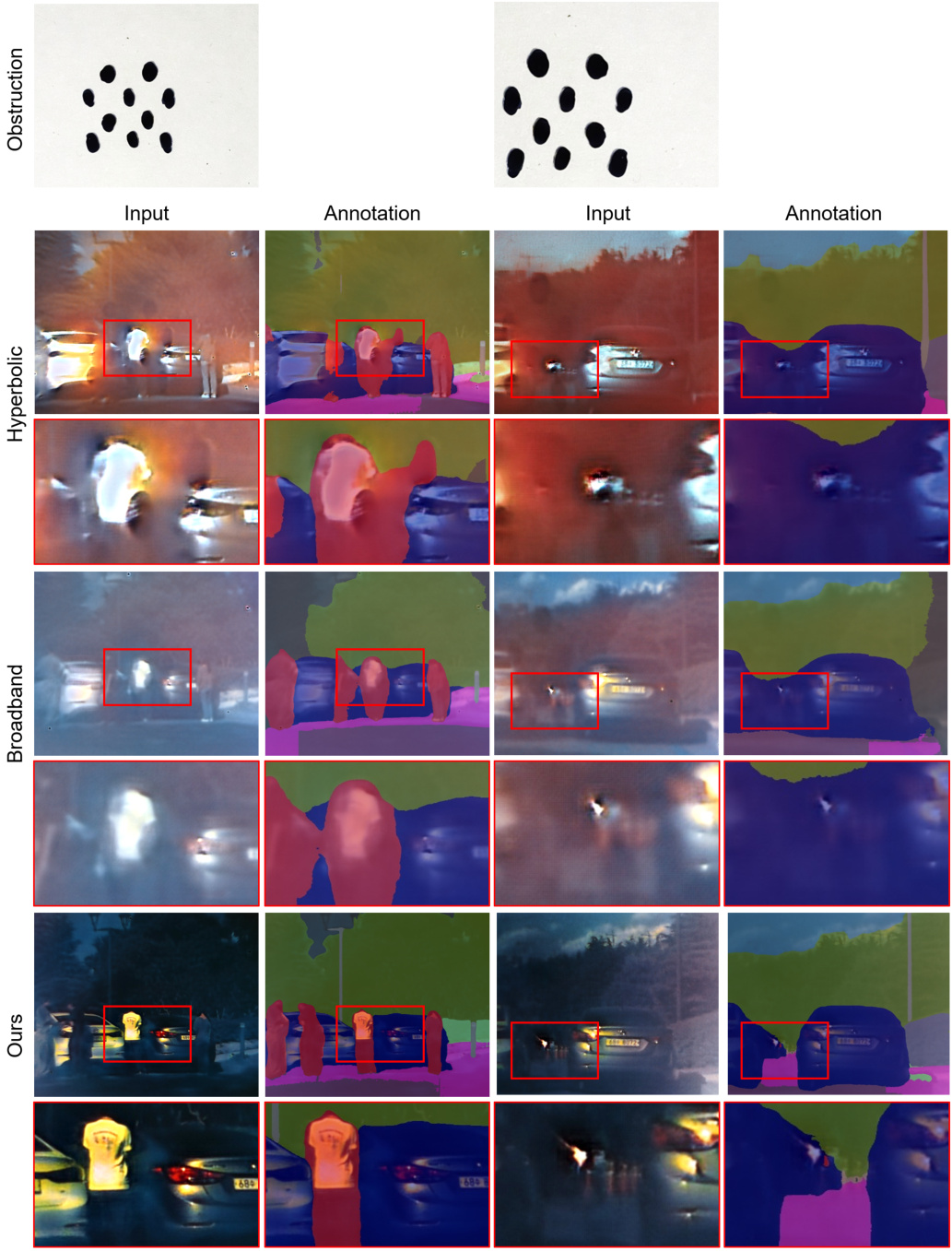}
		\caption{\textbf{Outdoor real-scene downstream vision comparison under practical conditions.}
For each metalens, we show the reconstructed input image and the corresponding semantic segmentation prediction from an off-the-shelf model. Compared with the Hyperbolic and Broadband baseline metalenses, the de-occluding broadband metalens more effectively suppresses obstruction traces and preserves semantically meaningful scene structures, leading to qualitatively more reliable semantic segmentation predictions. These outdoor real-scene results complement the controlled printed-benchmark evaluations in Supplementary Note~11. Ambient illumination and stray-light conditions could not be controlled as strictly as in the benchtop printed-target benchmark and degraded the image quality.}
        \label{fig:supp_outdoor_vision_task}
\end{figure}

Representative results are shown in Figure~\ref{fig:supp_outdoor_vision_task}. Across these outdoor examples, the de-occluding broadband metalens more effectively suppresses near-depth obstruction traces and preserves semantically meaningful scene structures compared with the Hyperbolic and Broadband baseline metalenses. As a result, the semantic segmentation predictions obtained from the proposed system are qualitatively more stable and more consistent with the visible scene content.

\begin{figure}[h]
	\centering
		\includegraphics[width=0.9\textwidth]{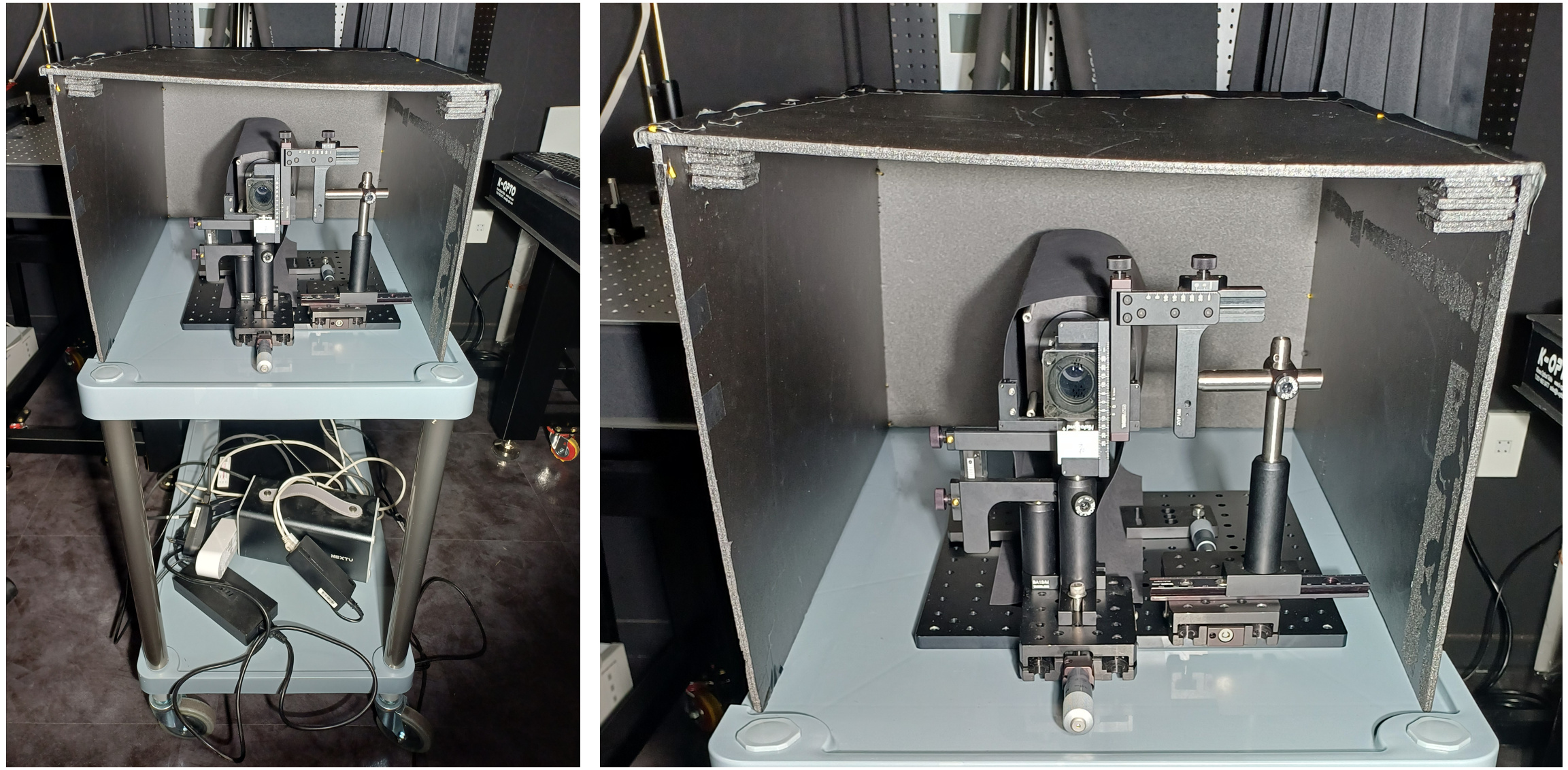}
		\caption{\textbf{Setup used for the outdoor real-scene comparison capture.}
Photographs of the portable shielded capture setup used for the outdoor real-scene semantic-segmentation demonstration. The setup allowed sequential acquisition through the Hyperbolic, Broadband, and de-occluding broadband metalenses while maintaining the near-depth obstruction configuration in front of the metalens path. The shielded enclosure reduced stray ambient light, but complete ambient-light blocking was not possible because the metalenses had to be exchanged sequentially and the optical path had to remain accessible during outdoor acquisition. Therefore, the outdoor captures are used as qualitative practical demonstrations, while the controlled printed-target experiments remain the primary quantitative benchmark.}
		\label{fig:supp_outdoor_vision_task_setup}
\end{figure}

The outdoor capture setup is shown in Figure~\ref{fig:supp_outdoor_vision_task_setup}. Because the three metalenses were exchanged sequentially in an outdoor acquisition environment, ambient illumination and stray-light conditions could not be controlled as strictly as in the benchtop printed-target benchmark. In particular, the need to exchange metalenses while maintaining access to the optical path prevented complete ambient-light sealing, which can reduce image contrast and make the captured images appear noisier or less uniform than the controlled benchmark captures. For this reason, the outdoor real-scene results are intended as qualitative practical validation, whereas the printed-benchmark experiments remain the primary quantitative downstream-vision evaluation.

\begin{figure}[h]
	\centering
		\includegraphics[width=0.9\textwidth]{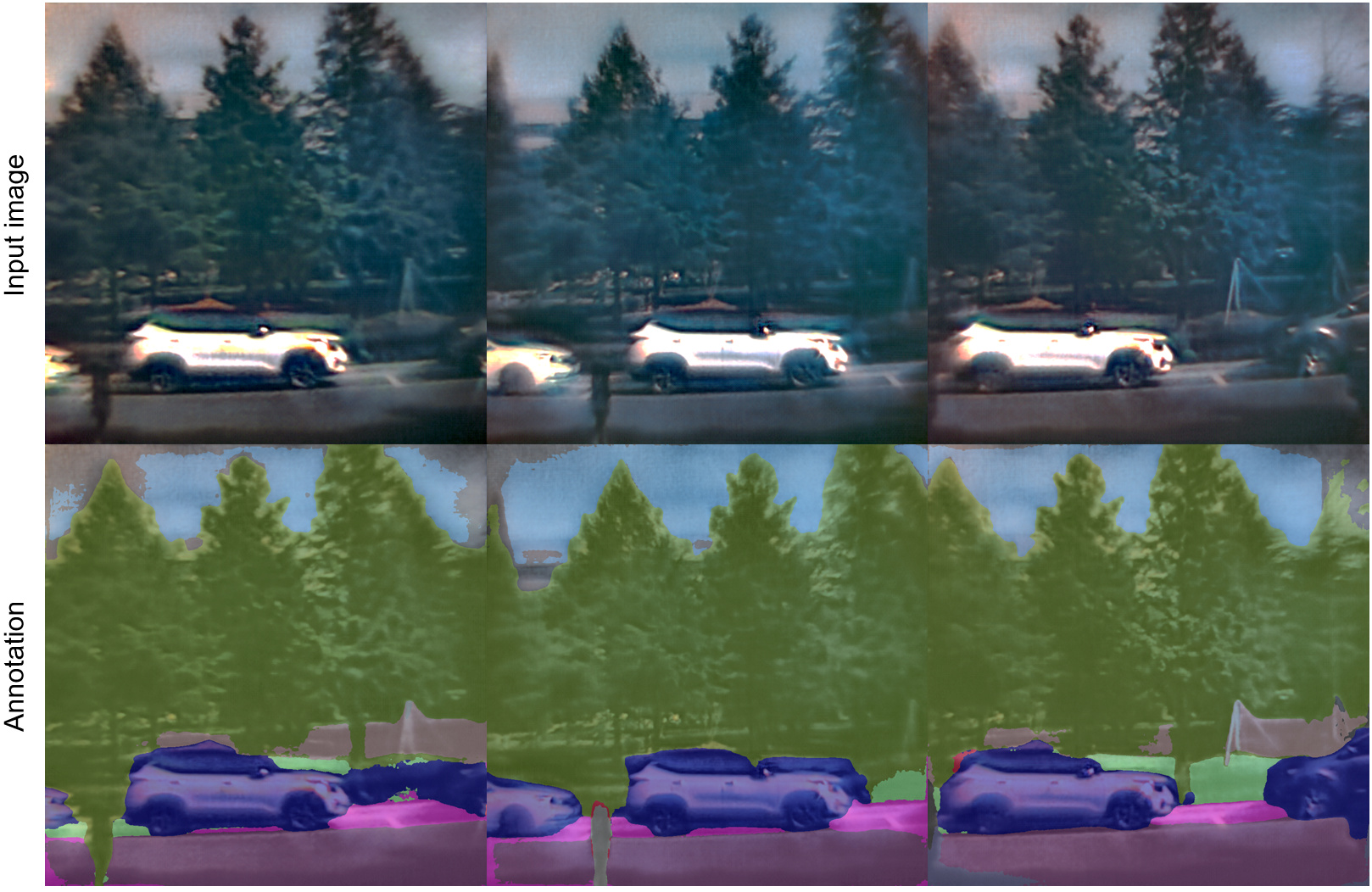}
		\caption{\textbf{Outdoor image capture and vision results using the compact prototype without obstruction.}
Representative outdoor scenes capture and vision task results using the compact prototype with an ambient-light-shielded housing.}
		\label{fig:supp_compact_camera_vision_clean}
\end{figure}

\begin{figure}[h]
	\centering
		\includegraphics[width=0.9\textwidth]{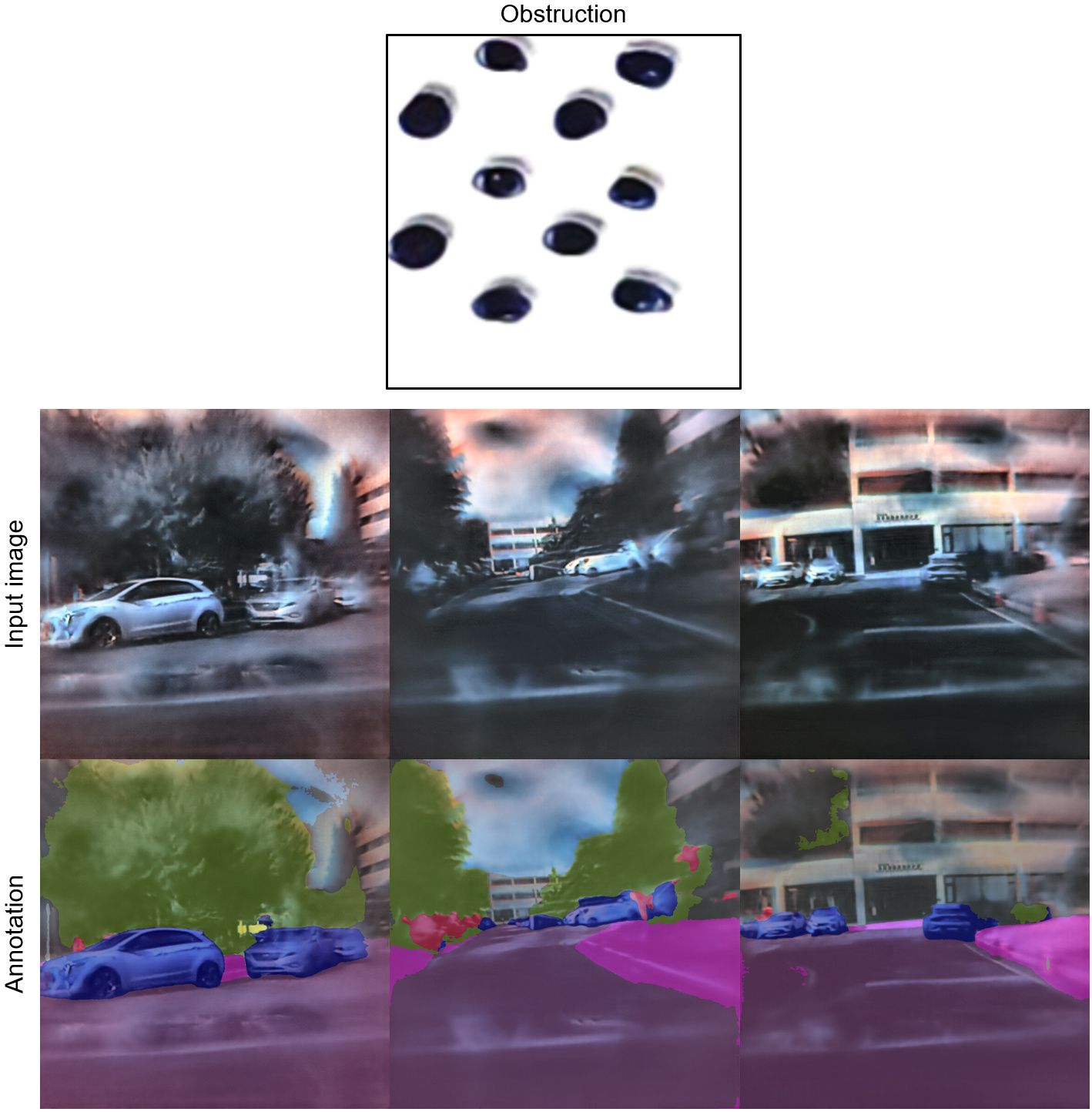}
		\caption{\textbf{Outdoor image capture and vision results using the compact prototype with obstruction.}
Representative outdoor scenes capture and vision task results using the compact prototype with an ambient-light-shielded housing.}
		\label{fig:supp_compact_camera_vision_obstructed}
\end{figure}

\clearpage 

\bibliographystyle{naturemag}
\bibliography{references}